\documentclass[11pt]{article}

\usepackage[margin=0.75in]{geometry}

\usepackage{amsmath, amssymb, bigints, bm}
\usepackage{graphicx,psfrag,epsf}
\usepackage{color, mathtools, graphics}
\usepackage{enumerate}
\usepackage{booktabs}
\usepackage{natbib}
\usepackage{url} % not crucial - just used below for the URL 
\usepackage[font=small,skip=0pt]{caption}

\newcommand{\bfT} {\mbox{\boldmath $\Theta$}}
\newcommand{\bft} {\mbox{\boldmath $\theta$}}

\newcommand{\bfe} {\mbox{\boldmath $\epsilon$}}

\newcommand{\bfmu} {\mbox{\boldmath $\mu$}}

\DeclareMathSymbol{\Theta}{\mathord}{operators}{"02}

\newcommand{\hsig} { h_\sigma }
\newcommand{\nOutcome}{  D }
\newcommand{\FactorLoadings} {\mbox{\boldmath $\Gamma$}}
\newcommand{\LowRankDiagCov} {\mbox{\boldmath $\Psi$}}
\newcommand{\nLatentFactor} {\mbox{$B$}}
\newcommand{ \SubjectLatentFactor} {\mbox{\boldmath $\eta$}}
\newcommand{ \subjectLatentFactor} {\mbox{ $\eta$}}
\newcommand{ \RedundantTerm} {\mbox{\boldmath $\Xi$}}
\newcommand{ \redundantTerm} {\mbox{ $\xi$}}
\newcommand{ \concParLagPDPM }{ {\color{black} \alpha_1}}

\newcommand{ \concParLagrgPDPM }{ {\color{black} \alpha_{1}}}
\newcommand{ \clustIndLagPDPM }{ {\color{black} {h_1} }}

\newcommand{ \clustIndLagrgPDPM }{ {\color{black} {h_{1,d'} } }}
\newcommand{ \nuLagPDPM }{ {\color{black} \nu_{\clustIndLagPDPM}}}

\newcommand{ \nuLagrgPDPM }{ {\color{black} \nu_{\clustIndLagrgPDPM}}}
\newcommand{ \nuAltLagPDPM }{ {\color{black} \nu_{l_1}}}
\newcommand{ \clustMembLagPDPM }{ {\color{black} H_1}}

\newcommand{ \clustMembLagrgPDPM }{ {\color{black} H_{1d'} }}
\newcommand{ \SetclustMembLagPDPM }{ {\color{black} \mathcal{H}_{\clustIndLagPDPM} }}

\newcommand{ \SetclustMembLagrgPDPM }{ {\color{black} \mathcal{H}_{\clustIndLagrgPDPM} }}
\newcommand{ \clustMembSubjLagPDPM }{ {\color{black} H_{1,i} }}

\newcommand{ \clustMembSubjrgLagPDPM }{ {\color{black} H_{1d',i} }}
\newcommand{ \piLagPDPM }{ {\color{black} \pi_{\clustIndLagPDPM}}}

\newcommand{ \piLagrgPDPM }{ {\color{black} \pi_{d',\clustIndLagrgPDPM}}}
\newcommand{ \altpiLagrgPDPM }{ {\color{black} \pi_{d', \clustIndLagrgPDPM'}}}
\newcommand{ \piSubjClustLagPDPM }{ {\color{black} \pi_{\clustMembSubjLagPDPM}}}

\newcommand{ \piSubjClustLagrgPDPM }{ {\color{black} \pi_{d', \clustMembSubjrgLagPDPM}}}
\newcommand{\uiLagPDPM}{ \color{black} u_{1, i} }

\begin{document}

\title{ Flexible Bayesian Product Mixture Models for Vector Autoregressions}

%\thanks{
%    The authors gratefully acknowledge support from NIH awards R01AG071174 and R01MH120299}
\title{\bf Non-parametric Bayesian Vector Autoregression using Multi-subject Data}
  \author{Suprateek Kundu\thanks{
    The authors gratefully acknowledge support from NIH awards R01AG071174 and R01MH120299}\hspace{.2cm}\\
    Department of Biostatistics, The University of Texas MD Anderson Cancer Center\\
    and \\
    Joshua Lukemire \\
    Department of Biostatistics and Bioinformatics, Emory University}
  \maketitle

\begin{abstract}%   <- trailing '%' for backward compatibility of .sty file
Bayesian non-parametric methods based on Dirichlet process mixtures have seen tremendous success in various domains and are appealing in being able to borrow information by clustering samples that share identical parameters. However, such methods can face hurdles in heterogeneous settings where objects are expected to cluster only along a subset of axes or where clusters of samples share only a subset of identical parameters. We overcome such limitations by developing a novel class of product of Dirichlet process location-scale mixtures that enable independent clustering at multiple scales, which result in varying levels of information sharing across samples. First, we develop the approach for independent multivariate data. Subsequently we generalize it to multivariate time-series data under the framework of multi-subject Vector Autoregressive (VAR) models that is our primary focus, which go beyond parametric single-subject VAR models. We establish posterior consistency and develop efficient posterior computation for implementation. Extensive numerical studies involving VAR models show distinct advantages over competing methods, in terms of estimation, clustering, and feature selection accuracy. Our resting state fMRI analysis from the Human Connectome Project reveals biologically interpretable connectivity differences between distinct intelligence groups, while another air pollution application illustrates the superior forecasting accuracy compared to alternate methods.
\end{abstract}

\noindent%

\section{Introduction} \label{sec:intro}

Multivariate time-series data routinely arise in diverse application areas such as finance \citep{cramer1978multivariate}, econometrics \citep{engle1981one}, air pollution forecasting \citep{nath2021long} and medical imaging \citep{kundu2021scalable}, among other domains. %In such paradigms, standard multivariate techniques that are designed for independent and identically distributed (i.i.d.) outcomes may not be suitable. 
In order to tackle such data, a rich body of work on modeling autocorrelations and temporal cross-correlations between variables with multivariate outcomes has been developed, of which vector autoregressive (VAR) models are widely used \citep{lutkepohl2005new}. Our focus in this paper is on Bayesian VAR modeling, which was initially heavily motivated by econometric research \citep{doan1984forecasting}, and has since seen a rich development \citep{korobilis2013var}. %while VAR models incorporating variable selection were developed by \cite{korobilis2013var} that induced sparsity in the transition matrix via indicator variables and by \cite{gefang2014bayesian} who induced shrinkage on the elements of the autocovariance matrix via Bayesian Lasso approach based on doubly adaptive elastic-net prior. A class of Bayesian VAR models relying on graphical prior distributions were proposed by  \cite{wang2010sparse} and \cite{ahelegbey2016bayesian, ahelegbey2016sparse}.  
More recently, Bayesian VAR models have been adopted with increasing prominence in biomedical research including patient-level predictive modeling  \citep{lu2018bayesian} and {\color{black}functional Magnetic Resonance Imaging (fMRI) applications} \citep{gorrostieta2013hierarchical, chiang2017bayesian} in neuroimaging studies. However, existing Bayesian VAR literature has primarily focused on methodological and computational developments, with limited theoretical investigations. Recently, \cite{ghosh2018high} addressed this gap by establishing posterior consistency for the autocovariance matrix in  parametric Bayesian VAR models based on single subject data.

The vast majority of the Bayesian VAR literature involves Gaussian assumptions and parametric prior specifications that may not be sufficiently flexible in characterizing the underlying probability distributions with non-regular features. For example, it is known that the nature of shocks in econometric analysis may not always be Gaussian \citep{weise1999asymmetric}. Similarly, flexible VAR modeling is necessary for analyzing heterogeneous multi-subject data in neuroimaging studies, where parametric VAR models may prove inadequate (see our Human Connectome Project (HCP) application in Section 6). Non-Gaussianity is also observed in air pollution data captured via sensors \citep{kim2013sensor}, where it is often of interest to perform forecasting using VAR models \citep{hajmohammadi2021multivariate}.  {\color{black}Such parametric VAR models may result in inaccurate performance when parametric assumptions are violated or even mis-specified.} To bypass parametric constraints in VAR models, some recent articles relaxed Gaussianity assumptions \citep{jeliazkov2013nonparametric}. Recently, Bayesian nonparametric VAR models were proposed by \cite{kalli2018bayesian} involving single subject data, where the mixing weights of the transition density depend on the previous lags. On the other hand, \cite{billio2019bayesian} proposed Dirichlet process mixture of normal-Gamma priors on the VAR autocovariance elements. Unlike for the parametric case, the non-parametric methods are more robust to mis-specification and can potentially cater to a large class of models. However, the above approaches were applied to small or moderate dimensional data with limited or no emphasis on pooling information across samples and with negligible or no theoretical investigations.

Existing literature has largely ignored the problem of developing provably flexible non-parametric Bayesian VAR methodology to model heterogeneous multi-subject time-series data, to our knowledge. Such approaches are desirable over single-subject VAR analyses in terms of being able to pool information across samples in a flexible manner that can accommodate arbitrary probability distributions. It also facilitates robust and reproducible parameter estimates and provides a natural foundation to conduct inferences to test for differences across samples via credible intervals, which may not be straightforward under single-subject analysis. Although there is some literature on parametric VAR modeling of multi-subject data, these existing approaches typically require {\it apriori} knowledge of class labels \citep{gorrostieta2013hierarchical, chiang2017bayesian, kook2021bvar}. Hence, they have a limited ability to accommodate heterogeneity within each class and may result in poor performance when the class labels are mis-specified due to no clear distinction between groups. {\color{black} Moreover, they clearly suffer from the aforementioned pitfalls of parametric methods.} %{\color{black} In general, parametric models may often be susceptible to mis-specifications, that are often naturally bypassed in non-parametric modeling.}

 Motivated by the above discussions, we propose a broad class of novel Bayesian non-parametric models that specify Dirichlet process (DP) mixture priors independently on mutually exclusive subsets of model parameters. Our specification results in a product of Dirichlet process mixture (PDPM) priors. %that is related to the product of experts model \citep{hinton2002training} in the machine learning literature. 
 A key feature of the proposed approach is the ability to allow differential clustering at multiple scales, which enables clusters of samples that share only a subset of common model parameters resulting in greater flexibility. We develop several variants of the proposed approach that encourage differential degrees of heterogeneity via different modes of multiscale clustering by altering the manner in which the parameter space is partitioned. First, we develop the PDPM approach in the generic setting of multivariate density estimation for kernel mixtures of the form $\int K(x; \Theta)dP(\Theta)$, and establish posterior consistency properties. We also provide a toy example that illustrates the distinct numerical advantages of the product mixture models compared to a traditional DP mixtures in terms of clustering accuracy. Subsequently, we generalize the proposed PDPM approach to multivariate time-series data under the framework of a VAR model, which is our primary focus in this article. In such settings,  the multiscale clustering approach becomes even more relevant given a large number of parameters in the autocovariance matrix whose dimension grows quadratically with the vector dimension. Starting from a VAR model that allows for limited differences in clustering across multiple scales and greater model parsimony, we eventually develop a variant that is able to independently cluster row-specific parameters, which provides greater flexibility in practical applications. By specifying appropriate base measures in the DP prior, it is possible to enable appropriate shrinkage for the autocovariance elements that facilitates feature selection. Additional dimension reduction is also possible via a low rank representation for the residual covariance.

By designing non-parametric Bayesian VAR models based on heterogeneous multi-subject data, we are able to relax the parametric assumptions and provide a more flexible characterisation of heterogeneity via unsupervised clustering. The proposed methods are particularly desirable in terms of being able to bypass any restrictive assumptions such as the presence of replicated samples, which is routinely assumed in Bayesian non-parametric literature  \citep{tokdar2006posterior, durante2017nonparametric}. Replicated samples are structured to share fully identical sets of model parameters within a given cluster, which may not be realistic in applications where heterogeneous samples are often effectively clustered only along a subset of directions with the remaining axes being uninformative/redundant for clustering \citep{agrawal2005automatic}. The assumption of replicated samples may also be inadequate to characterize heterogeneity when there is reason to suspect differential clustering at different scales at the level of mean and covariance parameters. {\color{black} Another appealing feature of the proposed approach is the associated posterior consistency properties for density estimation, as the number of samples ($n$) grow to $\infty$. %The proposed non-parametric Bayesian modeling results in greater flexibility by establishing  theoretical guarantees for a wider class of true models and bypasses issues related to model mis-specification that may arise in parametric settings.
We note that such theoretical results for VAR models involving multivariate time-series data represent non-trivial extensions of the rich theoretical properties established in the Bayesian non-parametric literature for independent multivariate and scalar outcomes \citep{tokdar2006posterior, canale2017posterior}.  We resolve the significant challenges arising from the non-parametric Bayesian theoretical analysis by establishing Kullback-Leibler properties for VAR models, and constructing carefully designed sieves that are shown to satisfy certain entropy bounds and tail prior probability conditions under the product of DP priors. Moreover, we show that the theoretical results hold for commonly used base measures that enable straightforward posterior computation.

% The latter setting may be better suited for time-series data with higher temporal resolution that is not applicable for fMRI applications of interest. 

We develop an efficient and scalable Markov chain Monte Carlo (MCMC) implementation for the proposed class of models in the VAR framework}. In addition, we illustrate the sharp numerical advantages under the proposed non-parametric Bayesian VAR approach in terms of accurately recovering the true model parameters, inferring the sparsity structure for the autocovariance matrix, and identifying the true clustering patterns, compared to competing state-of-the-art methods. Our analysis of resting state fMRI data from a subset of individuals in the HCP study infers several effective connectivity differences between the high and low fluid intelligence groups that are supported by existing evidence in literature. Moreover, the analysis under the proposed approach produces biologically reproducible estimates that are consistent across repeated neuroimaging scans from the same samples. In contrast, a single subject VAR analysis is able to identify only one effective connectivity difference across groups, which seems biologically implausible. {\color{black} Using a second data application example involving air pollution data from the Environment Protection Agency (EPA), we illustrate the considerable advantages in forecasting accuracy under the proposed approach compared to a parametric VAR model even when the dimension of the outcome is small.}

 The rest of the article is structured as follows. {\color{black} Section 2 develops the product of DP mixtures for independently distributed multivariate data and establishes posterior consistency properties. Section 3 extends the methodology to multivariate time-series data under a VAR framework, along with illustrating theoretical properties. Section 4 describes the posterior computation scheme, Section 5 reports results from extensive simulation studies involving VAR models. Sections 6 and 7 describe our analysis of the neuroimaging data from the HCP as well as air pollution data from the EPA. Section 8 contains additional discussions. Appendices are provided that contain other relevant details.}

%
% Methods
%

\section{Product of DP Mixtures for Multivariate Data}
\subsection{A Primer on DP mixture approaches}
Consider i.i.d. random vectors ${\bf x}_i,i,=1,\ldots,n,$ each of dimension $D\times 1$, and denote the collection of vectors as $X_n=\{ {\bf x}_1,\ldots,{\bf x}_n\}$. Non-parametric Bayesian literature has often focused on modeling these vectors under a DP location mixture or location-scale framework. Such approaches \citep{escobar1995bayesian} often specify ${\bf x}_i\sim N({\bfmu}_i, \Sigma_i), ({\bfmu}_i, \Sigma_i)\sim P, P\sim DP(MP_0), i=1,\ldots,n,$ where $\Sigma_{i}\in S_{\nOutcome\times \nOutcome}$ denotes the covariance  for subject $i$ and  $S_{\nOutcome\times \nOutcome}$ denotes the space of all $\nOutcome \times \nOutcome$ symmetric positive definite matrices, $P_0$ denotes the base measure of the DP and $M$ is the precision parameter. We note that alternative choices other than the Gaussian kernel may also be used but are not considered here for simplicity. The resulting DP location-scale mixture induces the unknown probability density $f_P({\bf x})=\int \phi_\Sigma({\bf x} - \bfmu) dP({\bfmu}, \Sigma)$, where  $\phi_{\Sigma}(\cdot - \bfmu)$ denotes the density of a $\nOutcome$-dimensional normal distribution with mean $\bfmu$ and covariance $\Sigma$. Given that $P\sim DP(MP_0)$, the proposed method results in probability distributions on the class of densities $\mathcal{F}=\{ f_P\}$, which can also be seen from the result $f_P({\bf x})=\sum_{h=1}^\infty \pi_h \phi_{\Sigma_h} \big({\bf x} - \bfmu_h \big), $ where $(\bfmu_h,\Sigma_h)\sim P_0$ and $\pi_h=\nu_h\prod_{l=1}^{h-1}(1-\nu_h), \nu_h\sim Be(1,M),$ using Sethuraman's (1994) stick-breaking representation.

The above commonly used DP mixture specification results in clusters of replicated samples that share identical sets of parameters $({\bfmu}, \Sigma)$, which allows for pooling of information across samples resulting in robust learning. While such a clustering mechanism is often backed by posterior consistency guarantees and typically has good numerical performance for moderate to large dimensions, it may not be well-equipped to succeed for more heterogeneous settings where the clustering is dictated by a small number of axes or subspaces, with the other axes being irrelevant to clustering. In such settings, the clustering accuracy under typical DP mixture models may be compromised resulting in inferior performance for practical applications. A more flexible approach is to allow differential clustering at multiple scales that does not constrain samples to share fully identical parameter sets, but instead allows subsets of parameters to cluster independently resulting in partially overlapping clusters. Such a multi-scale clustering approach results in a more accurate characterization of heterogeneity that is expected to improve finite sample performance. The above arguments form the basis of the proposed product mixture priors in this article.

\subsection{Proposed Methodology and Properties}
We propose a class of novel product mixture priors that is equipped to perform differential clustering at multiple scales. Consider equally sized mutually exclusive and exhaustive subsets of the full parameter set denoted as  $\bfmu=\cup_{m_1=1}^{\mathcal{M}_{\mu}} \bfmu_{m_1},\mbox{ } {\bm\tilde{\bm\sigma}} = \cup_{m_2=1}^{\mathcal{M}_{\sigma}} {\bm\sigma_{m_2}}$, where ${\tilde{\bm\sigma}}$ denotes the vectorized upper triangular matrix of $\Sigma$, and  $\{\bfmu_1,\ldots,\bfmu_{\mathcal{M}_\mu} \}$ represent subsets of equal cardinality, and similarly for $\{{\bm\sigma_{1}},\ldots,{\bm\sigma_{\mathcal{M}_\sigma}} \}$. Consider specifying the following product of DP priors on the parameters:
\begin{eqnarray}
&&\bfmu_{m_1}\stackrel{indep}{\sim} P_\mu, \mbox{ } m_1=1,\ldots, \mathcal{M}_{\mu}, \mbox{ } {\bm\sigma_{m_2}} \stackrel{indep}{\sim} P_{\sigma}, \mbox{ } m_2=1,\ldots, \mathcal{M}_{\sigma}, \mbox{ } \Sigma\in S_{\nOutcome\times\nOutcome},\nonumber \\
&&P_\mu\sim DP(\alpha_1P^*_1), \mbox{ } P_{\sigma}\sim DP(\alpha_2 P^*_2), \label{eq:productDP}
\end{eqnarray}
where each component is assigned independent priors $\bfmu_{m_1}\stackrel{indep}{\sim} P_\mu,{\bm\sigma_{m_2}} \stackrel{indep}{\sim} P_{\sigma},$ that follow Dirichlet process with base measures $P^*_1$ and $P^*_2$ respectively, with corresponding precision parameters $\alpha_1,\alpha_2$. %We note that the subsets \{$\bfmu_{1},\ldots, \bfmu_{\mathcal{M}_\mu}$ \} are required to have equal size for specification (\ref{eq:prodDP}) to hold. A similar requirement holds for subsets \{${\bm\sigma}_{1},\ldots, {\bm\sigma}_{\mathcal{M}_\sigma}$\}. 
The specification (\ref{eq:productDP}) results in a product of DP priors on the original parameters $(\bfmu,\Sigma)$ that is denoted by $\Pi^*$, where the exact prior depends on the way in which the partitions are defined. Hence, one can obtain a class of product mixture priors by tweaking the partition structure to reflect the most appropriate setting for the data at hand. The product priors in (\ref{eq:productDP}) induce priors $\Pi$ on $\mathcal{F}$ via the relationship:
\begin{align}
&f_{P}({\bf x}) = \int \int \phi_{\Sigma}\big({\bf x} - \bfmu\big) d\Pi^*(\bfmu,\Sigma) \\ %= \int \int \phi_{\Sigma}\big({\bf x} - \bfmu\big) \prod_{m_1=1}^{\mathcal{M}_\mu}\prod_{m_2=1}^{\mathcal{M}_\sigma} \Pi^*(\bfmu_{m_1},{\bm\sigma_{m_2}}), \nonumber \\
&= \sum_{h_{11}=1}^\infty \ldots \sum_{h_{1\mathcal{M}_\mu}=1}^\infty\sum_{\hsig=1}^\infty \pi_{1,h_{11}}\ldots \pi_{\mathcal{M}_{\mu},h_{1 \mathcal{M}_\mu}}\pi_{\sigma,\hsig}\prod_{t=1}^T \phi_{\Sigma_{\hsig}}\bigg({\bf x} -  ({\bm\mu}^T_{1,h_{11}},\ldots,{\bm\mu}^T_{\mathcal{M}_\mu,h_{1\mathcal{M}_\mu}})^T \bigg), 
\label{eq:productDPM}
\end{align}
where the second equality is obtained by Sethuraman's (1994) stick breaking representation  with $ \pi_{h_{1m_1}}=\nu_{h_{1m_1}}\prod_{l_1<h_{1m_1}}(1-\nu_{l_1})$, $\nu_{l_1}\sim Be(1,\alpha_1)$,  $\pi_{\sigma,\hsig}=\nu_{\sigma,\hsig}\prod_{l_2<\hsig}(1-\nu_{\sigma,l_2})$, $\nu_{\sigma,\hsig}\sim Be(1,\alpha_2)$, and further $\small \Sigma_{\hsig} \sim P^{*}_2, \mbox{ } {\bm \mu}_{m_1} \sim P^{*}_1$,  and $\mathcal{M}_\sigma$ is assumed to be one in the above expression. The choice of $\mathcal{M}_\sigma=1$ is guided by practical considerations in VAR models that is our primary focus (next section) where the residual covariance matrix often has a sparse or even diagonal structure after regressing out the lag effects of previous time points. However our treatment can be generalized to $\mathcal{M}_\sigma>1$ in a straightforward manner. We note that the above form in (\ref{eq:productDPM}) follows the generic kernel mixture representation $\int K(x; \Theta)dP(\Theta)$ that is commonly considered in non-parametric Bayesian density estimation literature \citep{wu2008kullback}. We denote the resulting class of priors on $\mathcal{F}$ arising from (\ref{eq:productDPM}) as the product of Dirichlet process mixtures (PDPM).

\begin{figure}
\centering
\includegraphics[width=1.0\linewidth, height=2.5in]{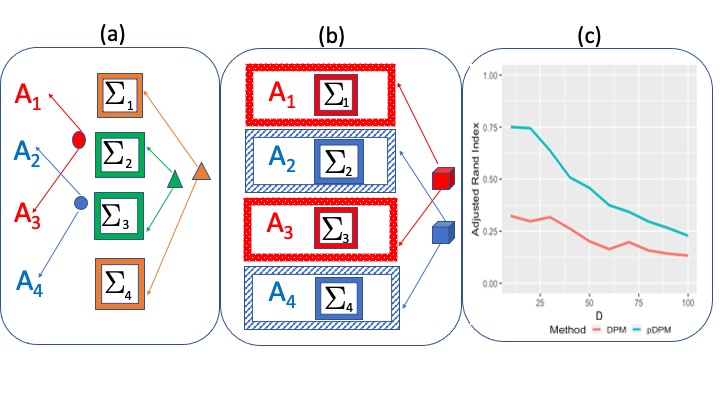}
\caption{\footnotesize{A schematic representation of the product of DP mixture prior. Panel (a) illustrates the product prior that separately clusters the mean (represented by A) into red and blue clusters and the covariance ($\Sigma$) into green and saffron clusters. Panel (b) represents the traditional DP mixture prior that forms clusters containing samples having identical values for both the mean and covariance parameters. Panel (c) illustrates the results from the toy example under the traditional DPM and the proposed PDPM, in terms of the change in clustering accuracy with varying dimension when the clustering is dictated by a subset of axes.}}
\label{fig:schema}
\end{figure}

The most straightforward case of the prior in (\ref{eq:productDP}) that can be applied in practice is given as $\Pi^*(\bfmu,\Sigma) = P_{\mu}(\bfmu)\times P_{\sigma}(\Sigma)$, which specifies independent priors on the mean and covariance parameters without further partitioning these parameters (i.e. $\mathcal{M}_\mu=1,\mathcal{M}_\sigma=1$). The PDPM operates by clustering the mean and covariance parameters independently, which suggests separate modes of pooling information across samples for the mean and covariance. Such a multiscale clustering approach results in greater flexibility and a more accurate characterization of heterogeneity by allowing for dedicated clusters of samples that share common mean signatures (but not necessarily for the covariance), along with independently constructed subgroups of samples that share common patterns in the covariance (but not necessarily for the mean). The proposed mode of information learning via multiscale clustering is clearly more advantageous compared to the standard DPM priors, i.e. $(\bfmu,\Sigma)\sim P, P\sim DP(M P_0)$ that constrain clusters of samples to share identical mean {\it and}  covariance parameters, which may result in spurious clusters when subgroups of samples only share subsets of common parameters instead of having fully identical parameter sets. As a more flexible generalization, one can consider the {\it generalized PDPM (gPDPM)} model that specifies independent DP priors for each element of the mean vector, i.e. $\Pi^*(\bfmu,\Sigma) = \prod_{m=1}^{D}P_{\mu}(\mu_m)\times P_{\sigma}(\tilde{\bm\sigma})$. The gPDPM approach allows separate clustering for each element of $\bfmu$ across samples, which enables differential clustering along various axis and provides a more granular approach for pooling information. In addition, the covariance parameters are clustered independently from the mean as before. Such an approach is expected to excel in settings where the clustering is dictated by a subset of axes in the mean with the other axes being redundant towards clustering, which may often be the case in practical applications. The above discussions highlight the advantages of the multiscale clustering aspect under the proposed product mixture modeling methodology, and provides the central motivation for this article. A schematic representation of ideas is presented in Figure \ref{fig:schema}. The left panel (Figure \ref{fig:schema}a) illustrates the clustering mechanism under the PDPM approach that clusters the mean and covariance parameters independently, while the middle panel (Figure \ref{fig:schema}b) illustrates the clustering under the standard DPM approach that allocated fully identical parameter sets to clusters of samples.

{\noindent \underline{\bf Toy Example:}} We illustrate the advantages of the multiscale clustering approach using a toy example. Multivariate data  $Y_i \sim N_{D}(\bfmu_i, \Sigma_i)$, for $i = 1, \ldots, 250$, was generated such that the mean across samples where identical except the first $d$ elements, where  $d \approx \frac{D}{3}$. For the first $d$ elements of $\bfmu$, there were 5 clusters, each with a corresponding $d-$vector of $\bfmu$ values. Similarly, 5 clusters were generated for  $\Sigma(D\times D)$ that were constructed independent of the mean. We used the standard DPM and the proposed PDPM to fit these data, and evaluate the clustering performance across varying dimensions. The posterior computation steps for both approaches are just simplified versions of the posterior computation for the PDPM-VAR model that will be introduced in the sequel, and so they are omitted here for conciseness. For each method, we evaluated the adjusted Rand index for clustering the mean vector at each MCMC iteration, and the corresponding performance measure is the average adjusted Rand index \citep{rand1971objective} across all MCMC iterations. The third panel (Figure \ref{fig:schema}c) illustrates the clustering accuracy for this toy example.  Unsurprisingly, the PDPM offers significant improvement over the DPM for such a heterogeneous clustering setup across varying dimensions $D$, which clearly illustrates the considerable advantages of the proposed approach. The standard DPM approach allocates identical mean and covariance parameters for all samples within each cluster, which can not tackle the differential clustering allocations between the mean and covariance and hence results in spurious clusters that adversely affect the clustering accuracy.

%$\mu_i = \mu_{i'}$ for all $i, i'$ {\it except} for the first $d$ elements of $\mu$, with $d \approx \frac{D}{3}$. For the first $d$ elements of $\mu$, there were 5 clusters, each with a corresponding $d-$vector of $\mu$ values. Similarly, $\Sigma_i$ belonged to one of 5 clusters, with $\Sigma_i = \Sigma_{i'}$ for $i, i'$ in the same covariance cluster. The means and covariances cluster memberships were generated separately. 

%We consider the problem of identifying the $\mu$ cluster membership for each $Y_i$. We fit a standard DPM model, with $\mu_i$ and $\Sigma_i$ clustered jointly. Similarly, we fit a product of DPMs, with $\mu_i$ and $\Sigma_i$ clustered separately. For each approach, we evaluate the adjusted Rand index at each MCMC iteration, and the corresponding performance measure is the average adjusted Rand index across all MCMC iterations. This represents the clustering performance within the MCMC, which will in turn determine the accuracy of the posterior estimation. We vary $D$ from 10 to 100 in increments of 10. For each simulation, $d \approx \frac{D}{3}$, corresponding to a third of the elements of the mean vector being different. The clustering performance is shown in Figure \ref{randDemoSim}. Clearly the product of DPMs offers significant improvement over the standard DPM for such a heterogeneous clustering setup. This motivates our following work extending this idea to VAR models.

{\noindent \underline{\bf Theoretical Properties:}} From a theoretical perspective, it is possible to show that the proposed product of DP approach leads to posterior consistency for Kullback-Leibler neighborhoods, under reasonable assumptions on the true density $f_0$ that are routinely assumed in multivariate density estimation literature \citep{wu2008kullback}. This is not surprising given that the density follows a generic kernel mixture representation $\int K(x; \Theta)dP(\Theta)$ commonly encountered in literature. Some additional notations are provided below.  Denote the Euclidean norm for a vector as $|| \cdot||$, and denote the spectral norm of a matrix as $||\cdot||_2$. Further, denote the eigenvalues of a $\nOutcome\times \nOutcome$ positive definite matrix $\Sigma$ in decreasing order as $\lambda_1(\Sigma)\ge \ldots \ge \lambda_\nOutcome(\Sigma)$. Let $a\lesssim b$ imply that $a$ is less than $b$ upto a constant, and $\lfloor \cdot \rfloor$ denote the floor operator. Denote the Kullback-Leibler (KL) divergence between densities $f,g\in \mathcal{F}$ as $KL(f,g)=\int \log(f/g)f$.  Denote the set of natural numbers as $\mathbb{N}$.

We now establish our result on positive prior support for Kullback-Leibler neighborhoods under the product of DP prior below, based on the following assumptions.

{\noindent \it (C1)} $0<f_0({\bf x})<M$ for some constant $M$ and all ${\bf x}\in \Re^D$;\\
{\noindent \it (C2)} $\int f_0({\bf x})\log(f_0({\bf x}))d{\bf x} < \infty$; \\
{\noindent \it (C3)} $\int f_0({\bf x})\log(f_0({\bf x})/\phi_\delta({\bf x}))d{\bf x}<\infty$ where $\phi_\delta({\bf x}) = \inf_{||{\bf t - x}||<\delta}f_0({\bf t}) $ for some $\delta>0$; \\
{\noindent \it (C4)} for some $\eta>0$, $\int ||{\bf x} ||^{2(1+\eta)}f_0({\bf x}) d{\bf x} <\infty$.\\

%Our result on positive prior support for Kullback-Leibler neighborhoods under the product of DP priors is stated below.

 \vskip 6pt

{\noindent \bf Lemma 1:} {\it Let $f_0\in \mathcal{F}$ and assume conditions {\it (C1)-(C4) hold}. Then for the prior defined in (\ref{eq:productDP}), we have $\Pi\big(f\in \mathcal{F}:\int \log\big(f_0/f \big)f_0 \le \eta^* \big)\ge 0$, for any $\eta^*>0$.}

 \vskip 6pt

{\noindent \bf Remark 1:} Lemma 1 provides weak consistency guarantees by establishing positive prior support for arbitrarily small Kullback-Leibler neighborhoods of $f_0$, as per \cite{schwartz1965bayes}.%which will play an integral role in establishing the strong consistency properties of the induced posterior distributions in the sequel. 
 
 \vskip 6pt

%measures deviations in terms of L-1 norms. %and provides tighter bounds compared to Kullback-Leibler divergence as evident using Csiszar's inequality that states $||f_1 - f_2||_1^2\le 2KL(f_1,f_2)$.

 An outline of the proof is provided in the Appendix. We note that although weak consistency suggests the ability of the proposed method to accurately recover the true density, in many cases Kullback-Leibler neighborhoods contain densities that may show non-negligible deviations from $f_0$. Hence it is of interest to investigate strong consistency for the proposed approach, which renders the desirable feature that the posterior distribution concentrates in arbitrarily small $L_1$ neighborhoods of the true density. The next result states that under certain tail conditions on the base measure of the DP priors, it is possible to derive strong posterior consistency corresponding to the PDPM priors.
 
  \vskip 6pt

 {\noindent \bf Theorem 1:} {\it Suppose $f_0$ satisfies the conditions of Lemma 1. Then the posterior is strongly consistent at $f_0$ under the PDPM and gPDPM priors $\Pi$ in (\ref{eq:productDPM}) with base measures that satisfy the conditions: (i) $\small P^*_2(\lambda_1(\Sigma^{-1}_{\hsig})>x^*)\lesssim \exp(-c_1 (x^*)^{c_2}), P^*_2(\lambda_\nOutcome(\Sigma^{-1}_{\hsig})<1/x^*)\lesssim(x^*)^{-c_3}, P^*_2\big(\frac{\lambda_1(\Sigma^{-1})}{\lambda_\nOutcome(\Sigma^{-1}_{\hsig})}>x^*\big)\lesssim (x^*)^{-\kappa}$, for some constants $c_1,c_2,c_3,\kappa$, and all clusters $h_\sigma$; (ii)  $P^*_1(||\bfmu_{m_1,h_{1m_1}} ||>x^*)\lesssim (x^*)^{-2(r+1)}$ for all clusters $h_{1m_1}$, where $m_1=1,\ldots,\mathcal{M}_\mu$.} %{\color{black}\it finalize proof in SM \ldots}
 
 \vskip 6pt
 
The proof of the above result is provided in the Appendix. %The above result assumes $\mathcal{M}_\sigma=1$ for simplicity, but can be potentially extended to more general cases with higher $m_2$. 
Theorem 1 assumes routinely used tail conditions on the base measures that are very reasonable and  hold for commonly used distributions (such as Gaussian and Laplace) on the mean, as well as inverse-Wishart distribution on the covariance (see Lemmas 2-3 in the sequel).  %Such tail conditions are typically used for establishing consistency in multivariate density estimation literature \citep{canale2017posterior}. 
The procedure for proving the strong consistency result in Theorem 1 corresponding to non-compact space of densities $\mathcal{F}$ relies on carefully designed sieves $\mathcal{F}_n$ that are compact subsets of $\mathcal{F}$ but that grow with $n$ to eventually cover all of $\mathcal{F}$ as $n\to\infty$. A careful choice for the sieve needs to be made to ensure that the metric entropy grows slowly with $n$, while the prior probability of the complement of $\mathcal{F}_n$ (denoted as $\mathcal{F}^c_n$) decreases exponentially fast, such that a summability condition holds. This is sufficient to guarantee strong consistency as per ideas in Theorem 5 of \cite{ghosal2007posterior}. These ideas regarding sieve construction were encapsulated in \cite{shen2013adaptive} for location mixtures, and later in \cite{canale2017posterior} for location-scale mixtures for multivariate density estimation. For clarity, we provide the following result involving sufficient conditions on the sieves that is borrowed from Theorem 1 in \cite{canale2017posterior}, and will be used in our proofs. Denote the entropy of a space of densities $\mathcal{G}\subset \mathcal{F}$  as $N(\epsilon,\mathcal{G},d)$, which is defined (in terms of the metric $d$) as the minimum integer $N$ for which there exists densities $f_1,\ldots,f_N \in \mathcal{F}$ satisfying $\mathcal{G} \subset \cup_{j=1}^N \{f:d(f,f_j)<\epsilon \}$. The distance metric used to study convergence in the space of densities $\mathcal{F}$ is evaluated in terms of the Hellinger distance (defined $d(f,g)=\big[\int(\sqrt{f} - \sqrt{g})^2 \big]^{1/2}$), as well as the  $L_1$ metric (defined as $|| f-g||_1 = \int |f-g |$).

%In addition, it is also possible to specify low rank structures on the residual covariance along with appropriate priors that  satisfy the tail conditions for the base measure $P^*_2$ (see Lemma 3 in the sequel). Such low rank representations are routinely used for dimension reduction in the factor model literature \citep{ghosh2009default} and are particularly useful for developing scalable posterior computation strategies. In summary, Theorem 1 extends the strong consistency result for multivariate density estimation to the case of product of DP mixture priors.

 %The following result, which is similar to Theorem 1 in \cite{canale2017posterior} for multivariate density estimation and is stated here for brevity, captures the sufficient conditions characterizing the sieves.
 
 \vskip 6pt

{\noindent \bf Theorem 2:} {\it Consider sieves $\small\mathcal{F}_n\subset \mathcal{F}$ with $\small\mathcal{F}_n\uparrow \mathcal{F}$ as $n\to\infty$,  where $\small\mathcal{F}_n=\cup_{j} \mathcal{F}_{n,j}$, such that (2A) $\small\Pi(\mathcal{F}_n^c)\lesssim e^{-bn}$ for $b>0$; and (2B) $\small\sum_{j}\sqrt{N(2\epsilon,\mathcal{F}_{n,j},d)}\sqrt{\Pi(\mathcal{F}_{n,j})} e^{-(4-c)n\epsilon^2}\to 0,$ for $c,\epsilon>0$. Then $\small\Pi(f:d(f_0,f)>8\epsilon \mid X_n)\to 0$ in $F^n_0-$probability for any $f_0$ in the weak support of $\Pi$ defined in (\ref{eq:productDP}).}

\vskip 6pt

In Theorem 2, condition ({\it 2A}) suggests that the sieve should grow with the sample size such that only small neighborhoods with exponentially small prior probabilities are excluded. On the other hand, condition ({\it 2B}) reflects the summability condition that involves smaller subsets $\mathcal{F}_{n,j}$ that cover the sieve $\mathcal{F}_n$ under the union operation. It places constraints on the growth rate of the metric entropy in a manner that the weighted sum of the square root of metric entropy of $\mathcal{F}_{n,j}$ (weighted by the corresponding square root of prior probabilities) go towards zero with increasing $n$. We construct such sieves in the proof of Theorem 1 in the Appendix, and illustrate that the conditions (2A) and (2B) are satisfied, which results in strong consistency. %Readers can refer \cite{canale2017posterior} for the proof.
 
 %More discussion on the summability condition for the sieves as well as the actual construction of the sieve under the product of DPM mixtures for multivariate density estimation is provided in the sequel as well as Appendix.

Model (\ref{eq:productDP}) lays the foundation for the novel PDPM priors, that potentially has a wide array of applications, and can likely be generalized to any framework that involves clustering under Dirichlet process mixtures. We are now well positioned to turn our focus on the primary goal in this article, which is to develop a provably flexible non-parametric Bayesian methodology for multivariate time-series data modeled under a VAR framework, which is one of the first such set of results in literature, to our knowledge.

%The proposed product of DP priors models the heterogeneity between samples via multiscale clustering, i.e. separately clustering the autocovariance and the residual covariance matrices. In addition to fully distinct clusters with no overlap that is the hallmark of typical mixture modeling approaches, the multiscale clustering approach also results in partially overlapping clusters that share either common autocovariance parameters or residual covariance matrices but not both. The proposed approach ensures greater flexibility, since the clustering of the residual covariance does not directly interfere with the clustering for the autocovariance elements and vice-versa, which is expected to translate to increased accuracy. %in settings where estimating autocovariance parameters is of primary interest \citep{han2015direct,ghosh2018high}. 
%In contrast, a standard DP mixture specification, i.e.  (\mathcal{\Theta},\Sigma)\sim P_{\mathcal{\Theta},\mathcal{S}},  P_{\mathcal{\Theta},\mathcal{S}}\sim DP(\alpha P^*)$, yields clusters of replicated samples that share identical autocovariance {\it and} residual covariance parameters, which may become restrictive in larger dimensions since the number of parameters increase quadratically with $D$. The above discussions highlight the advantages of the multiscale clustering aspect and the associated partially overlapping clusters, which provide a central motivation for the development of the product of DP priors in this article.

\section{Extension to Vector Autoregressive Models}
\subsection{Proposed Model}

Consider the data matrix $X_i=({\bf x}_{i1},\ldots,{\bf x}_{iT})$, where ${\bf x}_{it}$ represents the $(\nOutcome \times 1 )$ temporally dependent multivariate measurement for the $i$-th subject at the $t$-th time point ($i=1,\ldots,n,t=1,\ldots,T$). Note that our model can easily accommodate subject-specific scan lengths ($T_i$); however we will assume $T_i = T$ from hereon in, to ease the exposition. Throughout, we will also assume a fixed dimension ($\nOutcome$), and a pre-specified number of time scans ($T$), which is consistent with the routinely used fixed dimensional assumptions in the literature on non-parametric modeling of location-scale mixtures. Consider the VAR model:
\begin{eqnarray}
{\bf x}_{it} = \sum_{k=1}^{\min\{t-1,K\}} A_{ik}{\bf x}_{i,t-1} + \bfe_{it}, \mbox{ } \bfe_{it}\sim N({\bf 0}, \Sigma_{i}), \mbox{ } i=1,\ldots,n, \mbox{ } t=1,\ldots,T, \label{eq:base}
\end{eqnarray}
where $A_{ik}$ denotes the $\nOutcome\times \nOutcome$ matrix of autocovariance parameters for subject $i$ at lag $k$ ($k=1,\ldots,K$), $\Sigma_{i}\in S_{\nOutcome\times \nOutcome}$ denotes the time-invariant residual covariance  for subject $i$, and the lag order ($K$) is pre-specified as per standard practice in the VAR model literature \citep{ghosh2018high}. Model (\ref{eq:base}) implies that the mean of ${\bf x}_{t}$ depends on ${\bf x}_{t-1},\ldots,{\bf x}_{1}$ when $t\le K$ and on ${\bf x}_{t-1},\ldots,{\bf x}_{t-K}$ for $t>K$, with ${\bf x}_{i1}\sim N(0,\Sigma_{i})$ as per convention. %We note that it is also possible to fit the model after shaving off the first $K$ measurements. However, such an approach using a truncated time-series would lead to loss of information, and is therefore not considered in this paper. 
As is common in practice, the intercept term is fixed to be zero and not included in (\ref{eq:base}). %which is reasonable for our motivating neuroimaging applications involving centered and pre-processed fMRI data. % in which case one may write $E({\bf x}_{it})=\sum_{k=1}^{K} A_{ik}{\bf x}_{i,t-1}$ for all $t>K$. 

In order to understand the properties of (\ref{eq:base}), it is imperative to note that the likelihood for the $i$-th sample can be written as a product of conditional densities as 
\begin{eqnarray}
L(X_i \mid \mathcal{\Theta}_i, \Sigma_i) = 
\prod_{t=2}^T \phi_{\Sigma_{i}}\bigg({\bf x}_{it}- \sum_{k=1}^{\min\{t-1,K\}} A_{ik}{\bf x}_{i,t-k} \bigg)\times \phi_{\Sigma_{i}}\big({\bf x}_{i1}\big), \mbox{ } i=1,\ldots,n, \label{eq:lik}
\end{eqnarray}
where  $\mathcal{\Theta}_i$ denotes the collection of autocovariance matrices for sample $i$ across lags. For example, the likelihood for the $i$th sample under a VAR(2) model may be written as $\small \prod_{t=3}^T\phi_{\Sigma_{i}}\big({\bf x}_{it}- \sum_{k=1}^{2} A_{ik}{\bf x}_{i,t-k} \big)\times \phi_{\Sigma_{i}}\big({\bf x}_{i2}- A_{i1}{\bf x}_{i,1} \big)\times \phi_{\Sigma_{i}}\big({\bf x}_{i1}\big)$. The above likelihood in (\ref{eq:lik}) will be used throughout in our treatment of VAR models. We note that (\ref{eq:lik}) is a different way of representing the likelihood compared to the linear regression framework that is often used in single subject VAR models \citep{ghosh2018high}.

Our goal involves multi-subject VAR analysis by proposing suitable priors on $(\mathcal{\Theta}_i,\Sigma_i)$ in (\ref{eq:base}) to leverage common patterns of information across samples in an unsupervised and flexible manner. A natural framework for pooling information across subjects is via clustering, which also inherently results in model parsimony that is particularly important in our settings where the number of parameters grow with $n$.  Such a clustering approach should enable straightforward posterior computation and result in theoretical guarantees. To this end, we extend the PDPM methodology to the case of multivariate time-series data that imposes independent DP mixture priors separately on  $\mathcal{\Theta}$ and $\Sigma$ to induce multiscale clustering. Depending on the manner of the DP prior specification on the autocovariance elements, one can obtain different variants of the proposed method that allow for varying degrees of model parsimony and varying levels of information sharing within samples, via different patterns of autocovariance clusters. Such a multi-scale clustering approach is particularly relevant in the context of VAR models where the dimension of the autocovariance matrix increases quadratically with the outcome dimension $D$, making it imperative to avoid the assumption of replicated samples that is embedded in typical mixture modeling approaches. The resulting PDPM approach leads to a more fitting characterization of heterogeneity and greater accuracy, as illustrated via extensive numerical studies involving VAR models in the sequel. In addition, appropriate base measures in the DP can be chosen to encourage shrinkage in the autocovariance elements that facilitate feature selection, as well as to induce low rank decomposition for the residual covariance resulting in additional model parsimony.

\vskip 6pt

%In addition, appropriate base measures in the DP can be chosen to encourage shrinkage in the autocovariance elements that facilitate feature selection, as well as to induce low rank decomposition for the residual covariance resulting in additional model parsimony. We note that all the of advantages of the PDPM approach described in Section 2, automatically carryover to the VAR setting. For example, it is possible to bypass the assumptions of replicated samples by allowing varying degrees of information sharing across samples via multi-scale clustering that is particularly relevant given the large dimensions of the autocovariance matrix containing $D^2$ parameters. The resulting PDPM approach leads to a superior characterization of heterogeneity and greater accuracy, as illustrated via extensive numerical studies involving VAR models in the sequel.
%We first introduce the product of DP priors below, followed by generalizations to additional variants.

{\noindent \bf \underline{Product of DP mixtures for VAR models:}}
In the following specifications, we will omit subscript $i$ where appropriate, for notational convenience and as per convention \citep{wu2008kullback,canale2017posterior}. We propose the following PDPM prior 
\begin{eqnarray}
\mathcal{\Theta} = \{vec(A_1),\ldots,vec(A_K)\}\sim P_{\mathcal{\Theta}}, \mbox{ } P_{\mathcal{\Theta}}\sim DP(\alpha_1 P^{*}_1),
\Sigma \sim P_{\mathcal{S}}, \mbox{ } P_{\mathcal{S}}\sim DP(\alpha_2P^{*}_2),\label{eq:prodDP}
\end{eqnarray}
where %$vec(A)$ represents the operator that stacks the columns of  $A_{\nOutcome\times \nOutcome}$ to form a $\nOutcome^2\times 1$ vector, 
$\alpha_1,\alpha_2,$ represent precision parameters in the Dirichlet process, the base measure $P^{*}_1$ belongs to the space of probability measures $\mathcal{P}_1$ on $\mathcal{D}_1 =\underbrace{\Re^{\nOutcome^2\times 1}\times \ldots \times \Re^{\nOutcome^2\times 1} }_\text{K}$, and the base measure $P^{*}_2$ belongs to the space of probability measures $\mathcal{P}_2$ on $\mathcal{D}_2=S_{\nOutcome\times \nOutcome}$. Model (\ref{eq:prodDP}) specifies unknown distributions $P_{\mathcal{\Theta}}$ and $P_{\mathcal{S}}$ on model parameters, that are modeled under independent DP priors. The resulting product of DP priors in (\ref{eq:prodDP}) is defined on the space of densities $\mathcal{P}$ with domain $\mathcal{D}_1\times\mathcal{D}_2$ and may be expressed as  $\Pi^*(\mathcal{\Theta},\Sigma)= P_{\mathcal{S}}(\Sigma)\times P_{\mathcal{\Theta}}(\mathcal{\Theta})$.
%Throughout the article, we will use the shorthand notation $(A_1,\ldots,A_K)\sim f_{\mathcal{\Theta}}$ in place of  $(vec(A_1),\ldots,vec(A_K))\sim f_{\mathcal{\Theta}}$ as appropriate, to denote the prior specification on the autocorrelation matrices. 
This prior specification translates to a {\it product of DP mixture of VAR (PDPM-VAR)} models that induces a prior $\Pi$ on the space of probability densities $\mathcal{F}$ for the data matrix $X$ as follows:
\begin{align}
&f_{P}(X) = \int\int \prod_{t=1}^T \phi_{\Sigma}\bigg({\bf x}_{t} -  \sum_{k=1}^{\min\{t-1,K\}} A_{k}{\bf x}_{t-k} \bigg)  dP_{\mathcal{\Theta}}(\mathcal{\Theta})dP_{\mathcal{S}}(\Sigma) \nonumber \\
&= \sum_{h_1=1}^\infty \sum_{\hsig=1}^\infty \pi_{h_1}\pi_{\sigma,\hsig}\prod_{t=1}^T \phi_{\Sigma_{\hsig}}\bigg({\bf x}_{t} -  \sum_{k=1}^{\min\{t-1,K\}} A_{k,h_1}{\bf x}_{t-k} \bigg), \mbox{ }
\label{eq:stick}  
\end{align}
where $ \pi_{h_1}=\nu_{h_1}\prod_{l_1<h_1}(1-\nu_{l_1})$, $\nu_{h_1}\sim Be(1,\alpha_1)$, $ \pi_{\sigma,\hsig}=\nu_{\sigma,\hsig}\prod_{l_2<\hsig}(1-\nu_{\sigma,l_2})$, $\nu_{\sigma,\hsig}\sim Be(1,\alpha_2)$, and further $\small \Sigma_{\hsig} \sim P^{*}_2, \mbox{ } (vec(A_{1,h_1}),\ldots, vec(A_{K,h_1})) \sim P^{*}_1$. We consider a broad class of base measures  to study theoretical properties (Section \ref{sec:theory}), but for implementation we focus on specific choices for $(P_1^*,P^*_2)$ that facilitate posterior computations (Section \ref{sec:postcomp}).

While (\ref{eq:prodDP}) provides a greater degree of flexibility in terms of accommodating  heterogeneity compared to existing DP mixture approaches, there is further scope for generalizing this approach to accommodate additional heterogeneity in lag-specific and row-specific relationships. Such generalizations become particularly important when clusters of samples tend to share common autocovariance elements for some but not all lags or have identical elements for only a subset of rows/nodes in the autocovariance matrices in practical applications. For example, the latter scenario arises when the effective clustering for the autocovariance elements is confined to a subset of rows in the matrix $A$,  with the remaining rows being irrelevant with respect to clustering. Such aspects are routinely encountered in heterogeneous and high-dimensional clustering problems \citep{agrawal2005automatic}, such as our VAR settings of interest where the number of autocovariance parameters increase quadratically with the outcome dimension ($D$). %Such a scenario is reminiscent of the subspace clustering literature, and is motivated by neuroimaging applications where only a subset of nodes in the brain atlas exhibit reproducible clustering patterns (see Section 5). 
We now generalize the PDPM-VAR method below to account for such heterogeneous settings.

\vskip 6pt

{\bf \noindent \underline{Generalization across autocovariance rows:}} It is  possible to generalize the PDPM-VAR model in (\ref{eq:prodDP}) in a manner that relaxes the restriction to have fully identical autocovariance matrices for all samples within a given autocovariance cluster. In particular, consider an approach that specifies independent priors on the VAR model parameters corresponding to each row of the autocovariance matrices, which results in row-specific clustering patterns. In particular, denote $A_{k,d'\bullet}$ as the $d'$-th row of $A_k$ and consider the following specification 
\begin{eqnarray}
 vec\{ A'_{1,d'\bullet},\ldots, A'_{K,d'\bullet}\} \stackrel{indep}{\sim} P_{\mathcal{\Theta}_{d'}}, \mbox{ } P_{\mathcal{\Theta}_{d'}}\sim  DP(\alpha^*_{d'} P^{**}_{1d'}), \mbox{ }  \Sigma \sim P_{\mathcal{S}}, \mbox{ } P_{\mathcal{S}}\sim DP(\alpha_2P^{*}_2),  d'=1,\ldots, \nOutcome,
\label{eq:prodDP3}
\end{eqnarray}
where $A'$ denotes the transpose of $A$, and the row-specific priors $P_{\mathcal{\Theta}_{d'}}(vec\{A'_{1,d'\bullet},\ldots, A'_{K,d'\bullet} \})$ are specified independently for each row and jointly across lags. The product of DP prior in (\ref{eq:prodDP3}) is expressed as $\Pi^*(\mathcal{\Theta},\Sigma)= P_{\mathcal{S}}(\Sigma) \times \prod_{d'=1}^\nOutcome P_{\mathcal{\Theta}_{d'}}(vec\{A'_{1,d'\bullet},\ldots, A'_{K,d'\bullet}\})$, and results in the row-generalized PDPM-VAR (rgPDPM-VAR) model that induces priors on $\mathcal{F}$ via
\begin{align}
f_{P}(X) = \sum_{h_{1,1}=1}^\infty .. \sum_{h_{1,D}=1}^{\infty}  \sum_{\hsig=1}^\infty (\pi_{\sigma,\hsig} \prod_{d'=1}^\nOutcome\pi^*_{d',h_{1d'}})\prod_{t=1}^T \phi_{\Sigma_{\hsig}}\big({\bf x}_{t} -  \sum_{k=1}^{\min\{t-1,K\}} A_{k,h_{11},\ldots,h_{1\nOutcome}}{\bf x}_{t-k} \big),
\label{eq:stick3}  
\end{align}
%{\color{blue}
%\begin{align}
%f_{P}(X) = \sum_{h_{1,1}=1}^\infty \cdots \sum_{h_{1,D}=1}^{\infty}  \sum_{\hsig=1}^\infty \pi_{\sigma,\hsig} \pi^*_{1,h_{1,1}} \cdots \pi^*_{D,h_{1,D}} \prod_{t=1}^T \phi_{\Sigma_{\hsig}}\bigg({\bf x}_{t} -  \sum_{k=1}^{\min\{t-1,K\}} A_{k,h_{1,1},\ldots,h_{1,\nOutcome}}{\bf x}_{t-k} \bigg),
%\label{eq:stick3}  
%\end{align}
%}
where $A_{k,h_{11},\ldots,h_{1D}}$ denotes the autocovariance matrix at lag $k$ that assigns the $h_{1d'}$-th mixture component to the $d'$-th row with prior probability  $\pi^*_{d',h_{1d'}}=\nu^*_{d',h_{1d'}}\prod\limits_{l_{1d'}<h_{1d'}}(1-\nu^*_{l_{1d'}})$, where $\nu_{d',h_{1d'}}^*\sim Be(1,\alpha^*_{d'})$,  $vec\{A'_{1,d'\bullet,h_{1,d'}},\ldots,A'_{K, d'\bullet,h_{1,d'}} \} \stackrel{indep}{\sim} P^{**}_{1d'}$ and $A_{k, d'\bullet,h_{1,d'}} $ denotes the $d'$-th row for the matrix $A_k$ that takes values from the $h_{1,d'}$-th mixture component. Further,  $\Sigma_{\hsig}\sim P^*_2$ with prior probability $\pi_{\sigma,\hsig}=\nu_{\sigma,\hsig}\prod\limits_{l_2<\hsig}(1-\nu_{\sigma,l_2})$ and $\nu_{\sigma,\hsig}\sim Be(1,\alpha_2)$. The above prior results in independent clustering of autocovariance rows collectively across all lags, which results in greater flexibility compared to the PDPM-VAR in (\ref{eq:prodDP}) that constrains samples within autocovariance clusters to share entirely identical $D\times D$ autocovariance matrices.

%Further, the rgPDPM-VAR specifies that the rows of the concatenated matrix  $(A_{1,h_{11},\ldots,h_{1\nOutcome}},\ldots,A_{K,h_{11},\ldots,h_{1\nOutcome}})$ having dimension $D\times DK$ are randomly drawn from the prior distribution  $\prod_{d'=1}^D P^{**}_{1,d'}(vec\{A'_{1,d'\bullet},\ldots, A'_{K,d'\bullet} \})$.

In the scenario when multiple rows have identical clustering configurations, the rgPDPM-VAR model is able to identify clusters of samples that share identical autocovariance elements corresponding to a subset of nodes only, but exhibit variations corresponding to the remaining autocovariance rows. We note that for our motivating neuroimaging applications, this scenario translates to identical effective connectivity corresponding to a subset of brain regions within a autocovariance cluster, while the remaining brain regions are allowed to exhibit varying connectivity profiles within this cluster. By allowing row-specific clustering patterns in the autocovariance matrix, the rgPDPM-VAR approach results in a more complete characterization of heterogeneity compared to the PDPM-VAR. Additional generalizations are also possible; for example, one may extend specification (\ref{eq:prodDP3}) to impose row- and lag-specific priors. However, such extensions may result in a rapid rise in parameters that presents potential computational issues, and hence are not considered further. 
%Hence, we do not consider such additional generalizations further in this article. %and restrict our attention to the PDPM-VAR, lgPDPM-VAR and rgPDPM-VAR variants.

%In particular, PDPM-VAR does not address scenarios where there exist overlapping subgroups of samples that have common autoregressive parameters for some, but not all, lags. Similarly, the PDPM-VAR does not account for the possibility of clusters of samples that share common elements for only a subset of rows/vertices in the autocovariance matrices. {\color{black}This latter scenario corresponds to settings where effective clustering for the autocovariance elements are confined to a subset of rows in the matrix $A$,  with the remaining directions being irrelevant with respect to clustering. Such a scenario is reminiscent of  similar settings in the subspace clustering literature, and is motivated by neuroimaging applications where only a subset of nodes in the brain atlas exhibit reproducible clustering patterns (see Figure XXXXX).} In order to design a more flexible model for such heterogeneous scenarios that are expected to arise in practice, we generalize the product of DP priors in (\ref{eq:prodDP}) by proposing two distinct variants of the PDPM-VAR approach that are described below.

\vskip 6pt

{\noindent \bf \underline{Generalization across lags}:} For the second extension, we specify independent DP priors for the autocovariance matrices at each lag, which results in lag-specific clustering as follows: %generalized product of DP prior specification for such cases as follows:
\begin{eqnarray}
 && vec(A_{k}) \stackrel{indep}{\sim}  P_{\mathcal{\Theta}_k}, \mbox{ }  P_{\mathcal{\Theta}_k}\sim DP(\alpha_{1k} P^{*}_{1k}),  \mbox{ } \Sigma \sim P_{\mathcal{S}}, \mbox{ } P_{\mathcal{S}}\sim DP(\alpha_2P^{*}_2),  \mbox{ } k=1,\ldots,K,
\label{eq:prodDP2}
\end{eqnarray}
where $P_{\mathcal{\Theta}_k}$ denotes the unknown density for $vec(A_k)$ that is modeled under a DP prior with base measure $P^*_{1k}$ and precision parameter $\alpha_{1k} (k=1,\ldots,K)$, and the prior on the residual covariance parameters is defined similarly to (\ref{eq:prodDP}), but with the understanding that $\alpha_2$ and $P^*_2$ in the DP priors in (\ref{eq:prodDP2}) and (\ref{eq:prodDP}) are allowed to be distinct. The resulting product of DP priors in (\ref{eq:prodDP2}) %is defined on the space of densities $\mathcal{P}$ with domain $\mathcal{D}_1\times\mathcal{D}_2$ and 
may be expressed as  $\Pi^*(\mathcal{\Theta},\Sigma)=P_{\mathcal{S}}(\Sigma) \times \prod_{k=1}^K P_{\mathcal{\Theta}_k}(A_k)$.
As under the PDPM-VAR, specification (\ref{eq:prodDP2}) induces a prior on the space of densities $\mathcal{F}$ via 
\begin{align}
f_{P}(X) = \sum_{h_{11}=1}^\infty\ldots\sum_{h_{1K}=1}^\infty \sum_{\hsig=1}^\infty \pi_{\sigma,\hsig} \bigg(\prod_{k=1}^K\pi_{k,h_{1k}}\bigg)\prod_{t=1}^T \phi_{\Sigma_{\hsig}}\bigg({\bf x}_{t} -  \sum_{k=1}^{\min\{t-1,K\}} A_{k,h_{1k}}{\bf x}_{t-k} \bigg),
\label{eq:stick2}  
\end{align}
where $\pi_{k,h_{1k}}=\nu_{k,h_{1k}}\prod\limits_{l_{k,1k}<h_{k,1k}}(1-\nu_{k,l_{1k}}) \mbox{ }(k=1,\ldots,K),\mbox{ }\pi_{\sigma,\hsig}=\nu_{\sigma,\hsig}\prod\limits_{l_2<\hsig}(1-\nu_{\sigma,l_2})$ and $\nu_{k,h_{1k}}\sim Be(1,\alpha_{1k}), \nu_{\sigma,\hsig}\sim Be(1,\alpha_2)$, and further $vec(A_{k,h_{1k}})\sim P^*_{1k},\Sigma_{\hsig}\sim P^*_2$ for $k=1,\ldots,K$, using the stick-breaking construction in \cite{sethuraman1994constructive}. We denote the model under (\ref{eq:base}) and (\ref{eq:prodDP2}) as the lag-generalized product of DP mixture of VAR (lgPDPM-VAR) model and note that this model reduces to the PDPM-VAR for lag 1 models. This approach is expected to be less flexible compared to the rgPDPM-VAR method in general, but may exhibit some advantages when the clustering patterns are distinct across lags.

\vskip 6pt

\subsection{Theoretical Properties}\label{sec:theory}

{\noindent \underline{Notations and Definitions:}} In this section we will establish posterior consistency properties of the proposed product of DP mixture of VAR models. We will assume that the $\nOutcome\times T$ data matrices $X_1,\ldots,X_n,$ are i.i.d. under some true density $f_0\in \mathcal{F}$. We will study the convergence of the posterior around this true density with respect to the product measure $F^n_0$ associated with $f_0$ as $n$ grows to $\infty$, under some reasonable regularity conditions on $f_0$ and the conditions on the tail behavior of the base measures of the DP priors. We note that the theoretical derivations corresponding to VAR models involving multi-variate time-series data are more involved than the independent multivariate outcome settings in Section 2 that has been the focus of existing non-parametric Bayesian density estimation literature.  Moreover, our theoretical results assume fixed $T$ (finite time set-up) with growing number of samples, which is in contrast to theoretical settings in parametric VAR analysis for single subjects that rely on growing $T$ \citep{ghosh2018high}. 

%and by $\mathbb{N}^*$ the set of whole numbers.
%{\color{black} \it insert other definitions as needed \ldots}

Throughout the article, we will assume the following reasonable regularity conditions on $f_0$, which are adapted from standard assumptions made in multivariate density estimation literature. Let $f_0({\bf x}_t\mid X_{1:(t-1)})$ denote the true conditional density of ${\bf x}_{t}$ that depends on previous time scans upto a certain known lag ($K$).  %We assume that the true lag order $K$ is known that is standard in the VAR modeling literature \citep{ghosh2018high}. 
Consider the following assumptions. %for the case where $\nOutcome,T,$ are pre-specified and fixed.

{\noindent \it (A0)} The form of the true density satisfies $f_0(X)=\big\{\prod_{t=1}^T f_0({\bf x}_t\mid {\bf x}_{t-1},\ldots,{\bf x}_1)\big\}=\big\{\prod_{t=1}^T f_0({\bf x}_t\mid X_{1:(t-1)})\big\}$, for all $X\in \Re^{\nOutcome \times T}$.\\ %and where the conditional density reduces to $f({\bf x}_1)$ for $t=1$.\\
{\noindent \it (A1)} $0<f_0(X)<M$ for some constant $M$ and for all $X\in \Re^{\nOutcome \times T}$.\\
{\noindent \it (A2)}  $|\int f_0\big({\bf x}_t\mid X_{1:(t-1)}\big)\log\big(f_0({\bf x}_t\mid X_{1:(t-1)})\big)d{\bf x}_t|<\infty$, point-wise for $X_{1:(t-1)}$ for all $t$.\\
{\noindent \it (A3)} For all $t$ and some $\delta>0$, $\int f_0\big({\bf x}_t\mid X_{1:(t-1)}\big)\log\big(\frac{f_0({\bf x}_t\mid X_{1:(t-1)})}{\phi^*_\delta({\bf x}_t\mid X_{1:(t-1)})} \big)d{\bf x}_t<\infty,$ where $\phi^*_\delta({\bf x}_t\mid X_{1:(t-1)})=\inf_{||{\bf r} - {\bf x}_t||<\delta} f_0({\bf r}\mid X_{1:(t-1)}) $, point-wise for $X_{1:(t-1)}$.\\
{\noindent \it (A4)} For all $t$ and some $\eta>0$, $\int || {\bf x}_t||^{2(1+\eta)}f_0({\bf x}_t \mid X_{1:(t-1)})d{\bf x}_t < \infty$, point-wise for $X_{1:(t-1)}$.

Condition { \it (A0)} expresses the true density as a product of conditional densities, subject to a known $K$, where the true conditional density only depends on  ${\bf x}_{t-1},{\bf x}_{t-2},\ldots,{\bf x}_{t-K},$ when $t>K$ and depends on ${\bf x}_{t-1},{\bf x}_{t-2},\ldots,{\bf x}_{1}$ for $t\le K$. Condition {\it (A1)} assumes that the true density is bounded. Assumptions {\it (A2)-(A4)} impose regularity conditions on the conditional densities that are similar to standard assumptions made in nonparametric Bayesian literature for marginal densities \citep{wu2008kullback}. In the special case when the true density corresponds to a VAR structure, {\it (A0)-(A4)} would imply (among other things) that the true VAR parameters are well-behaved and satisfy stability conditions so that the true density does not blow up to $\infty$ or attenuate to zero. 

The following Theorem formally states the result on positive prior support under the above assumptions. The proof is provided in the Appendix and uses key results in \cite{wu2008kullback} for multivariate density estimation under DP mixtures. %However our result is distinct from these standard multivariate density estimation settings since it involves matrix-variate outcomes with temporally dependent columns and are modeled under a VAR framework. %We note that (A0)-(A4) are designed to satisfies key assumptions in that paper in order to prove the Kullback-Leibler property.

{\noindent \bf Theorem 3:} {\it Suppose assumptions ${\it (A0)-(A4)}$ are satisfied. Then the product of DP mixture priors $\Pi$ specified in (\ref{eq:prodDP}),  (\ref{eq:prodDP3}), and (\ref{eq:prodDP2}) satisfies the Kullback-Leibler property, i.e. $\Pi\bigg(f\in \mathcal{F}:\int \log\big(f_0/f \big)f_0 \le \eta^* \bigg)\ge 0$, for any $\eta^*>0$.}

\vskip 6pt

% {\noindent \bf Remark 2:} Theorem 3 involves multivariate time series data and  requires a much more involved proof that go beyond the relatively straightforward proof for the case of independent multivariate data presented in Lemma 1. %which will play an integral role in establishing the strong consistency properties of the induced posterior distributions in the sequel. 

%measures deviations in terms of L-1 norms. %and provides tighter bounds compared to Kullback-Leibler divergence as evident using Csiszar's inequality that states $||f_1 - f_2||_1^2\le 2KL(f_1,f_2)$.
 The next goal is to establish strong consistency for the proposed approach. We will pursue the same approach as earlier that relies on the summability conditions in Theorem 2 and suitable sieve constructions, {\color{black}noting that the result in Theorem 2 for multivariate data also holds for matrix variate outcomes since any matrix can be represented as a vector under the $vec(\cdot)$ operation.} However in practice, it may not be straightforward to construct such sieves for the matrix-variate density estimation case, since the metric entropy depends on a number of terms including the sample size $n$, dimension $\nOutcome$, as well as $T$ (see Theorem 4).  %Although it is not possible to directly adapt the sieves used for  multivariate density estimation in Section 2 to our settings of interest that involve matrix-variate data modeled under a VAR framework, 
 One can still leverage the stick-breaking representation to construct appropriate sieves inspired by the ideas implemented in \cite{shen2013adaptive},  as outlined below.

\vskip 6pt

{\noindent \underline{\bf Sieve Constructions:}} The sieves are constructed so as to allow the norm of the elements in the autocovariance matrices, as well as the condition number of the residual covariance matrices, to increase with sample size at an appropriate rate that satisfies the conditions in Theorem 2.  We note that the condition number of a matrix frequently appears in the random matrix literature \citep{edelman1988eigenvalues} and is defined as the ratio of the largest to the smallest eigen values, i.e.  $\lambda_1(\Sigma)/\lambda_\nOutcome(\Sigma) = \lambda_1(\Sigma^{-1})/\lambda_\nOutcome(\Sigma^{-1})$.  For our purposes, we construct the following sieves corresponding to the PDPM-VAR model in (\ref{eq:base}) and (\ref{eq:prodDP}) as:
\begin{eqnarray}
\small
\mathcal{F}_n &=& \bigg\{f_p: P = \sum\limits_{h_1\ge 1}\sum\limits_{\hsig\ge 1}\pi_{h_1}\pi_{\sigma,\hsig}\delta_{\mathcal{\Theta}_{h_1},\Sigma_{\hsig}}: \sum_{h_1>H_n}\pi_{h_1}<\epsilon_1, \sum_{\hsig>H_n}\pi_{\sigma,\hsig}<\epsilon_2,  \mbox{and for } \nonumber \\
&& \hsig\le H_n, \mbox{ } \underline{\sigma}_n^2\le \lambda_{\nOutcome,\Sigma_{\hsig}}\le  \lambda_{1,\hsig} \le \underline{\sigma}_n^2(1 + \epsilon/\sqrt{\nOutcome})^{M_n},\mbox{ } 1<\frac{\lambda_{1,\hsig}}{\lambda_{\nOutcome,\hsig}}\le  n^{H_n}  \bigg\}, \mbox{ }  \nonumber\\
\mathcal{F}_{n,{\bf jl}}&=& \bigg\{f_p\in \mathcal{F}_n: \mbox{for } h_1,\hsig\le H_n,
\underline{a_{h_1,j}}\le ||vec(A_{k,h_1}) ||\le \bar{a}_{h_1,j} \mbox{ } \forall k, \mbox{ } \underline{u_{\hsig,l}}\le \frac{\lambda_{1,\hsig}}{\lambda_{\nOutcome,\hsig}}\le u_{\hsig,l} \bigg\} \quad \quad, \label{eq:sieves-PDPMmm}
\end{eqnarray}
where $\delta_\theta$ denotes the probability measure degenerate at $\theta$, $\lambda_{d',\Sigma_{\hsig}}$ is a shorthand for  $\lambda_{d'}(\Sigma_{\hsig}),$ i.e. the eigen values corresponding to $\Sigma_{\hsig}$, $j,l$ are integers that are $\le H_n$ for a given $n$, the sequences $\{H_n \}, \{ M_n\}$, $\{\underline{\sigma}_n \} \{\underline{a_{h_1,j}}\}$, $\{\bar{a}_{h_1,j} \}, \{ \underline{u_{\hsig,j}}\},\{ u_{\hsig,j}\}$ grow to $\infty$ with $n$ and are chosen appropriately such that $\mathcal{F}_n \subset \cup_{{\bf j,l}}\mathcal{F}_{n,{\bf jl}}$, and further, $\mathcal{F}_n \uparrow \mathcal{F}$ as $n\to\infty$. Moreover, the sieves corresponding to {\color{black}rgPDPM-VAR} in (\ref{eq:base}) and (\ref{eq:prodDP3}) are constructed as:
\begin{eqnarray}
\small
&&\mathcal{F}_n = \bigg\{f_p: P =  \sum_{h_{1,1}=1}^\infty .. \sum_{h_{1,D}=1}^{\infty}  \sum_{\hsig=1}^\infty (\pi_{\sigma,\hsig} \prod_{d'=1}^\nOutcome\pi^*_{d',h_{1d'}})\delta_{\mathcal{\Theta}_{h_{1d'}},\Sigma_{\hsig}}: \sum_{h_{1d'}>H_n}\pi^*_{d',h_{1d'}}<\epsilon_1, \mbox{ } \forall d'\le \nOutcome, \nonumber \\
&& \sum_{\hsig>H_n}\pi_{\sigma,\hsig}<\epsilon_2, \mbox{ and for } \hsig\le H_n, \mbox{ }
 \underline{\sigma}_n^2\le \lambda_{\nOutcome,\hsig}\le \lambda_{1,\hsig} \le \underline{\sigma}_n^2(1 + \epsilon/\sqrt{\nOutcome})^{M_n}, \mbox{ } 1<\frac{\lambda_{1,\hsig}}{\lambda_{\nOutcome,\hsig}}\le  n^{H_n}  \bigg\}, \mbox{ }  \nonumber  \\
&& \mathcal{F}_{n,{\bf jl}} =\bigg\{f_p\in \mathcal{F}_n: {\color{black}\underline{a}_{h_{1d'},j}}\le ||vec(A_{k,d'\bullet,h_{1,d'}}) ||\le {\color{black}\bar{a}_{h_{1d'},j}} \mbox{ for } h_{11},\ldots,h_{1\nOutcome}\le H_n, \nonumber \\
&&\mbox{and } d'=1,\ldots,\nOutcome, \mbox{ and }\underline{u}_{\hsig,l}\le \frac{\lambda_{1,\hsig}}{\lambda_{\nOutcome,\hsig}}\le u_{\hsig,l}, \mbox{ for } \hsig\le H_n  \bigg\}, \label{eq:sieves-rgPDPM}
\end{eqnarray}
and the sieves for the lgPDPM-VAR model in (\ref{eq:base}) and (\ref{eq:prodDP2}) are constructed similarly as:
\begin{align}
\small
&\mathcal{F}_n = \bigg\{f_p: P =\sum_{h_{11}=1}^\infty\ldots\sum_{h_{1K}=1}^\infty \sum_{\hsig=1}^\infty \pi_{\sigma,\hsig} \big(\prod_{k=1}^K\pi_{k,h_{1k}}\big)\delta_{\mathcal{\Theta}_{h_{1k}},\Sigma_{\hsig}}: \sum_{h_{1,1k}>H_n}\pi_{k,h_{1k}}<\epsilon_1,\forall k=1,\ldots, K, \nonumber \\
& \sum_{\hsig >H_n}\pi_{\sigma,\hsig}<\epsilon_2,  \mbox{ and for } \hsig\le H_n, \mbox{ } \underline{\sigma}_n^2\le \lambda_\nOutcome(\Sigma_{\hsig})\le \lambda_1(\Sigma_{\hsig}) \le \underline{\sigma}_n^2(1 + \epsilon/\sqrt{\nOutcome})^{M_n}, \mbox{ } 1<\frac{\lambda_{1,\hsig}}{\lambda_{\nOutcome,\hsig}}\le  n^{H_n} \bigg\},  \nonumber  \\
& \mathcal{F}_{n,{\bf jl}} =\bigg\{f_p\in \mathcal{F}_n:   {\color{black}\underline{a}_{h_{1k},j}}\le ||vec(A_{k,h_{1k}}) ||\le {\color{black}\bar{a}_{h_{1k},j}} \mbox{ for all } h_{1k}\le H_n, \underline{u}_{\hsig,l}\le \frac{\lambda_{1,\hsig}}{\lambda_{\nOutcome,\hsig}}\le u_{\hsig,l}, \mbox{ } \hsig\le H_n \bigg\}, \label{eq:sieves-lgPDPM}
\end{align}
 and 
where $\mathcal{F}_n \subset \cup_{{\bf j,l}}\mathcal{F}_{n,{\bf jl}}$ and $\mathcal{F}_n \uparrow \mathcal{F}$ as $n\to\infty$, and it is understood that the sequences $\{H_n \}$, $\{ M_n\}$, $\{\underline{\sigma}_n \}, \{\underline{a_{h_1,j}}\}, \{\bar{a}_{h_1,j} \},\{ \underline{u_{\hsig,l}}\},\{ u_{\hsig,l}\}$ are chosen appropriately and can be specific to sieves corresponding to {\color{black} PDPM-VAR, lgPDPM-VAR or rgPDPM-VAR}. %The sieves in (\ref{eq:sieves-PDPMmm}), (\ref{eq:sieves-lgPDPM}), and (\ref{eq:sieves-rgPDPM}), are similar in terms of construction, but differ in terms of how the support of the autocovariance matrices is allowed to grow with the sample size. 
The following results establish entropy bounds that are vital to establishing strong consistency. 

 \vskip 6pt

{\noindent \bf Theorem 4:} {\it The entropy bound for sieves (\ref{eq:sieves-PDPMmm}) satisfies $\small N(\epsilon,\mathcal{F}_{n,{\bf jl}},|| \cdot||_1)
\lesssim \big(\frac{M^{\nOutcome}}{\epsilon^{-C_1}}\bigg)^{H_n} \times \\
\prod_{\hsig\le H_n} \big\{\frac{2\nOutcome u_{\hsig,l}}{\epsilon^2} \big\}^{\nOutcome(\nOutcome-1)/2}\times  \prod_{h_1\le  H_n}\bigg\{\bigg(\frac{C^*_{h_{1,j},h_{\sigma,l}}\bar{a}_{h_1,j}}{\underline{\sigma}_n\epsilon} +1\bigg)^{\nOutcome^2} - \bigg(\frac{C^*_{h_{1,j},h_{\sigma,l}}\underline{a}_{h_1,j}}{\underline{\sigma}_n\epsilon} -1 \bigg)^{\nOutcome^2} \bigg\}^K $, where constants $C_1>0$ and $C^*_{h_{1,j},h_{\sigma,l}} >0$ that depends on $(\nOutcome,t,K)$.
} 
 
 \vskip 10pt
 
 {\noindent \bf Corollary 1:} {\it The entropy for sieves in (\ref{eq:sieves-rgPDPM}) and (\ref{eq:sieves-lgPDPM}) corresponding to lgPDPM-VAR and rgPDPM-VAR respectively, satisfy $\small N(\epsilon,\mathcal{F}_{n,{\bf jl}},|| \cdot||_1)
\lesssim \mathcal{K}^* \bigg(\frac{M^\nOutcome}{\epsilon^{-C_1}}\bigg)^{H_n}\times \prod_{\hsig\le H_n} \big\{\frac{2\nOutcome u_{\hsig,l}}{\epsilon^2} \big\}^{\nOutcome(\nOutcome-1)/2}$, where $C_1$ is understood to vary depending on the specific variant of the PDPM model used.}

\vskip 10pt 

{\noindent \bf Remark 2:}  We have $\mathcal{K}^*=\prod_{h_{11}\le H_n}\cdots \prod_{h_{1K}\le H_n}\bigg\{\big(\frac{C^{**}_{h_{1,j},h_{\sigma,l}}{\color{black}\bar{a}_{h_{1k},j}}}{\underline{\sigma}_n\epsilon} +1\big)^{\nOutcome^2} - \big(\frac{C^{**}_{h_{1,j},h_{\sigma,l}}{\color{black}\underline{a}_{h_{1k},j}}}{\underline{\sigma}_n\epsilon} -1 \big)^{\nOutcome^2} \bigg\}$ corresponding to sieves (\ref{eq:sieves-lgPDPM}), and $\mathcal{K}^*= \prod_{h_{11}\le H_n}\cdots \prod_{h_{1D}\le H_n} \bigg\{\big(\frac{\tilde{C}^*_{h_{1,j},h_{\sigma,l}}\bar{a}_{h_{1d'},j}}{\underline{\sigma}_n\epsilon_2} +1\big)^{\nOutcome} - \big(\frac{\tilde{C}^*_{h_{1,j},h_{\sigma,l}}\underline{a}_{h_{1d'},j}}{\underline{\sigma}_n\epsilon_2} -1 \big)^{\nOutcome} \bigg\}^K$ corresponding to sieves (\ref{eq:sieves-rgPDPM}) in Corollary 1, where $C^{**}_{h_{1,j},h_{\sigma,l}}>0$, and $\tilde{C}^*_{h_{1,j},h_{\sigma,l}}>0$ depends on $(\nOutcome,T,K)$. %The constant $C_1$ in the entropy bounds is understood to be different corresponding to sieves in (\ref{eq:sieves-PDPMmm}), (\ref{eq:sieves-lgPDPM}), and (\ref{eq:sieves-rgPDPM}).}
 
\vskip 6pt 

%{\noindent \bf Corollary 2:} {\it The entropy corresponding to sieves (\ref{eq:sieves-rgPDPM}) satisfies
%$\small N(\epsilon,\mathcal{F}_{n,{\bf jl}},|| \cdot||_1)
%\lesssim  \bigg(\frac{M^{\nOutcome}}{\epsilon_2^{C_1}}\bigg)^{H_n} \times \prod_{\hsig\le H_n} \big\{\frac{2\nOutcome u_{\hsig,l}}{\epsilon_2^2} \big\}^{\nOutcome(\nOutcome-1)/2}  \prod_{d'=1,h_{1d'}\le H_n}^\nOutcome \Bigg\{\bigg(\frac{\tilde{C}^*_{h_{1,j},h_{2,l}}\bar{a}_{h_{1d'},j}}{\underline{\sigma}_n\epsilon_2} +1\bigg)^{\nOutcome} - \bigg(\frac{\tilde{C}^*_{h_{1,j},h_{2,l}}\underline{a}_{h_{1d'},j}}{\underline{\sigma}_n\epsilon_2} -1 \bigg)^{\nOutcome} \Bigg\}^K$, where $C_1>0$ and $\tilde{C}^*_{h_{1,j},h_{2,l}}>0$ that depends on $(\nOutcome,t,K)$ {\color{black} check $\tilde{C}^{*}$ notation in proof\ldots}}}.\\

%{\noindent \bf Corollary 2:} {\it The entropy corresponding to sieves (\ref{eq:sieves-rgPDPM}) satisfies
%$\small N(\epsilon,\mathcal{F}_{n,{\bf jl}},|| \cdot||_1)
%\lesssim  \bigg(\frac{M^{\nOutcome}}{\epsilon_2^{C_1}}\bigg)^{H_n} \times \prod_{\hsig\le H_n} \big\{\frac{2\nOutcome u_{\hsig,l}}{\epsilon_2^2} \big\}^{\nOutcome(\nOutcome-1)/2}  \prod_{d'=1,h_{1d'}\le H_n}^\nOutcome \Bigg\{\bigg(\frac{\tilde{C}^*_{h_{1,j},h_{\hsig,l}}\bar{a}_{h_{1d'},j}}{\underline{\sigma}_n\epsilon_2} +1\bigg)^{\nOutcome} - \bigg(\frac{\tilde{C}^*_{h_{1,j},h_{\hsig,l}}\underline{a}_{h_{1d'},j}}{\underline{\sigma}_n\epsilon_2} -1 \bigg)^{\nOutcome} \Bigg\}^K$, where $C_1>0$ and $\tilde{C}^*_{h_{1,j},h_{2,l}}>0$ that depends on $(\nOutcome,t,K)$ {\color{black} check $\tilde{C}^{*}$ notation in proof\ldots}}.\\

Having established the entropy bounds, the next step is to propose sensible base measures that satisfy the tail conditions and summability constraints in Theorem 2. These base measures are characterized via conditions on the tail probabilities that are elaborated below, and include some commonly used choices as discussed in the sequel. %These conditions will be directly relevant for verifying the sufficient conditions in Theorem 2 for establishing strong consistency.

%We note that these assumptions are relatively mild and are similar to those used already used extensively in non-parametric multivariate density estimation literature.\\
{\it \noindent (B1)} The base measures corresponding to $P_{\mathcal{S}}$ in (\ref{eq:prodDP}), (\ref{eq:prodDP3}) and (\ref{eq:prodDP2}) satisfy $\small P^*_2(\lambda_1(\Sigma^{-1}_{h_\sigma})>x^*)\lesssim \exp(-c_1 (x^*)^{c_2})$, $P^*_2(\lambda_\nOutcome(\Sigma^{-1}_{h_\sigma})<1/x^*)\lesssim(x^*)^{-c_3}$, $P^*_2\big(\frac{\lambda_1(\Sigma^{-1}_{h_\sigma})}{\lambda_\nOutcome(\Sigma^{-1}_{h_\sigma})}>x^*\big)\lesssim (x^*)^{-\kappa}$, for some positive constants $c_1,c_2,c_3,\kappa$, and corresponding to the cluster $h_\sigma$.\\
{\it \noindent (B2)}  The base measure corresponding to $P_{\mathcal{\Theta}}$ specifies independence across lags, and satisfies the following tail conditions: (i) under PDPM-VAR, $\small  P^*_1(||vec(A_{k,h_1}) ||>x^*)\lesssim (x^*)^{-2(r+1)}$  for cluster $h_1$; (ii) under lgPDPM-VAR, $\small P^*_{1k}(||vec(A_{k,h_{1k}}) ||>x^*)\lesssim (x^*)^{-2(r+1)}$ for cluster $h_{1k}$; and (iii) under rgPDPM-VAR, $P^*_{1d'}(||vec\{A'_{1,d'\bullet,h_{1,d'}},\ldots,A'_{K,d'\bullet,h_{1,d'}}\} ||>x^*)\lesssim (x^*)^{-2(r^*+1)}$ corresponding to cluster $h_{1d'}$, for some constants $r,r^*>0$, and $d'=1,\ldots,D$.

\vskip 4pt

The above conditions on the base measures are very reasonable and  hold for commonly used distributions on autocovariance matrices (such as Gaussian and Laplace), as well as inverse-Wishart distribution corresponding to $P^*_2$.  These tail conditions are also satisfied by certain low rank decompositions for the covariance, such as a factor model form ($\Sigma = \Lambda\Lambda^T + \Omega$ where the $D\times B$ matrix $\Lambda$ contains $B<<D$ factors), which is particularly suitable for scaling up the approach to higher dimensions. Such low rank representations are routinely used for dimension reduction in the factor model literature \citep{ghosh2009default}. Denote $\mathcal{A}_{d',h_{1,d'}}=vec\{A'_{1,d'\bullet,h_{1,d'}},\ldots,A'_{K,d'\bullet,h_{1,d'}}\}$ and let $DE$ denote a double exponential prior. The following Lemmas formalize the above discussions on the base measures. 

\vskip 3pt

{\noindent \bf Lemma 2}: {\it Condition {\it(B2)} holds when  $ P^*_1(vec(A_{k,h_1}))$ is specified as $ N_{\nOutcome^2}(vec(A_k);\bfmu,\Lambda)$ with $\small \Lambda\sim IW(\Lambda_0,\nu_\lambda)$ corresponding to PDPM-VAR, and for a similar choice of $P^*_{1k}(vec(A_{k,h_{1k}}))$ under lg-PDPM-VAR. It is also satisfied when $P^{**}_{1d'}(\mathcal{A}_{d',h_{1,d'}})$= $\prod_{k=1}^K N_{\nOutcome}(A_{k,d'\bullet,h_{1,d'}};\bfmu,\Lambda_{d'}), \Lambda_{d'}\sim IW(\Lambda_{0d'},\nu_{\lambda,d'})$, under rgPDPM-VAR. Further, {\it(B2)} also holds if the above base measures are changed to a product of independent DE($\lambda$) priors with suitably large $\lambda$.}

\vskip 6pt

{\noindent \bf Lemma 3}: {\it Condition {\it(B1)} holds when for cluster $h_\sigma$, $\small P^*_2(\Sigma_{h_\sigma})= IW(\Sigma_{h_\sigma};\Sigma_0,\nu_\sigma)$, as well as under the low rank representation $\Sigma_{h_\sigma}=\Gamma_{h_\sigma}\Gamma_{h_\sigma}^T + \Omega_{h_\sigma}$ where $\Gamma_{h_\sigma}$ is $D\times B$ and $\Omega_{h_\sigma}=diag(\sigma^2_{1,h_\sigma},\ldots,\sigma^2_{\nOutcome,h_\sigma})$, and  $P^*_2(\Sigma_{h_\sigma})$= $\big\{\prod_{d'=1}^\nOutcome\prod_{m'=1}^B N(\gamma_{d'm',h_\sigma};0,1)\big\}\big\{\prod_{j=1}^\nOutcome Ga(\sigma^{-2}_{j,h_\sigma};a_\sigma,b_\sigma) \big\}$.}

\vskip 6pt

The proof of Lemma 2 is provided in the Appendix, while that of Lemma 3 follows directly from Corollaries 1 and 2 in \cite{canale2017posterior}. We note that $B<< D$ in Lemma 3 ensures a reduced rank structure on the residual covariance matrix. 

One can now use the entropy bounds derived in Theorem 4 and Corollary 1 along with  tail conditions in {\it (B1)-(B2)} to establish our strong consistency under a broad class of base measures, by applying Theorem 2. Our strong consistency result is stated below.

\vskip 6pt

{\noindent \bf Theorem 5:} {\it Suppose Theorem 3 holds, and {\it (B1)-(B2)} are satisfied. Then for suitably large constants $ r,r^*, \kappa$, the posterior distribution corresponding to the PDPM-VAR, lgPDPM-VAR and rgPDPM-VAR are strongly consistent at $f_0$ under suitable choice of sequences $\{H_n \}, \{ M_n\}, \{\underline{\sigma}_n \} \{\underline{a_{h_1,j}}\}, \{\bar{a}_{h_1,j} \}, \{ \underline{u_{\hsig,j}}\},\{ u_{\hsig,j}\}$ in the sieves (\ref{eq:sieves-PDPMmm}), (\ref{eq:sieves-rgPDPM}), and (\ref{eq:sieves-lgPDPM}).}

\vskip 6pt

{\noindent \bf Remark 3:} In mathematical terms, strong posterior consistency can be written as $\Pi(\{ f: d(f,f_0)>\epsilon_f \} \mid X^{(1)},\ldots, X^{(n)}) \to 0 $ as $n\to \infty$ in $F_0^n$ probability for any $\epsilon_f>0$.

\vskip 6pt

 {\noindent \bf Remark 4:} While Theorem 5 is stated in terms of general class of base measures that satisfy {\it (B1)-(B2)}, we rely on commonly used base measures outlined in Lemmas 2-3 for implementing the proposed approach. We elaborate these choices in the next section.

%{\noindent \bf Remark 3:} {\color{black}The strong consistency result in Theorem 4 corresponding to PDPM-VAR is satisfied when the sequences for $\mathcal{F}_{n,{\bf j}, {\bf l}}$ are chosen as: $M_n=\underline{\sigma}_n^{-2c_2}=n$, $H_n=\lfloor Cn\epsilon^2/\log(n) \rfloor$ for some positive constants $C$ and $c_2$, $\underline{a}_{h_1,j}= n^{H_n}(j_{h_1}-1), \bar{a}_{h_1,j}= n^{H_n}j_{h_1}$,  $u_{\hsig,l}=n^{l_{\hsig}}$, $\underline{u}_{\hsig,l}=n^{l_{\hsig}-1}$. The choices for these sequences are made similarly for the other variants.}

\section{Posterior Computation}\label{sec:postcomp}

We outline the posterior computation steps to fit all proposed VAR models that is the main focus of this work. Our approach alternates between sampling parameters related to the autocovariance matrices and the residual covariance matrix.  For all models, we update the autocovariance parameters row-wise for one outcome at a time. For the PDPM-VAR, rgPDPM-VAR, and lgPDPM-VAR we use a hierarchical representation of  Laplace base measures \citep{park2008bayesian}. Under these base measures, these autocovariance elements follow independent $DE(\lambda)$ distributions \citep{park2008bayesian}. Explicit details are provided in Appendix Section 3.

In order to scale up the implementation of the proposed method to high dimensional applications, we use a reduced rank factor model representation for the residual covariance matrix in our implementation, which provides a desired balance between computational scalability and theoretical flexibility. In particular, such a low rank structure on the residual covariance does not adversely impact the accuracy of parameter estimates compared to an unstructured covariance matrix, in our experience involving extensive numerical experiments with true unstructured residual covariances. Further, it is considerably more flexible and results in greater accuracy compared to a diagonal residual covariance that is routinely used in VAR literature \citep{kook2021bvar} but may be restrictive in practical applications. In particular, we specify $	\Sigma_i = \FactorLoadings_i  \FactorLoadings_i'  +\LowRankDiagCov_i$,
where $\FactorLoadings_i$ is a $D \times \nLatentFactor$ factor loadings matrix with $\nLatentFactor (<< D)$ factors, and  $\LowRankDiagCov_i$ is $\text{diag}\{ \sigma_{i,1}^2, \ldots,  \sigma_{i,D}^2 \}$.To facilitate posterior computation, we use the following parameter expanded version of the model, 
\begin{align*}
	\mathbf{x}_{i,t} &= \sum_{k=1}^{\min\{t-1, K\}} A_{ik} \mathbf{x}_{i,t-k} + \FactorLoadings_i^* \SubjectLatentFactor_{i,t}^* + \boldsymbol{\epsilon}^*_{i,t}, \;
	\SubjectLatentFactor_{i,t}^* \sim N(0, \RedundantTerm_i), \;
	\boldsymbol{\epsilon}^*_{i,t} \sim N(0, \LowRankDiagCov_i),
\end{align*}
where $\RedundantTerm_i = \text{diag} \{   \redundantTerm_{i,1}, \ldots,  \redundantTerm_{i,\nLatentFactor} \}$. Under the low rank representation, we impose DP mixture priors on $(\FactorLoadings_i^*$, $\RedundantTerm_i$, $\LowRankDiagCov_i$) leading to a mixture prior on $\Sigma_i$. This corresponds to the prior $\Sigma_i \sim \sum_{h_\sigma=1}^{\infty} \pi_{\sigma,h_\sigma}\delta_{(\FactorLoadings^*_{h_\sigma},\RedundantTerm_{h_\sigma},\LowRankDiagCov_{h_\sigma})}$, where $(\FactorLoadings^*_{h_\sigma},\RedundantTerm_{h_\sigma},\LowRankDiagCov_{h_\sigma})\sim P^*_2\equiv P_{\FactorLoadings^*} \times P_{\RedundantTerm} \times P_{\LowRankDiagCov}$. Here $P_{\FactorLoadings^*}$ is a product of independent standard normal distributions, $P_{\RedundantTerm}$ is a product of independent $Gamma(1/2, 1/2)$ distributions yielding a half-Cauchy prior on the diagonal elements of $\FactorLoadings$ and a Cauchy prior on the lower-off-diagonal elements as in \cite{ghosh2009default}, and the inverse of the diagonal elements of $\LowRankDiagCov$ have independent $Gamma(\alpha_\sigma, \beta_\sigma)$ priors. %Note that here, by diagonal elements of $\FactorLoadings_i^*$, we refer to elements $\Gamma_{i, 1, 1}, \ldots, \Gamma_{i, B, B}$. 

%This combination of priors induces a half Cauchy prior for the diagonal elements of $\FactorLoadings$ and Cauchy priors for its lower-triangular elements - {\it \color{red} requires more explanation}.

\section{Simulation Studies}

%We conduct extensive simulations to investigate the performance of the proposed PDPM-VAR variants. 
We compared the performance under the proposed approaches to a state-of-the-art single-subject VAR approach, as well as an ad-hoc clustering extension of the single subject VAR model that is able to borrow information across samples. We generated data for $n = 100$, $200$, $D = 100$, $T_i =250$, and different levels of sparsity within the autocovariance matrices were considered (75\% and 90\%). For each data generation setting we generate 25 simulation replicates, and for all settings the true VAR model involved $K=2$ lags.  We consider four settings for generating the subject-level autocovariance matrices that differ with respect to the clustering structure. Settings 1-3 represent the PDPM-VAR, lgPDPM-VAR and rgPDPM-VAR scenarios respectively, while Setting 4 represents a more heterogeneous setting that is obtained via introducing additional random noise to the autocovariance elements generated under Setting 3. For Setting 1, we use 3 autocovariance clusters, for Setting 2 we use 3 clusters for lag 1 and 2 clusters for lag 2, and for Settings 3-4 we vary the true number of clusters randomly (between $2-5$) across rows of the autocovariance matrix, and the elements of these matrices are generated randomly in order to ensure a stable time series. In Setting 3, subjects within a cluster share the exact same elements for the corresponding rows of the autocovariance matrix, whereas in Setting 4 the subject-level rows of the autocovariance matrix within a cluster are random deviations from a shared mean row. Each cluster's residual covariance matrix was generated from an Inverse Wishart distribution with $D$ degrees of freedom and diagonal scale matrix with elements equal to $D/2$. Subject level time courses were obtained by starting with random values for the multivariate observation at the first time point, and subsequently generating future observations from the assumed true VAR model. For a subject, an additional 5 time scans were generated after the initial $T_i$ observations to evaluate forecasting accuracy.

\subsection{Approaches and Performance Metrics} 

We compare the proposed approaches to the single subject Bayesian VAR (SS-VAR) model developed in \cite{ghosh2018high}, which separately models the time courses for each subject. %under a matrix-normal prior on the autocovariance matrix, and a Wishart prior on the row covariance, and a noninformative prior on the column covariance. %Under a somewhat restrictive assumption that requires the column covariance to be identical to the residual covariance, their approach derives a computationally efficient Gibbs sampler for single VAR modeling that is not equipped to pool information across subjects and is hence expected to result in sub-optimal performance when there are shared patterns across samples. 
We also consider a two-stage clustering extension of this method, where we first estimated subject specific autocovariance and residual covariances under the single VAR approach by \cite{ghosh2018high} and then applied the k--means clustering separately to the vectorized autocovariance and residual covariance matrix estimates. We choose the number of clusters to maximize the silhouette score \citep{rousseeuw1987silhouettes} and we then allocate each sample to one of the k clusters that is based on both the autocovariance terms and the residual covariance estimates from the initial SS-VAR fit. We subsequently concatenate the time courses across all subjects within the same cluster in order to borrow information within cluster, and finally re-fit the SS-VAR model to this concatenated data separately for each cluster. Since the true clustering structure was assumed to be unknown when fitting the model, it was not possible to compare the performance with existing multi-subject VAR modeling methods that assume known groupings \citep{chiang2017bayesian,kook2021bvar}.

 %Posterior computation for all methods was carried out using MCMC with 1000 burn-in samples and 6000 MCMC samples. For both the single subject VAR models and our proposed PDPM-VAR variants, we obtain initial values for the autocovariance matrices by fitting {\color{blue}OLS regression models separately to each outcome across all time points. For time point $t$, the $D\times K$ predictor variables included in the model are the observed node time courses for time points $(t-1), \ldots, (t-K)$}. For our proposed PDPM-VAR models, we then obtain initial cluster memberships using k--means clustering with $k=10$. {\color{blue} We follow a similar procedure to obtain initial values for the residual covariance clusters. We first calculate subject-specific sample covariance estimates, which are used as the starting values for the single subject VAR models. These subject-specific estimates are then clustered using k--means clustering with $k=10$ to obtain initial cluster memberships for the PDPM-VAR models.} {\color{black} The above initialization approach for cluster indices resulted in better performance across all methods compared to randomly initializing the cluster memberships.} % {\color{black} \it I commented out this para since this information can be presented in SM or skipped if needed ...}

We evaluate performance in terms of (1) autocovariance estimation accuracy, (2) clustering accuracy, (3) feature selection for identifying structural zeros in the autocovariance matrices, and (4) forecasting accuracy. Following \cite{ghosh2018high}, we measure estimation accuracy using the relative L2 error of the estimates to the true estimates.  Clustering accuracy is measured using the adjusted Rand index \citep{rand1971objective},  which measures agreement between the assigned and true cluster labels, adjusted for chance agreement.  Feature selection performance is evaluated via area under the receiver operating characteristic (RoC) curve and precision recall curve (PRC).  To calculate both curves we considered a sequence of significance thresholds, and for each threshold, we examined the corresponding credible interval to infer the significance. The corresponding sequence of sensitivity versus 1-specificity values were plotted over varying thresholds in order to obtain the ROC curve, while the PRC was obtained by plotting the positive predictive value ($1 - FDR$) against sensitivity (FDR denotes the false discovery rate). Finally, forecasting accuracy is measured via the relative L2 error of the predicted time courses for time scans  $T_i+1,\ldots, T_i+5$. The MCMC chains converged for all methods as assessed using Dickey-Fuller tests of stationarity, %\citep{dickey1979distribution}
 although the results are not displayed due to space constraints.
 %{\color{black} We also report the computation times under the proposed variants as well as the competing approaches. {\it insert}} {\color{blue} - need to discuss first}

\subsection{Simulation Results}

\begin{figure}
\centering
\includegraphics[width=1.0\linewidth, height=3.5in]{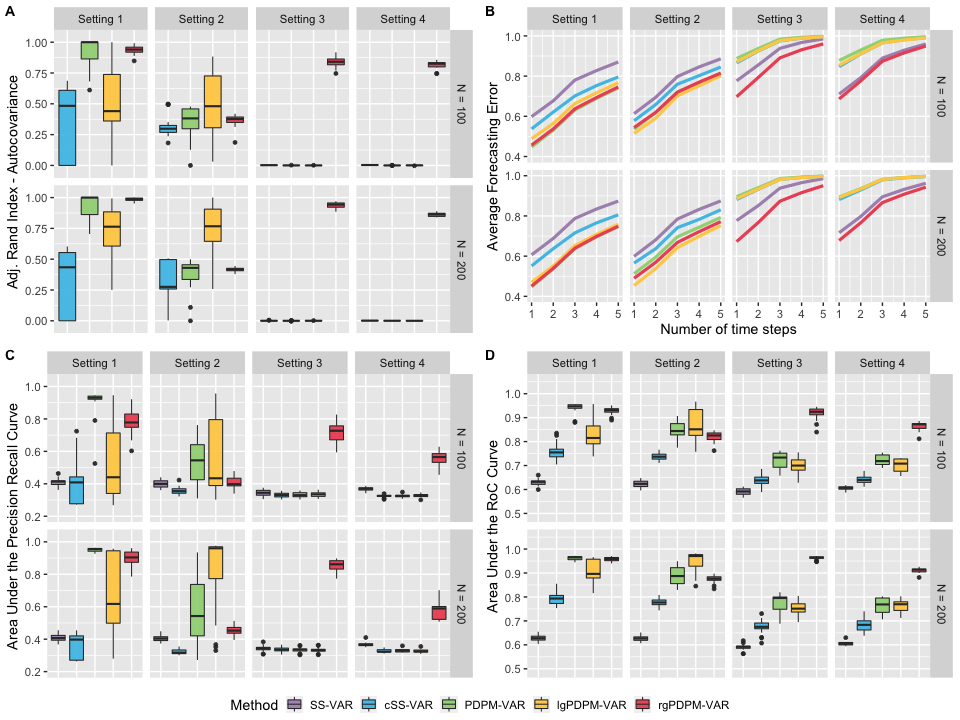}
\caption{\footnotesize{Simulation results for $D=100$, $T=250$ case with sparsity level $0.75$. Panel A displays the adjusted Rand index for clustering the autocovariance. Panel B displays the forecasting error. Panels C and D display the area under the PR and RoC curves for identifying autocovariance non-zero elements.}}
\label{sim:ARIPRCROC}
\vskip 10pt

\centering
\includegraphics[width=1.0\linewidth, height=2.5in]{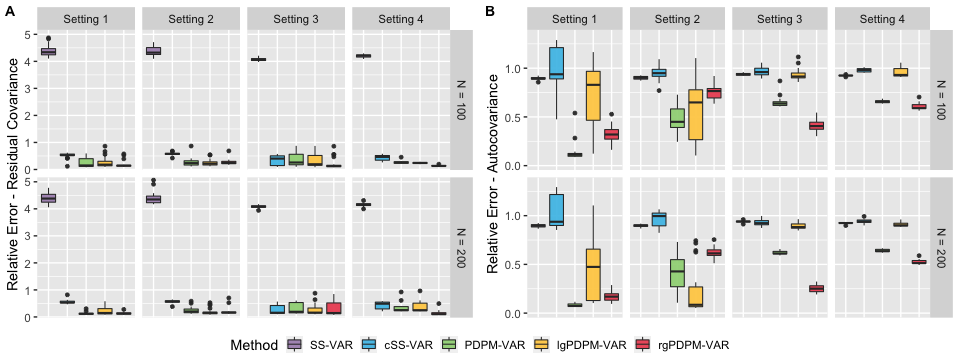}
\caption{\footnotesize{Relative L1 error for estimating the residual covariance (Panel A) and the subject-specific autocovariance matrices (Panel B) for $D=100$, $T=250$ case with sparsity level $0.75$. }}
\label{sim:L1}
\end{figure}

Simulation results are presented in Figures \ref{sim:ARIPRCROC}--\ref{sim:L1}. Due to space constraints, we provide the simulation results for the most challenging case ($D=100$, $T = 250$) at the 75\% sparsity level here and present the other cases in the Appendix. Several general patterns are clear from the results. First, the clustering performance for the autocovariance depends heavily on the true clustering structure (Figure \ref{sim:ARIPRCROC}, Panel A), with the PDPM-VAR, lg-PDPMVAR, and rgPDPM-VAR generally outperforming the other approaches when the data is generated from Settings 1-3 respectively. However, the rgPDPM-VAR often has close to optimal clustering performance when the PDPM-VAR is the true model and it also performs the best for the heterogeneous Setting 4, which reflects the generalizability of this variant. Critically, when there are differences in clustering across the different outcomes corresponding to more heterogeneous scenarios (e.g. Settings 3-4), only the rgPDPM-VAR is able to achieve a good clustering score.  Finally, across all settings, the SS-VAR with clustering has the worst performance, demonstrating that the ad hoc two-stage analysis procedure is not able to accurately pool information across subjects.

The areas under the ROC and PR curves (Figure \ref{sim:ARIPRCROC}, Panels C and D) illustrate a consistently superior feature selection performance under the three proposed variants compared to the single subject VAR model with and without clustering. As expected, the PDPM-VAR, lgPDPM-VAR and rgPDPM-VAR approaches have higher area under the ROC and PR curves when the data is generated from Settings 1-3 respectively. In addition, the rgPDPM-VAR often has comparable area under the curve with PDPM-VAR for $n=200$ under Setting 1 and the best performance under the more heterogeneous Setting 4. These results imply the ability of rgPDPM-VAR to accurately identify the sparsity structure of the autocovariance with a low risk of false discoveries for data with unknown clustering.

When estimating the residual covariance matrices (Figure \ref{sim:L1}, Panel A), all three proposed approaches are able to heavily outperform the SS-VAR model. The performance under the three proposed approaches is generally comparable, with the rgPDPM-VAR outperforming the others in the more heterogeneous Settings 3 and 4. In addition, the SS-VAR approach with initial clustering has a higher relative error compared to the rgPDPM-VAR for the vast majority of cases, although it occassionally has a slightly improved performance in Setting 3. We conjecture that this is due to the assumed full rank structure for the residual covariance that is modeled via an inverse-Wishart distribution under the SS-VAR, which aligns with the true data generation scenario, in contrast to the assumed low-rank structure on the PDPM-VAR. Unfortunately, the SS-VAR approach with clustering has extremely poor performance in terms of autocovariance estimation (Figure \ref{sim:L1} B), while the PDPM-VAR, lgPDPM-VAR, and rgPDPM-VAR approaches typically have the lowest errors when the data is generated from Settings 1-3 respectively. The rgPDPM-VAR method also has the best autocovariance estimation performance under the more heterogeneous Setting 4. 

%does not translate to autocovariance estimation (Figure \ref{sim:L1}, Panel B). This approach has by far the highest relative error for estimating the autocovariance elements, , with rgPDPM-VAR also performing the best under the more heterogeneous Setting 4. %Moreover, with the exception of Setting 2, lgPDPM-VAR has the worst performance among the three PDPM-VAR variants.

Figure \ref{sim:ARIPRCROC} Panel B displays the forecasting error for each of the autocovariance clustering setups, averaged over the sparsity level and the number of time points per subject. With the exception of Setting 2 where lgPDPM-VAR performs best, the rgPDPM-VAR approach has the best or close to optimal forecasting performance for other settings. Moreover, in the more heterogeneous Settings 3-4, the SS-VAR method with initial clustering has better forecasting accuracy compared to the PDPM-VAR and lgPDPM-VAR approaches, although it can not outperform the rgPDPM-VAR method. The relative forecasting performance levels off for greater than three time steps for all approaches, as expected.
 
{\noindent \underline{Synopsis of findings:}}
 Overall, the rgPDPM-VAR provides a desirable balance between model parsimony and accurate estimation and inference across various degrees of heterogeneity across samples. The advantages under the rgPDPM-VAR are most pronounced under the heterogeneous Settings 3 and 4, and it often has close to optimal performance in Setting 1 for larger $n$. This illustrates the advantages of pooling information across subjects, while accommodating varying levels of heterogeneity at the level of the rows of the autocovariance matrix. While the SS-VAR approach with ad-hoc clustering is also able to pool information, it is highly sensitive to the clustering accuracy in the first step, and it can not capture clustering uncertainty, resulting in inferior performance. 

\section{Analysis of Human Connectome Project Data}

\subsection{Analysis Description}

We use the rgPDPM-VAR approach to investigate effective connectivity differences between individuals with high and low fluid intelligence (FI) using a subset of resting-state fMRI  data from the Human Connectome Project. Preprocessing details for these data can be found in \citep{smith2013resting}. We adopt the 360-region Glasser atlas for parcellation as in \cite{akiki2019determining}, where each node has a corresponding time course with $T=1200$. We centered and scaled the subject level time courses for each node before analysis, and verified that each node's time course was stationary using Dickey-Fuller tests. We grouped the brain nodes into one of 6 well known functional brain networks \citep{akiki2019determining}, and fit the VAR model with lag-1 on each of these networks. We selected the lag 1 model following previous literature on VAR models applied to fMRI data \citep{kook2021bvar}, and based on the poor temporal resolution of the fMRI data. We restrict our analysis to a subset of samples with the highest 10\% and lowest 10\% fluid intelligence scores, with $n=306$ samples. We note that the grouping information was only used for post-model fitting comparisons in effective connectivity across groups. We used 1500 burn-in and 3500 MCMC iterations.

%Of these, previous evidence has implicated the XXX and XXX in fluid intelligence, and thus we restrict our attention to these brain networks for subsequent analyses. % note: if go with this justification we have evidence from ....

%We selected a subset of subjects with either low or high fluid intelligence (FI) for analysis. FI was measured using the Penn Matrix Reasoning Task A, or PMAT24\_A\_CR. We restricted our sample to subjects in the lowest $10\%$ and highest $10\%$ of FI. This corresponded to inclusion of subjects with FI $\leq 10$ in the ``low" group and FI $\geq 22$ in the ``high" group. The total number of subjects included in the analysis was 306. We apply the rg-PDPMVAR and the SS-VAR approaches to estimate the VAR(1) model parameters for each sample in this data. We fit the model for each network separately. We use XXX burnin iterations and XXX MCMC iterations for each method. 

To the best of our knowledge, our approach for analyzing fluid intelligence-related effective connectivity differences using heterogeneous multi-subject data is one of the first such attempts. Most existing approaches involve a single-subject VAR analysis, and subsequently these estimates are combined to estimate between-subject variations and examine group differences \citep{deshpande2009multivariate}. There are a handful of approaches for estimating effective connectivity by pooling information across multiple subjects, however they assume known groups \citep{chiang2017bayesian} with limited heterogeneity within groups, and have similar limitations as outlined in the Introduction. Our analysis using the rgPDPM-VAR model is able to compute effective connectivity for multiple samples without any given group labels and can account for heterogeneity in an unsupervised manner. We compare the performance with a SS-VAR approach that analyses each sample separately, and subsequently performs permutation tests to assess significant differences (10,000 permutations). For both methods, false discovery rate control was applied to obtained significant elements.

In addition to investigating effective connectivity differences, we are interested in the clustering reliability and biological reproducibility of our findings. We report clustering reliability over two distinct MCMC runs, that are designed to evaluate the reliability of the clusters discovered by rgPDPM-VAR. As discussed in the introduction, for heterogeneous multivariate measurements, one can expect a subset of nodes/rows to drive the clustering whereas for other nodes the clustering patterns likely hold little information. We calculate the ARI for the node-level clustering across the two MCMC runs to investigate this aspect of clustering reliability. To assess biological reproducibility, we conduct our VAR analysis for two scans collected from each individual using different phase-encodings (LR1 and RL1), with the expectation that the parameter estimates should be similar corresponding to the two scans. We examine the correlation of the estimated autocovariance elements across the two runs (LR1 and RL1) under both the SS-VAR and the rgPDPM-VAR, with high correlation providing evidence that the findings are reproducible.

%We assess the reproducibility of our approach using two techniques. As discussed in the introduction, it is likely that only some nodes/rows have informative clustering patterns across samples, whereas for other nodes the clustering patterns likely hold little information. One possible concern is if the results under the rg-PDPMVAR are reproducible across multiple runs of the MCMC due to the potentially ambiguous clustering of some autocovariance rows. We refer to this as clustering reliability, as we assess this by repeating the analysis a second time using the same data.  Second, to assess the biological reproducibility of the findings, we repeat the analysis using a different set of the HCP data. This data is collected from the same subjects at rest, but using a different phase-encoding direction. The original data were collected in left-to-right order (or LR1) and the new data are collected in right-to-left order (RL1). We examine the correlation of the estimated rows of the autocovariance matrices across the two runs (LR1 and RL1) under both the SS-VAR and the rgPDPM-VAR, with high correlation providing evidence that the findings are reproducible. 

%To assess the reproducibility of our approach, we conducted our analysis a second time using a different set of HCP resting state data from this same set of subjects collected using a different set of acquisition parameters and compared the results across the two analyses. % TODO comment of ARI depending on which LR1 LR2 RL1 RL2 setup we use.

\subsection{Results}

\begin{figure}
    \centering
    \begin{tabular}{ccc}
    \includegraphics[width=0.32\linewidth]{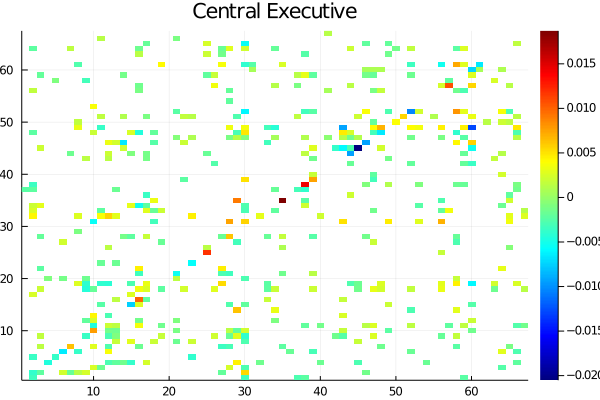} & \includegraphics[width=0.32\linewidth]{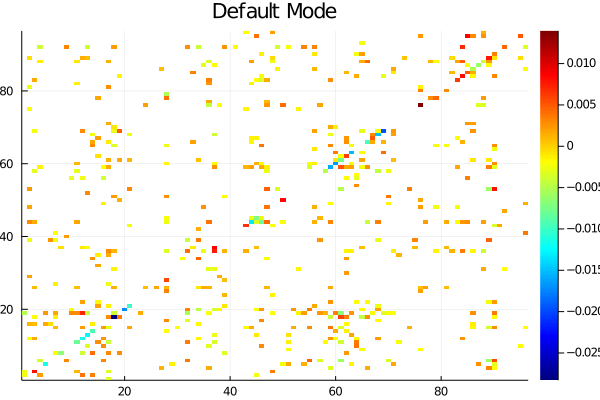} &
    \includegraphics[width=0.32\linewidth]{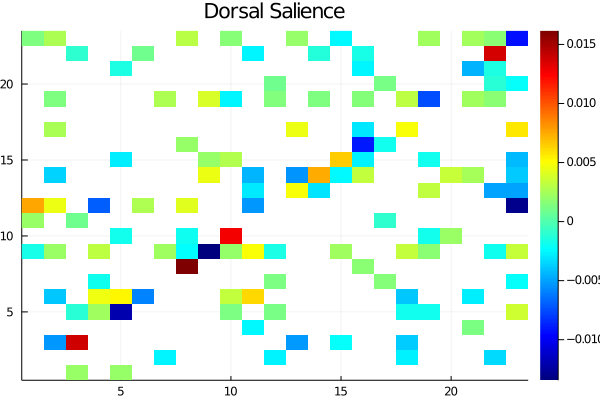} \\ \includegraphics[width=0.32\linewidth]{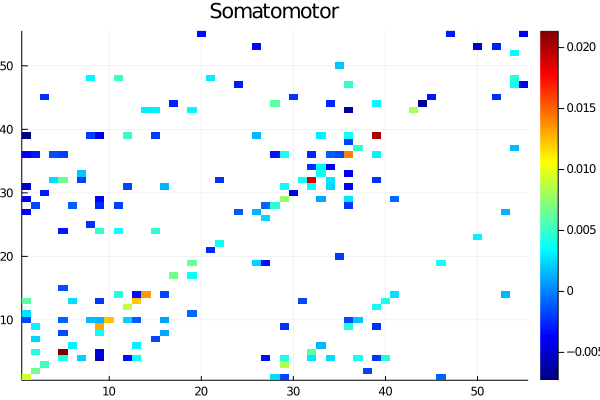} &
    \includegraphics[width=0.32\linewidth]{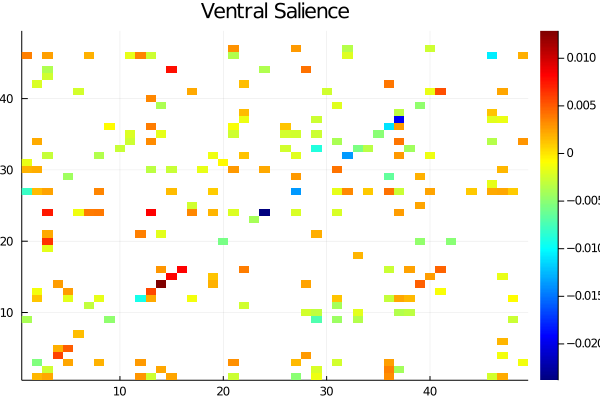} & \includegraphics[width=0.32\linewidth]{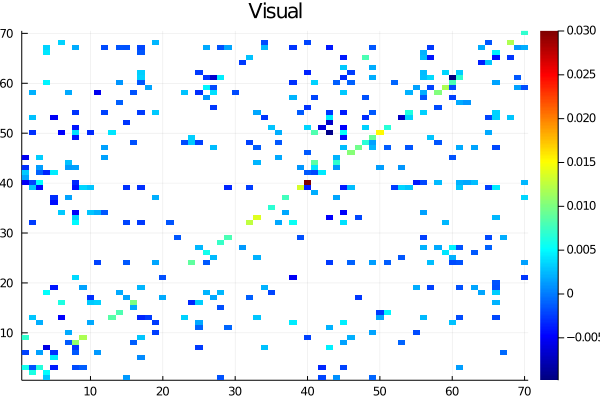}
    \end{tabular}
    \caption{Elements of the autocovariance matrices exhibiting significant differences between the low and high FI groups. The color of the element represents the strength of the mean difference between groups (high FI $-$ low FI), with white elements corresponding to non-significant elements.}
    \label{fig:rgpdpmvarSigNet}
\vskip 10pt

    \begin{tabular}{|c|cc|}
    \hline Clustering Reliability & \multicolumn{2}{c|}{Biological Reproducibility} \\ \hline
    \includegraphics[width=0.3\linewidth]{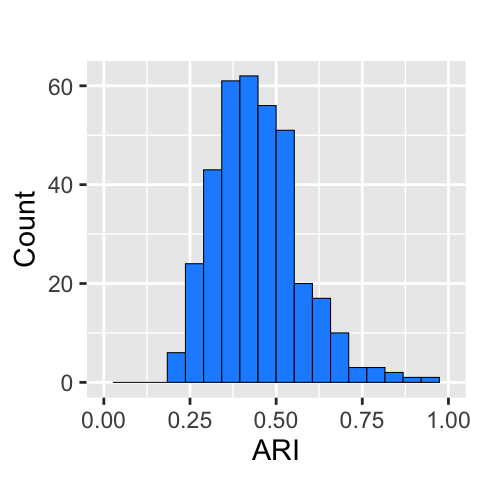} & \includegraphics[width=0.3\linewidth]{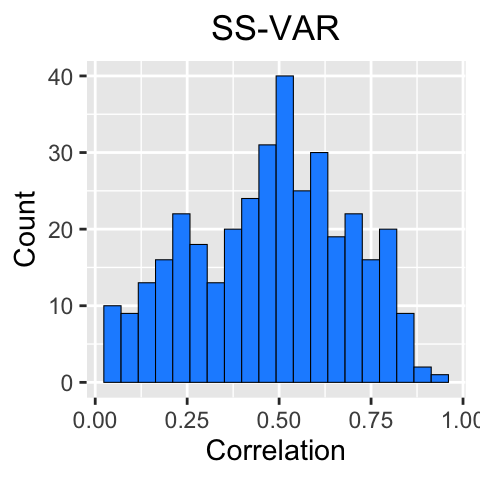}  & \includegraphics[width=0.3\linewidth]{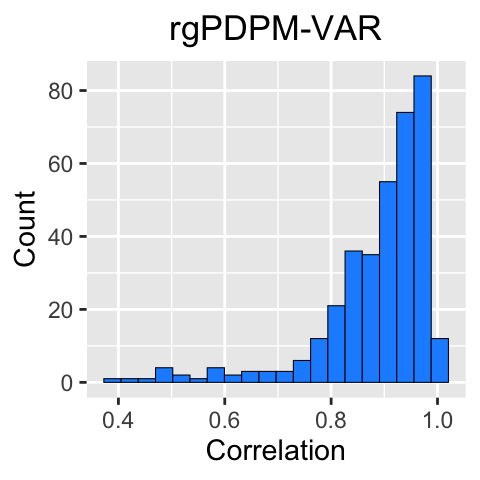} \\\hline
    \end{tabular}
        \setlength{\belowcaptionskip}{-10pt}
    \caption{Adjusted Rand index across two runs of the analysis of the HCP LR1 data for assessing clustering reliability (left). Correlation between the rows of $A_i$ across two different HCP data sets (right). The first run of the analysis was on the LR1 phase-encoding data and the second on RL1.}
    \label{fig:crossrunperf}
\end{figure}

% TODO = now using larger number of perms, basically everything dropping out for SS VAR
Figure \ref{fig:rgpdpmvarSigNet} displays heatmaps of the significant autocovariance differences between the low and high FI groups under the rg-PDPMVAR, after appropriate FDR control. Several patterns are clear from Figure \ref{fig:rgpdpmvarSigNet}. First, the rg-PDPMVAR is able to identify a large number of significant differences between the two groups  after FDR control. Second, the rg-PDPMVAR finds a large number of strong differences along the diagonal. These correspond to AR(1) coefficients, and it seems sensible that if there are differences between groups at Lag 1 that they would be strongly related to each nodes' own time course. Thirdly, the strongest differences were observed corresponding to the nodes in the Dorsal Salience network, as illustrated in Table \ref{table:selectednodes}. These nodes were identified by looking at  columns of the autocovariance matrix with a large proportion of significant elements, which accounts for the varying sizes for the 6 networks. These findings are consistent with previous evidence, which have suggested the dorsal salience and attention networks to be highly related to fluid intelligence \citep{santarnecchi2017network}. %Fourth, the posterior distributions for the autocovariance elements are multi-modal as reported in Figure \ref{fig:postdens}, which points to the importance of flexible priors on the autocovariance elements that can accommodate varying degrees of heterogeneity.
We note that in contrast, only one significantly different effective connectivity difference between the high and low fluid intelligence groups was reported under the SS-VAR approach. Such results are clearly biologically implausible. %since brain connectivity is known to be associated with FI \citep{finn2015functional}. 
Our overall findings point to the advantages of performing a multi-subject analysis accounting for heterogeneity, over a single subject analysis. %Moreover unlike the rgPDPM-VAR method that can use posterior distributions to infer significant differences, the SS-VAR approach requires additional computationally expensive permutation testing to assess significant differences that inflates the overall analysis speed. %Finally, the rgPDPM-VAR analysis for the Dorsal Salience network took {\color{black} \it xxxx minutes to run on an Intel machine with \ldots }

To examine biological reproducibility, the right--hand side of Figure \ref{fig:crossrunperf} displays histograms of the correlations of the rows of the autocovariance matrices across the two analyses corresponding to the LR1 and RL1 fMRI scans, under the SS-VAR and rgPDPM-VAR. The estimates under the rgPDPM-VAR exhibit a very high degree of correlation, almost entirely $>0.8$. On the other hand, the correlation for the majority of the elements is less than 0.5 under SS-VAR, with only 10 elements registering a correlation greater than 0.8, which implied considerably lower reproducibility overall compared to the multi-subject analysis. Moreover a non-negligible number of nodes had weak reproducibility with correlations less than 0.25 under SS-VAR. In addition, Figure \ref{fig:crossrunperf} (left) displays a histogram of the ARI for clustering each node across two separate runs of MCMC on the LR1 data, which illustrates clustering reliability. In general, most nodes exhibited 9--10 clusters. As hypothesized in the Introduction, we see a pattern in which a subset of nodes exhibited very high clustering reliability across runs ($>0.7$), which supports our hypothesis that only some nodes contribute  meaningfully towards clustering of samples. On the other hand, most of the nodes exhibited relatively moderate clustering reliability (ARI $\approx$ 0.5), which indicates much higher clustering than chance, but not fully consistent clustering across all subjects corresponding to these nodes. We note that given the strong biological reproducibility results, the moderate or low clustering reliability for a subset of nodes in our analysis should be attributed to the fact that these nodes are irrelevant to clustering. This provides further justification for using the rg-PDPMVAR, which is designed to accommodate exactly this kind of clustering structure.  Finally in terms of computation time under the rgPDPM-VAR, a single MCMC iteration on the dorsal salience network required approximately 1 second of computation time on an 8 core 2021 M1 Macbook Air.

%SS-VAR has a high degree of correlation for some nodes, but many of them exhibit very low correlations, indicating a lack of biological reproducibility. %Combined with the lack of biological interpretability in findings, the single subject analysis provides a contrast to the benefits accrued under the rgPDPM-VAR method.

%The figure clearly displays a high degree of correlation across the two runs. These findings are consistent at the level of the entire autocovariance matrix as well (Central Executive $\rho=0.967$, Default Mode $\rho = 0.960$, Dorsal Salience $\rho = 0.981$, Somatomotor $\rho = 0.970$, Ventral Salience $\rho = 0.970$, and Visual $\rho = 0.962$). Taken together, these correlation findings indicate that the rg-PDPMVAR results in highly reproducible estimates, even for cases in which the clustering structure is not particularly informative.

%Area R\_7Am of the Dorsal Salience network exhibited differences between FI groups in roughly $1/3$ of its connections. This brain region is located in the superior parietal cortex. 

%Area R\_8Ad of the Default Mode network had extremely high clustering accuracy (ARI $=0.93$) and a large number of significant differences across groups. This region is part of the dlPFC and one if its distinguishing features is that it is heavily mylinated \citep{glasser2016multi}. High amounts of mylination have been shown to be one of the primary anatomical drivers of intelligence CITE NAT NEURO PAPER.

% Version 2 - One method at a time

% Table of significant elements of the autocovariance matrix
\begin{table}
\centering
\scriptsize{
\begin{tabular}{llr}
  \toprule
Node & Network & Prop. FI Diff. \\ 
  \midrule
L\_PF & Dorsal Salience & 0.43 \\ 
  R\_7Am & Dorsal Salience & 0.35 \\ 
  L\_IFSa & Dorsal Salience & 0.35 \\ 
  L\_PHT & Dorsal Salience & 0.35 \\ 
  R\_PHT & Dorsal Salience & 0.30 \\ 
  R\_PF & Dorsal Salience & 0.30 \\ 
  L\_6a & Dorsal Salience & 0.30 \\ 
  L\_PFt & Dorsal Salience & 0.30 \\ 
  R\_PGs & Central Executive & 0.28 \\ 
  R\_IFSa & Dorsal Salience & 0.26 \\ 
   \bottomrule
\end{tabular}

\begin{tabular}{llr}
  \toprule
Node & Network & Prop. FI Diff. \\ 
  \midrule
  R\_PFt & Dorsal Salience & 0.26 \\ 
  L\_PEF & Dorsal Salience & 0.26 \\ 
  L\_TE2p & Dorsal Salience & 0.26 \\ 
  R\_V3A & Visual & 0.23 \\ 
  R\_PSL & Ventral Salience & 0.22 \\ 
  R\_7PL & Dorsal Salience & 0.22 \\ 
  R\_6r & Dorsal Salience & 0.22 \\ 
  L\_7Am & Dorsal Salience & 0.22 \\ 
  L\_6r & Dorsal Salience & 0.22 \\ 
  L\_V3A & Visual & 0.21 \\ 
   \bottomrule
\end{tabular}

}
    \setlength{\belowcaptionskip}{-10pt}
\caption{Table of the 20 nodes with large proportion of significant effects on other nodes within their network. Note that the proportion is used instead of the raw count to account for the different network sizes.}
\label{table:selectednodes}
\end{table}

% \begin{figure}
%     \centering
%     \begin{tabular}{ccc}
%     \includegraphics[width=0.3\linewidth]{figs/CentralExecutive_L_13l.png} & \includegraphics[width=0.3\linewidth]{figs/CentralExecutive_L_33pr.png} &
%     \includegraphics[width=0.3\linewidth]{figs/CentralExecutive_L_OFC.png} \\ 
%     \end{tabular}
%     \caption{Plot of the posterior density of the AR(1) coefficients for several selected nodes.}
%     \label{fig:postdens}
% \end{figure}

\section{Forecasting of Air Quality Data}

While the fMRI study described above focuses on connectivity between brain regions, fMRI studies are not generally concerned with forecasting accuracy. To demonstrate the forecasting accuracy of our method, we next apply the proposed methods to an open source air quality data set from the EPA (\url{https://www.epa.gov/outdoor-air-quality-data}). The data consist of daily measurements from air quality monitors spread across the United States. For our purposes, we consider the air quality index time series for nitrogen dioxide (NO$_2$), ozone (O$_3$), and carbon monoxide (CO). We used data from sensors having 100 days of consecutive data in 2000 from May 20th to August 27th. A simple kernel regression density plot for the data for each pollutant produced non-Gaussian curves (see Appendix), which motivates the use of non-parametric Bayesian analysis over parametric forecasting models. While our method does not require the data from each sensor to be overlapping, this restriction helps ensure that the forecasting results are due only to the method and not so some other data-dependent difference. Additionally, the relatively short time span helps reduce concerns about non-stationarity. For each time course, the time series were differenced (5 steps), demeaned, and outliers were replaced using the \texttt{tsoutliers} function in R \citep{lopez2019r}. The time courses were then checked for stationarity using a Dickey-Fuller test, and sensors for which the time courses were not stationary were removed from the data. After this procedure, we had 41 sensors with complete data.

We fit the rgPDPM-VAR model to this data using 1500 burn-in samples and 3500 MCMC samples. The concentration parameter was set to 5 to encourage more clusters to spawn. The resulting forecasting accuracy was measured in terms of the relative error and was 0.549 at step 1, 0.877 at step 2, and 0.929 at step 3. Thus the model shows considerable forecasting performance for 1 step forecasting, and naturally this performance degrades as the number of forecasting steps increases. As a comparison, we also used the SS-VAR approach of \cite{ghosh2018high} to analyse each sensor separately. The SS-VAR model illustrates decent performance, but is considerably worse that the rgPDPM-VAR, with a relative forecasting error of 0.598 at step 1, 0.897 at step 2, and 0.959 at step 3. This suggests that even for a small number of nodes, pooling information across sensors/subjects under the rgPDPM-VAR can still provide forecasting benefits. We note that other choice could have been made for the number of differencing steps. However, the overall relative performance between the two methods stays similar, with the proposed rgPDPM-VAR consistently performing better over different settings, as reported below.

% This part is fors supplement/appendix
\begin{table}[h]
\centering
\resizebox{1.0\textwidth}{!}
{
\begin{tabular}{|c|cc|cc|cc|}
\hline
 Differencing Level & \multicolumn{2}{c}{1 Step Forecasting} & \multicolumn{2}{c}{2 Step Forecasting} & \multicolumn{2}{c|}{3 Step Forecasting} \\
 \cline{2-7}
 & rgPDPM-VAR & SS-VAR & rgPDPM-VAR & SS-VAR & rgPDPM-VAR & SS-VAR \\ 
  \hline
1 & 1.202 (0.043) & 1.204 (0.120) & 1.022 (0.089) & 1.054 (0.146) & 1.002 (0.037) & 1.009 (0.053) \\ 
  2 & 0.718 (0.064) & 0.731 (0.118) & 0.995 (0.131) & 1.034 (0.140) & 1.022 (0.021) & 1.022 (0.041) \\ 
  3 & 0.644 (0.143) & 0.696 (0.254) & 0.913 (0.376) & 0.934 (0.411) & 1.057 (0.129) & 1.067 (0.178) \\ 
  4 & 0.616 (0.133) & 0.703 (0.226) & 0.889 (0.212) & 0.932 (0.256) & 0.962 (0.227) & 1.000 (0.498) \\ 
  5 & 0.549 (0.118) & 0.598 (0.167) & 0.877 (0.247) & 0.897 (0.237) & 0.929 (0.127) & 0.959 (0.150) \\ 
  6 & 0.638 (0.177) & 0.702 (0.267) & 0.909 (0.247) & 0.954 (0.292) & 0.951 (0.126) & 0.979 (0.106) \\ 
  7 & 0.580 (0.174) & 0.673 (0.264) & 1.059 (0.428) & 1.064 (0.52) & 0.958 (0.173) & 0.940 (0.186) \\ 
   \hline
\end{tabular}
}
\caption{Forecasting results under the rgPDPM-VAR and SS-VAR models for the air quality data with different levels of differencing. The values in the cells represent the relative L2 error between the model-based predictions and the true values, and the standard deviation is given in parenthesis.}
\label{extraForecastingResults}
\end{table}

% Interpretation

\section{Discussion}

In this work, we developed a non-parametric Bayesian framework that provides a fundamentally novel way to borrow information across samples, via a class of novel product of DP mixture priors. The proposed approach has the ability to flexibly borrow information across heterogeneous samples via unsupervised clustering at multiple scales that bypasses restrictive parametric assumptions and the requirement of replicated samples, and provides distinct advantages over existing methods. We developed the approach for multivariate outcomes, and subsequently extended the approach to multi-variate time series data under the VAR framework, where several variants of the method was developed to cater to varying degrees of heterogeneity. We illustrated posterior consistency properties for the proposed approach via novel theoretical results for multi-variate time series data modeled under the VAR framework, and implemented the approaches via an efficient MCMC sampling scheme. Through extensive simulations, we showed a superior numerical performance under the proposed methods compared to a single subject analysis, even when the latter is augmented via an additional adhoc clustering step that is designed to borrow information across similar samples. In addition, the rgPDPM-VAR variant often had superior performance compared to the other PDPM variants in heterogeneous data settings, and it 
produced biologically interpretable and reproducible parameter estimates in our analysis of the HCP data compared to single subject analysis. Similarly, we found that the rgPDPM-VAR could obtain superior forecasting performance to single-sensor analysis of air quality data, indicating that the approach has use as a predictive tool as well.  Overall, we found that the proposed rgPDPM-VAR, which clusters rows of the subject-level autocovariance matrices, was able to perform well in nearly all settings.

%produced reproducible estimates for our HCP application that were also biologically meaningful. %Our overall results indicated that the rgPDPM-VAR model was suitable for use in diverse settings when there is a high degree of heterogeneity in the data and where subgroups within the population are unknown.

%We demonstrated our proposed approach using resting-state data from the HCP. We showed that our approach is able to find more differences between high and low fluid intelligence groups, as compared to the single subject VAR models. Moreover, the regions of the brain with the most differences were localized in the dorsal salience network, which is consistent with previous studies of intelligence. Finally, through a reproducibility study, we showed that our findings were highly reproducible across a different data collection run using a different phase encoding direction, increasing our confidence in our findings.

While this work introduced several variants of the PDPM-VAR model, there are numerous potential extensions that lie within our class of models. In particular, future directions might investigate possible generalizations intended to induce sparsity in the parameter estimates. For example, a spike and slab prior could be used to model the autocovariance elements, with the slab component modeled using a DP mixtures. Additionally, the models could be generalized to accommodate even higher levels of heterogeneity, such as clustering individual autocovariance elements separately. However, such extensions may involve a massive computational burden. The proposed approaches, particularly the rgPDPM-VAR, seems to strike a desirable balance between computational complexity, clustering flexibility and model parsimony, with theoretical guarantees and appealing practical performance. Finally, we note that the proposed product of DP priors provides a viable improvement over  traditional DP mixture models and it should have wide applicability to other types of settings that go beyond the VAR framework, which is of immediate interest in this article. We expect to pursue these directions in future research.

%For example, a single MCMC iteration on the dorsal salience network required less than 20 seconds of computation time on a 2021 M1  Macbook with a {\color{black}2.4 GHz 8-Core Intel Core i9 processor}. %Such computational costs would increase dramatically if we instead required clustering each element of the autocovariance.

% \vskip 6pt 

% \begin{center}
% {\large\bf SUPPLEMENTARY MATERIAL}
% \end{center}

% {\bf \noindent Appendix:} PDF file containing additional proofs, lemmas, computation details, and simulation results.

{\it Acknowledgements:} The authors gratefully acknowledge support from NIH awards R01AG071174 and R01MH120299.% The code for this paper is available on request.}

\bibliographystyle{apalike}
\bibliography{arXiv_submission_2.bib}

\vskip 20pt

{\noindent \bf \Large Appendix}

\section{Proofs of Results}

{\noindent \bf Proof of Lemma 1:}  An outline for the proof of Lemma 1 is provided, which follows similar steps as the proof of Lemma 1 in \cite{canale2017posterior}. One may write $KL(f_0,f_P)= KL(f_0,f_{P_\epsilon})+KL(f_{P_\epsilon},f_P)$ and then show that each of the terms in the right hand side can be made exceedingly small with positive probability, for some compactly supported $P_\epsilon$. %The first term can be shown to be infinitesimally small with positive probability using results in \cite{wu2008kullback}. 
The compact support for $P_\epsilon$ is taken as $ [-\mu^*,\mu^*]^D \times \{\Sigma\in \mathcal{S}: \underline{\sigma}^2 < \lambda_l(\Sigma) < \bar{\sigma}^2, 1\le l\le p \},$ for some constants $\mu^*>0$ and $0<\underline{\sigma}<\bar{\sigma}$, and eigen values denoted as $\lambda_l,l=1,\ldots,p$, the second term can also be shown to be infinitesimally small with positive probability. %along similar lines as the proof of Lemma 1 in \cite{canale2017posterior}. Hence a detailed proof is omitted. 

%{\noindent \bf Proof of Lemma 1:} The proof follows similar steps as the proof of Lemma 1 in \cite{canale2017posterior}, and is omitted due to space constraints.

\vskip 12pt

{\noindent \bf Proof of Theorem 1:} The proof relies on two parts, i.e calculating the entropy bounds and calculating the prior probability of the constructed sieves, and then using them to show the summability condition in Theorem 2 holds.  We will illustrate the proof for the case of $\mathcal{M}_\mu=1$, and extensions to higher values of $\mathcal{M}_\mu$ are straightforward.
 
 First, we will construct sieves of the following form -
\begin{eqnarray}
\small
\mathcal{F}_n &=& \bigg\{f_p: P = \sum\limits_{h_1\ge 1}\sum\limits_{\hsig\ge 1}\pi_{h_1}\pi_{\sigma,\hsig}\delta_{\bfmu_{h_1},\Sigma_{\hsig}}: \sum_{h_1>H_n}\pi_{h_1}<\epsilon,  \sum_{\hsig>H_n}\pi_{\sigma,\hsig}<\epsilon, \mbox{ and for}\nonumber \\
&& \hsig\le H_n, \mbox{ } \underline{\sigma}_n^2\le \lambda_\nOutcome, \lambda_1 \le \underline{\sigma}_n^2(1 + \epsilon/\sqrt{\nOutcome})^{M_n}, \mbox{ } 1<\frac{\lambda_1(\Sigma_{\hsig})}{\lambda_\nOutcome(\Sigma_{\hsig})}\le  n^{H_n} \bigg\}, \\
\mathcal{F}_{n,{\bf j,l}} & =& \bigg\{f_p\in \mathcal{F}_n: \mbox{ for } h_1,\hsig\le H_n, 
n^{H^2_n}(j_{h_1} -1)=\underline{a}_{h_1,j}\le ||\bfmu_{h_1} ||\le \bar{a}_{h_1,j}=n^{H^2_n}j_{h_1}, \nonumber\\
&&\mbox{ }  k\in \{1,\ldots,K \},\mbox{ } 
 n^{l_{\hsig}-1} <\frac{\lambda_1(\Sigma_{\hsig})}{\lambda_\nOutcome(\Sigma_{\hsig})}\le n^{l_{\hsig}}  \bigg\}, \label{eq:sieves-PDPM}
\end{eqnarray}
where $M_n=\underline{\sigma}^{-2c_2}=n$ and $H_n=\lfloor Cn\epsilon^2/\log(n) \rfloor$ for some positive constant $C$, 
and clearly $\mathcal{F}_n\subset \cup_{{\bf j,l}} \mathcal{F}_{n,{\bf j,l}}$.  %Comparing to the notations used in the manuscript,  we note that $\underline{a}_{h_1,j}= n^{H_n}(j_{h_1}-1), \bar{a}_{h_1,j}= n^{H_n}j_{h_1}$, and $u_{\hsig,l}=n^{l_{\hsig}}$, for integers $(j_1,\ldots,j_{H_n})\in \mathbb{N}$ and $(l_1,\ldots,l_{H_n})\in \{1,\ldots, H_n \}$. 
 Using similar techniques used to derive (\ref{eq:prior-prob1}) in Lemma 6 in the Appendix, it is possible to show the tail condition (2A) holds in Theorem 2, i.e. $\Pi(\mathcal{F}^c_n)\le e^{-b^*n}$. Further, using similar steps as in the proof of Lemma 4 in the Appendix, the distance between the two densities $f_{P_1}$ and $f_{P_2}$ can be expressed as 
$\small ||f_{P_1} - f_{P_2} ||_1 \le \epsilon^2 + \sum_{h_1,\hsig<H_n}|\pi^{(1)}_{h_1}\pi^{(1)}_{\sigma,\hsig}-\pi^{(2)}_{h_1}\pi^{(2)}_{\sigma,\hsig}| + \epsilon \nonumber \\
+ \sum_{h_1,\hsig\le H_n}\pi^{(1)}_{h_1}\pi^{(1)}_{\sigma,\hsig}\bigg\|\phi_{\Sigma^{(1)}_{\hsig}}\big({\bf x} - \bfmu^{(1)}_{h_1}\big) - \phi_{\Sigma^{(2)}_{\hsig}}\big({\bf x}_t - \bfmu^{(2)}_{h_1}\big) \bigg\|_1, $
where $||\cdot ||_1$ denotes the $L_1$ norm. Using similar steps as in the proof of Lemma 2 in \cite{canale2017posterior}, it is possible to show that 
\begin{eqnarray}
&&\bigg\|\phi_{\Sigma^{(1)}_{\hsig}}\big({\bf x} - \bfmu^{(1)}_{h_1}\big) - \phi_{\Sigma^{(2)}_{\hsig}}\big({\bf x}_t - \bfmu^{(2)}_{h_1}\big) \bigg\|_1 \le 
\sqrt{\frac{2}{\pi}} \frac{|| \bfmu^{(1)}_{h_1}- \bfmu^{(2)}_{h_1}||}{\sqrt{\lambda_{\nOutcome}(\Sigma^{(2)}_{\hsig})}} + \nonumber \\
&&\bigg\{ \sum_{k=1}^\nOutcome \frac{\lambda_{k}(\Sigma^{(1)}_{\hsig})}{\lambda_{k}(\Sigma^{(2)}_{\hsig})} - \log \frac{\lambda_{k}(\Sigma^{(1)}_{\hsig})}{\lambda_{k}(\Sigma^{(2)}_{\hsig})} -1 \bigg\}^{1/2} + \bigg\{ 2\nOutcome || O^{(1)}_{\hsig} - O^{(2)}_{\hsig}||_2 \frac{\lambda_1(\Sigma^{(1)}_{\hsig})}{\lambda_d(\Sigma^{(2)}_{\hsig})}\bigg\},\label{eq:entbound}
\end{eqnarray}
 where $O{(j')}$ represents the matrix of orthogonal vectors in the spectral decomposition of $\Sigma^{(j')}, j'=1,2$. Finally, we need to establish an upper bound for the term $\sum_{h_1,\hsig<H_n}|\pi^{(1)}_{h_1}\pi^{(1)}_{\sigma,\hsig}-\pi^{(2)}_{h_1}\pi^{(2)}_{\sigma,\hsig}|$ in (\ref{eq:entbound}), which is given by {\color{black} Lemma 5} as 
$ \sum_{h_1,\hsig<H_n}\bigg|\tilde{\pi}^{(1)}_{h_1}\tilde{\pi}^{(1)}_{\sigma,\hsig}- \pi^{(2)}_{h_1}\pi^{(2)}_{\sigma,\hsig}\bigg| + \bigg|1-(1-\epsilon)^2 \bigg|, \mbox{ where } \tilde{\pi}=\frac{\pi_{h}}{(1-\sum_{h>H}\pi_{h})}.$

For a given $f_P\in \mathcal{F}_{n,{\bf jl}}$ with $P=\sum\limits_{h_1\ge 1}\sum\limits_{\hsig\ge 1}\pi^{(1)}_{h_1}\pi^{(1)}_{\sigma,\hsig}\delta_{\big(\Theta^{(1)}_{h_1},\Sigma^{(1)}_{\hsig}\big)}$ and $\Sigma_{h}=(O_{h}\Lambda_{h}O^T_{h})^{-1}$ where $\Lambda_{h}=diag(\lambda_{h,1},\ldots,\lambda_{h,\nOutcome})$, we will construct another density $f_{\hat{P}}$ with $\hat{P}= \sum\limits_{h_1\ge 1}\sum\limits_{\hsig\ge 1}\pi^{(1)}_{h_1}\pi^{(1)}_{\sigma,\hsig}\delta_{\big(\hat{\Theta}^{(1)}_{h_1},\hat{\Sigma}^{(1)}_{\hsig}\big)}$ within the $\epsilon$-net and then compute the cardinality of the $\epsilon$-net set to derive an upper bound for the entropy of sieve $\mathcal{F}_{n,{\bf jl}}$. 
To construct such a density, we will choose:
\begin{itemize}
    \item $\hat{\bfmu}_{h_1}\in \mathcal{\hat{R}}_{h_1},h_1=1,\ldots,H,$ where $\mathcal{\hat{R}}_{h_1}$ is a $\epsilon^*$-net of $\mathcal{R}_{h_1}:=\{\bfmu\in \Re^{\nOutcome}:\underline{\mu}_{h_1,j}\le ||\bfmu || \le \bar{\mu}_{h_1,j} \}$, such that $||\bfmu_{h_1} - \hat{\bfmu}_{h_1} ||\le \underline{\sigma}_n\epsilon$, $k=1,\ldots,K,$ where $\bar{\mu},\underline{\mu},$ and $\underline{\sigma}_n$ correspond to the sieve boundaries in (\ref{eq:sieves-PDPM}). %$\epsilon^*=\pi\underline{\sigma}_n\epsilon\times \frac{1}{2\big\{T (u_{\hsig,l})^{(T-k)(\nOutcome-1)/2} \big\} \times   \big\{ c(\nOutcome,T,K) + K^{T-2}\big(\frac{\bar{a}^2_{h_1,j}}{\underline{\sigma}^2_n} \big)^{T-2} \big\} }$, using (\ref{eq:kappa}).
 % which will reduce the upper bound for $\mathcal{K}^*$ in (\ref{eq:kappa}) to $\lesssim \epsilon$.
    \item $\{\hat{\pi}_{h_1}\hat{\pi}_{\hsig}, h_1,\hsig\le H_n \}\in \hat{\Delta}$, where $\hat{\Delta}$ is a $\epsilon$-net of a $H^2_n$ dimensional probability simplex such that $\sum_{h_1,\hsig\le H_n} |\tilde{\pi}_{h_1}\tilde{\pi}_{\hsig} - \hat{\pi}_{h_1}\hat{\pi}_{\hsig} | \le \epsilon$, and $\tilde{\pi}_h=\frac{\pi_h}{\sum_{h\le H_n}\pi_h}, h\le H_n$.
    \item $\hat{O}_{h}\in \mathcal{\hat{O}}_{h}$, where $\mathcal{\hat{O}}_{h}$ is a $\delta_h$-net of the set $\mathcal{O}_{h}$ defined as the set of $\nOutcome\times \nOutcome$ orthogonal matrices with respect to the spectral norm $||\cdot ||_2$ with $\delta_h=\epsilon^2/(2\nOutcome u_{h,l})$ such that $|| O_{h} - \hat{O}_{h}||_2\le T\delta_h$.
    \item $(m_{h,1},...,m_{h_\nOutcome})\in \{1,...,M\}^\nOutcome ,h = 1,...,H,$ such that $\hat{\lambda}_{h,l} = \{\underline{\sigma}^2(1 + \epsilon\sqrt{\nOutcome})^{m_{h,l}-1} \}^{-1}$ will satisfy $1\le \hat{\lambda}_{h,l}/\lambda_{h,l} < (1 + \epsilon/\sqrt{\nOutcome})$.
\end{itemize}
%Using this construction, the term in (\ref{eq:csis1}) is shown to be bounded by $\sqrt{T\sum_{d'=1}^\nOutcome \bigg\{\big(\frac{\hat{\lambda}_{h,\nOutcome-d'+1}}{\lambda_{h,\nOutcome-d'+1}} -1 \big)^2 \bigg\} }$. Moreover 
Under this construction, it can be shown that $||f_P - f_{\hat{P}} ||_1 < C^*\epsilon$ for some constant $C^*$, by employing some additional algebra and similar arguments as in the proof of Lemma 2 in \cite{canale2017posterior} and Theorem 4 in our paper. Further, the cardinality of the $\epsilon$-net can be computed by noting that 
$\#(\hat{\Delta})\lesssim \epsilon^{-H^2_n} \mbox{ for } j=1,2,\#(\mathcal{\hat{O}}_{h})\lesssim\delta_h^{-\nOutcome(\nOutcome-1)/2},\#(\mathcal{\hat{R}}_{k,h})\lesssim [(\frac{\bar{a}_{h}}{\epsilon^*}+1)^{\nOutcome} - (\frac{\underline{a}_{h}}{\epsilon^*}-1)^{\nOutcome} ]$. Using these quantities, one can write the upper bound for the exponential of the entropy bound as 
\begin{eqnarray}
&&(M)^{\nOutcome H_n}\epsilon^{-H^2_n}\times \prod_{h_1\le H_n} \big\{(\frac{\bar{a}_{h_1,j}}{\underline{\sigma}_n\epsilon}+1)^{\nOutcome} - (\frac{\underline{a}_{h_1,j}}{\underline{\sigma}_n\epsilon}-1)^{\nOutcome}  \big\} \prod_{\hsig\le H_n} \big\{\frac{2\nOutcome u_{\hsig,l}}{\epsilon^2} \big\}^{\nOutcome(\nOutcome-1)/2}\nonumber \\
&&\approx \exp\bigg\{DH_n\log(M) + H^2_n\log(\frac{1}{\epsilon}) + \frac{D(D-1)}{2}\log(n^{l_{\hsig}}) + \frac{D(D-1)}{2} \log(\frac{1}{\epsilon}) \bigg\},
\label{eq:entropy}
%&&\lesssim \prod_{\hsig\le H_n} \big\{\frac{2 \nOutcome u_{\hsig,l}}{\epsilon^2} \big\}^{\nOutcome(\nOutcome-1)/2}  \prod_{h_1\le H_n}\Bigg\{\bigg(\frac{C^*_{h_{1,j},h_{\sigma,l}}\bar{a}_{h_1,j}}{\underline{\sigma}_n\epsilon} +1\bigg)^{\nOutcome^2} - \bigg(\frac{C^*_{h_{1,j},h_{\sigma,l}}\underline{a}_{h_1,j}}{\underline{\sigma}_n\epsilon} -1 \bigg)^{\nOutcome^2} \Bigg\}^K 
% &\approx& (M)^{dH_n}\epsilon^{-C_1H_n} \prod_{h_2\le H_n} \big\{\frac{2d u_{h_2,l}}{\epsilon^2} \big\}^{d(d-1)/2}  \times \prod_{h_1\le H_n}\bigg(\frac{\bar{a}_{h_1}^{d^2} - \underline{a}_{h_1}^{d^2}}{\epsilon} \bigg)^{K} \times \prod_{h_2\le H_n}\prod_{h_1\le H_n} \sqrt{\lambda_d(\Sigma^{(2)}_{h_2})}
%\lesssim (M)^{dH_n}\epsilon^{-C_1H_n}\prod_{h_1\le H_n} \big\{(\frac{\bar{a}_{h_1}}{\epsilon^*}+1)^{d^2} - (\frac{\bar{a}_{h_1}}{\epsilon^*}-1)^{d^2}  \big\}^{K} \prod_{h_2\le H_n} \big(2d u_{h_2,l} \big)^{d(d-1)/2},
\end{eqnarray}
 when $n$ and $j_{h_1}$ is large, and using the definitions of $\underline{a}$ and $\bar{a}$ defined in (\ref{eq:sieves-PDPM}), and following similar steps as in the proof of Theorem 2 in Canale and De Blasi (2017) to show that 
  $\bigg[\big(\frac{\bar{a}_{h_1,j}}{\underline{\sigma}_n\epsilon/2}+1 \big)^{\nOutcome} - \big(\frac{\underline{a}_{h_1,j}}{\underline{\sigma}_n\epsilon/2}-1 \big)^{\nOutcome}  \bigg]\lesssim \bigg[\frac{n^{(H_n+\frac{1}{2c_2})\nOutcome}j_{h1}^{\nOutcome-1}}{(\epsilon)^{\nOutcome}} \bigg] $.

%Using similar steps as in the proof of Theorem 2 in Canale and De Blasi (2017), it is possible to show that 
%  $\bigg[\big(\frac{\bar{a}_{h_1,j}}{\underline{\sigma}_n\epsilon/2}+1 \big)^{\nOutcome} - \big(\frac{\underline{a}_{h_1,j}}{\underline{\sigma}_n\epsilon/2}-1 \big)^{\nOutcome}  \bigg]\lesssim \bigg[\frac{n^{(H_n+\frac{1}{2c_2})\nOutcome}j_{h1}^{\nOutcome-1}}{(\epsilon)^{\nOutcome}} \bigg] $,  when $n$ and $j_{h_1}$ is large, and using the definitions of $\underline{a}$ and $\bar{a}$ defined in (\ref{eq:sieves-PDPM}). The entropy bound in (\ref{eq:entropy}) may be rewritten as
%  \begin{eqnarray*}
% && \exp\bigg\{DH_n\log(M) + H^2_n\log(\frac{1}{\epsilon}) + \frac{D(D-1)}{2}\log(n^{l_{\hsig}}) + \frac{D(D-1)}{2} \log(\frac{1}{\epsilon}) \bigg\}.
% && \big\{\frac{2\nOutcome u_{\hsig,l}}{\epsilon^2} \big\}^{\nOutcome(\nOutcome-1)/2} \big\{ (u_{\hsig,l})^{(T-1)(\nOutcome-1)/2} \big\}^{K\nOutcome^2H_n} = \mathcal{C}_1 (u_{\hsig,l})^{\nOutcome(\nOutcome-1)/2 + K\nOutcome^2H_n(T-1)(\nOutcome-1)/2} \\
% &&\approx \mathcal{C}_1 \exp\bigg\{ \bigg(\frac{\nOutcome(\nOutcome-1)}{2} + K\nOutcome^2H_n(T-1)\frac{\nOutcome-1}{2}\bigg) \log(n^{l_{\hsig}})\bigg\}\times \big(\frac{1}{\epsilon}\big)^{\nOutcome(\nOutcome-1)},
 % \end{eqnarray*}
 
 Further using similar steps as in (\ref{eq:sieve-prob}) in the proof of Theorem 5, we have
\begin{eqnarray*}
&&\Pi(\mathcal{F}_{n,{\bf jl}})\le \prod\limits_{h_1\le H_n} P^*_1\big(||\bfmu_{h_1} ||>n^{H^2_n}(j_{h_1} -1)\big)\prod_{\hsig\le H_n}P^*_2\big( \lambda_1(\Sigma)/\lambda_\nOutcome(\Sigma)>n^{(l_{\hsig}-1)}\big), \nonumber\\
&\lesssim& \prod_{h_1\le H_n}\big\{\big( n^{H^2_n}(j_{h_1}-1)\big)^{-1_{(j_{h_1}\ge 2)}2(r+1)}\big\}\times\prod_{\hsig\le H_n} (n^{(l_{\hsig}-1)})^{-1_{(l_{\hsig}\ge 1)}\kappa} \nonumber\\
&\approx&  \bigg\{n^{-2H^3_n(r+1)} \prod_{h_1\le H_n} \big(j_{h_1}-1\big)^{-1_{(j_{h_1}\ge 2)}2(r+1)} \bigg\}\times \bigg\{\prod_{\hsig\le H_n} (n^{\kappa(l_{\hsig}-1)})^{-1_{(l_{\hsig}\ge 1)}} \bigg\}, \mbox{ for large } n. 
\end{eqnarray*}
 
 Finally, using similar steps as in Lemma 7 in the Appendix it is possible to show that the summability condition in Theorem 2 holds. This proves the strong consistency result for the product mixture of DP priors for multivariate density estimation.

\vskip 12pt

{\noindent \bf \underline{Proof of Theorem 3}:}
We will use the conditions in Lemmas 2-4 and Theorem 2 in Wu and Ghosal (2008) to prove our results.  Note that $f_0(X)=\prod_{t=1}^T f_0({\bf x}_t\mid X_{1:(t-1)})$, where $f_0({\bf x}_t\mid X_{1:(t-1)})=f_0({\bf x}_1)$ for $t=1$ by convention. For any $P\in \mathcal{P}$, note that the KL divergence can be expressed as $KL(f_0(X),f_P(X))=\int f_0(X)\log\bigg(\frac{f_0(X)}{f_P(X)} \bigg)d{\bf x}_1\ldots d{\bf x}_T$
\begin{eqnarray}
&=& \int \prod_{t=1}^T f_0\big({\bf x}_t\mid X_{1:(t-1)}\big)\log\bigg(\prod_{t=1}^T\frac{f_0({\bf x}_t\mid X_{1:(t-1)})}{f_P({\bf x}_t\mid X_{1:(t-1)})}\bigg) d{\bf x}_1\ldots d{\bf x}_T \nonumber \\
&=& \int \prod_{t=1}^T f_0\big({\bf x}_t\mid X_{1:(t-1)}\big) 
\times\bigg\{\sum_{t=1}^T\log\bigg(\frac{f_0({\bf x}_t\mid X_{1:(t-1)})}{f_P({\bf x}_t\mid X_{1:(t-1)})}\bigg)\bigg\} d{\bf x}_1\ldots d{\bf x}_T \nonumber \\
&=& \sum_{t=1}^T \bigg[\prod_{t^*> t}\underbrace{\int f_0\big({\bf x}_{t^*}\mid X_{1:(t^*-1)}\big)d{\bf x}_{t^*}}_\text{=1} 
\times \int\bigg\{\int  f_0\big({\bf x}_t\mid X_{1:(t-1)}\big)\log\bigg(\frac{f_0({\bf x}_t\mid X_{1:(t-1)})}{f_P({\bf x}_t\mid X_{1:(t-1)})}\bigg) d{\bf x}_t\nonumber \\
&& \times \prod_{t'< t} f_0\big({\bf x}_{t'}\mid X_{1:(t'-1)}\big)d{\bf x}_{t'}\bigg\} \bigg] \nonumber \\
&=& \sum_{t=1}^T \bigg\{\int KL\big( f_0\big({\bf x}_t\mid X_{1:(t-1)}\big), f_P\big({\bf x}_t\mid X_{1:(t-1)}\big)\big)\prod_{t'=1}^{t-1}f_0({\bf x}_{t'}\mid X_{1:(t'-1)})d{\bf x}_{t'-1}\ldots d{\bf x}_1 \bigg\},\label{eq:KLsum}
\end{eqnarray}
which is a sum of integrals involving Kullback-Leibler divergence of conditional densities. In the following derivations, we will use the shorthand notation $KL_{(f_0,f_P)}({\bf x}_t\mid X_{1:(t-1)})$ to denote $KL\big( f_0\big({\bf x}_t\mid X_{1:(t-1)}\big), f_P\big({\bf x}_t\mid X_{1:(t-1)}\big)\big)$ where convenient. We will prove that $KL_{(f_0,f_P)}({\bf x}_t\mid X_{1:(t-1)})$ is infinitesmall $(<\epsilon)$ with positive probability pointwise for $X_{1:(t-1)}$ for $t=1,\ldots,T$, which will imply that the above sum on the right hand side of the equality also becomes  infinitesmall for fixed $T$, and hence we will prove our Theorem.

As a first step, we will define $P_\epsilon$ on a compact set {\color{black}$\{\bfT: -a\le A_k(j,j') \le a$ , $\mbox{ } k=1,\ldots,K$, $\mbox{ and } ||\sum_{k=1}^K A_k{\bf x}_{t-k} ||<m \}\times \{\Sigma\in \mathcal{S}:h_m=m^{-\eta}\le\lambda_\nOutcome(\Sigma)<\ldots<\lambda_1(\Sigma)\le \bar{M},\mbox{ } t=1,\ldots,T \}=\mathcal{D}^*_1\times\mathcal{D}^*_2$}, such that $P_\epsilon(\mathcal{D}^*_1\times\mathcal{D}^*_2)=1$, for some $m,\eta>0$. We will construct $P_\epsilon$ such that it ensures that the upper bounds for the terms in the right hand side of the above equality are arbitrarily small with positive prior probability:
\begin{eqnarray}
KL_{(f_0,f_P)}\big({\bf x}_t\mid X_{1:(t-1)}\big)\big)=KL_{( f_0, f_{P_{\epsilon}})}\big({\bf x}_t\mid X_{1:(t-1)}\big) + KL_{( f_{P_\epsilon}, f_{P})}\big({\bf x}_t\mid X_{1:(t-1)}\big). \label{eq:KLbase}
\end{eqnarray}
Write ${\bf r}=\sum_k A_k{\bf x}_{t-k} $, and $\small t^{-1}_m=\int\limits_{||{\bf r} ||<m}  f_0\big( {\bf r}\mid X_{1:(t-1)}  \big)d( {\bf r})$ and define $t^{-1}_m f_{P_{\epsilon}}\big({\bf x}_t\mid X_{1:(t-1)}\big)$  
\begin{align*}
&\small = \int\limits_{||{\bf r} ||<m} \phi_{\Sigma}\big({\bf x}_t - {\bf r}\mid X_{1:(t-1)} \big) f_0\big({\bf r}_{t-k}\mid X_{1:(t-1)}  \big)d({\bf r})
\ge \int\limits_{||{\bf r} ||<m}\phi_{h_m I_\nOutcome}\big({\bf x}_t - {\bf r}\mid X_{1:(t-1)} \big)  f_0\big({\bf r}\mid X_{1:(t-1)}  \big)d({\bf r}) \\
&\times \big(\frac{\lambda_\nOutcome(\Sigma)}{\lambda_1(\Sigma)}\big)^{(\nOutcome-1)/2} = t_m \int\limits_{||{\bf x}_{t}-\bft h_m ||<m}\phi_{h_m I_\nOutcome}\big(\bft\mid X_{1:(t-1)} \big) f_0\big({\bf x}_{t}-\bft h_m\mid X_{1:(t-1)}  \big)d\bft \times \big(\frac{\lambda_\nOutcome(\Sigma)}{\lambda_1(\Sigma)}\big)^{(\nOutcome-1)/2}
\end{align*}
 where $\bft=({\bf x}_t - \sum_{k=1}^{K} A_k {\bf x}_{t-k})/h_m$, and the inequality in the second last step is derived using the fact that $\big(\frac{\lambda_\nOutcome(\Sigma)}{\lambda_1(\Sigma)}\big)^{(\nOutcome-1)/2} \phi_{\lambda_\nOutcome(\Sigma) I_\nOutcome}\big({\bf x}\big)\le \phi_\Sigma\big({\bf x}\big)\le\big(\frac{\lambda_1(\Sigma)}{\lambda_\nOutcome(\Sigma)}\big)^{(\nOutcome-1)/2} \phi_{\lambda_1(\Sigma)I_D}\big({\bf x}\big)$.
 Using the above, we have $KL_{(f_0,f_{P_\epsilon})}\le \big(\frac{\lambda_1(\Sigma)}{\lambda_\nOutcome(\Sigma)}\big)^{(\nOutcome-1)/2}KL_{(f_0,f_{P_{\epsilon,\Sigma=h_m I_\nOutcome}})}$, where $f_{P_{\epsilon,\Sigma=h_m I_\nOutcome}}$ has the same form as $f_{P_\epsilon}$ but with $\Sigma$ set to $h_m I_\nOutcome$. Hence, the next step is to show that  $KL(f_0,f_{P_{\epsilon,\Sigma=h_m I_\nOutcome}})({\bf x}_t \mid X_{1:(t-1)})$ can be made arbitrarily small with positive prior probability, point-wise for all $X_{1:(t-1)}$, which will help establish that the first term on the right hand side of (\ref{eq:KLbase}) is negligible with positive prior probability. 
 
 Since when $m\to \infty$ and $h_m\to 0$, $\phi_{h_m I_\nOutcome}\big({\bf x}_t - \sum_{k=1}^{K} A_k {\bf x}_{t-k}\mid X_{1:(t-1)} \big)$ takes non-zero values only when  $\sum_{k=1}^{K} A_k {\bf x}_{t-k} \to {\bf x}_t $, we obtain $\phi_{h_m I_\nOutcome}\big({\bf x}_t - \sum_{k=1}^{K} A_k {\bf x}_{t-k}\mid X_{1:(t-1)} \big)  f_0\big(\sum_{k=1}^{K} A_k {\bf x}_{t-k}\mid X_{1:(t-1)}  \big)\to f_0({\bf x}_t\mid X_{1:(t-1)})$ as $m\to \infty$ and $h_m\to 0$, which also implies $f_{P_{\epsilon,\Sigma=h_m I_\nOutcome}}({\bf x}_t\mid X_{1:(t-1)})\to f_0({\bf x}_t\mid X_{1:(t-1)})$, point-wise for all $X_{1:(t-1)}$. We will combine the above convergence with the fact the $\log\big(\frac{f_0({\bf x}_t\mid X_{1:(t-1)})}{f_{P_{\epsilon,\Sigma=h_m I_\nOutcome}}({\bf x}_t\mid X_{1:(t-1)})}\big)$ is bounded and integrable (as shown in the sequel) and subsequently apply the DCT to achieve our result. As a next step, note $KL_{(f_0,f_{P_{\epsilon,\Sigma=h_m I_\nOutcome}})}({\bf x}_t \mid X_{1:(t-1)})=\int\limits_{||{\bf x}_t\le m ||}f_0({\bf x}_t\mid X_{1:(t-1)})\log\big(\frac{f_0({\bf x}_t\mid X_{1:(t-1)})}{f_{P_{\epsilon,\Sigma=h_m I_\nOutcome}}({\bf x}_t\mid X_{1:(t-1)})}\big)d{\bf x}_t$ + $\int\limits_{||{\bf x}_t> m ||}f_0({\bf x}_t\mid X_{1:(t-1)})\log\big(\frac{f_0({\bf x}_t\mid X_{1:(t-1)})}{f_{P_{\epsilon,\Sigma=h_m I_\nOutcome}}({\bf x}_t\mid X_{1:(t-1)})}\big)d{\bf x}_t$.
%\begin{align*}
%\int\limits_{||{\bf x}_t\le m ||}f_0({\bf x}_t\mid X_{1:(t-1)})\log\big(\frac{f_0({\bf x}_t\mid X_{1:(t-1)})}{f_{P_{\epsilon,\Sigma=h_m I_\nOutcome}}({\bf x}_t\mid X_{1:(t-1)})}\big)d{\bf x}_t+ \int\limits_{||{\bf x}_t> m ||}f_0({\bf x}_t\mid X_{1:(t-1)})\log\big(\frac{f_0({\bf x}_t\mid X_{1:(t-1)})}{f_{P_{\epsilon,\Sigma=h_m I_\nOutcome}}({\bf x}_t\mid X_{1:(t-1)})}\big)d{\bf x}_t.
%\end{align*}
Using similar arguments as in the Proof of Theorem 2 in \cite{wu2008kullback}, for $||{\bf x}_t ||>m$, we have $f_{P_{\epsilon,\Sigma=h_m I_\nOutcome}}({\bf x}_t\mid X_{1:(t-1)})\big)$
\begin{align*}
&\ge t_m\int\limits_{||\sum_k A_k{\bf x}_{t-k} ||<m} \phi_{h_m I_\nOutcome}\big({\bf x}_t + m{\bf x}_t/||{\bf x}_t ||\mid X_{1:(t-1)} \big)f_0\big({\bf r}\mid X_{1:(t-1)} \big) d(\sum_{k=1}^K A_k {\bf x}_{t-k})\\
&= \phi_{h_m I_\nOutcome}\big({\bf x}_t + m{\bf x}_t/||{\bf x}_t || \big)\times \bigg\{\underbrace{t_m\int\limits_{||\sum_k A_k{\bf x}_{t-k} ||<m} f_0\big({\bf r}\mid X_{1:(t-1)} \big) d(\sum_{k=1}^K A_k {\bf x}_{t-k})}_\text{=1}\bigg\}\\
&=\phi_{h_m I_\nOutcome}\big({\bf x}_t + m{\bf x}_t/||{\bf x}_t || \big) = m^{\eta}\phi_{ I_\nOutcome}\big(m^{\eta}{\bf x}_t + m^{1+\eta}{\bf x}_t/||{\bf x}_t || \big) \ge ||{\bf x}_t ||^{\eta}\phi_{I_\nOutcome}(2||{\bf x}_t ||{\bf x}_t).
\end{align*}
Similarly for $||{\bf x}_t ||\le m$, it is possible to show that given a constant $\delta>0$, $f_{P_{\epsilon,\Sigma=h_m I_\nOutcome}}({\bf x}_t\mid X_{1:(t-1)})\big)\ge c\inf\limits_{||{\bf r}-{\bf x}_t||<\delta}f_0({\bf r}\mid X_{1:(t-1)})$, using similar steps as in the proof of Theorem 2 in \cite{wu2008kullback}. Hence for a given constant $R<m$, we have 
\begin{align}
    \log\bigg(\frac{f_0({\bf x}_t\mid X_{1:(t-1)})}{f_{P_{\epsilon,\Sigma=h_m I_\nOutcome}}({\bf x}_t\mid X_{1:(t-1)})}\bigg) =\xi({\bf x}_t;X_{1:(t-1)})\le 
     \left\{
    \begin{array}{ll}
      r^*_1=\log\bigg(\frac{f_0({\bf x}_t\mid X_{1:(t-1)})}{c\inf\limits_{||{\bf r}-{\bf x}_t||<\delta}f_0({\bf r}\mid X_{1:(t-1)})} \bigg),   & ||{\bf x}_t ||<R,\\
     \max\bigg\{\log\bigg(\frac{f_0({\bf x}_t\mid X_{1:(t-1)})}{||{\bf x}_t ||^{\eta}\phi_{I_\nOutcome}(2||{\bf x}_t ||{\bf x}_t)} \bigg), r^*_1 \bigg\} & ||{\bf x}_t ||\ge R.
    \end{array}
\right. \label{eq:uppbound1}
\end{align}
Further note that $ f_{P_{\epsilon,\Sigma=h_m I_\nOutcome}}({\bf x}_t\mid X_{1:(t-1)}) \le M t_1$ which implies that $\log\bigg(\frac{f_0({\bf x}_t\mid X_{1:(t-1)})}{f_{P_{\epsilon,\Sigma=h_m I_\nOutcome}}({\bf x}_t\mid X_{1:(t-1)})} \bigg)\ge$ $\log\bigg(\frac{f_0({\bf x}_t\mid X_{1:(t-1)})}{M\phi^*_1}\bigg)$, where $\phi^*_1=\int\limits_{||{\bf x}_t||<1} f_0({\bf x}_t \mid X_{1:(t-1)})$ the lower bound is $<0$. Combining this fact with the upper bound in (\ref{eq:uppbound1}), it is possible to write 
\begin{eqnarray*}
KL\bigg(\frac{f_0({\bf x}_t\mid X_{1:(t-1)})}{f_{P_{\epsilon,\Sigma=h_m I_\nOutcome}}({\bf x}_t\mid X_{1:(t-1)})}\bigg)\le \int f_0({\bf x}_t\mid X_{1:(t-1)}) \max\bigg\{\xi({\bf x}_t;X_{1:(t-1)}),\bigg|\log\big(\frac{f_0}{M \phi^*_1}\big)\bigg| \bigg\}.
\end{eqnarray*}
Using assumptions {\it (A2)-(A4)} and similar algebraic steps as in the proof of Theorem 2 in \cite{wu2008kullback}, it is possible to show that the right hand side is bounded, point-wise for all $X_{1:(t-1)}$. Hence using DCT, the term $KL_{(f_0,f_{P_{\epsilon,\Sigma=h_m I_\nOutcome}})}\big({\bf x}_t\mid X_{1:(t-1)}\big)$ can be made arbitrarily small with positive prior probability. 
%Moreover the same applies for the term of the left hand side of the inequality  $KL_{(f_0,f_{P_\epsilon})}\big({\bf x}_t\mid X_{1:(t-1)}\big)\le \big(\frac{\lambda_1(\Sigma)}{\lambda_\nOutcome(\Sigma)}\big)^{(\nOutcome-1)/2}KL_{(f_0,f_{P_{\epsilon,\Sigma=h_m I_\nOutcome}})}\big({\bf x}_t\mid X_{1:(t-1)}\big)$.
 Therefore, we see that for any $\epsilon>0$, there exists $m_\epsilon$ such that $KL_{(f_0,f_{P_\epsilon})}\big({\bf x}_t\mid X_{1:(t-1)}\big)\le \big(\frac{\lambda_1(\Sigma)}{\lambda_\nOutcome(\Sigma)}\big)^{(\nOutcome-1)/2}KL_{(f_0,f_{P_{\epsilon,\Sigma=h_{m_\epsilon} I_\nOutcome}})}\big({\bf x}_t\mid X_{1:(t-1)}\big)<=\epsilon/2$. Hence the first term in (\ref{eq:KLbase}) is bounded by $\epsilon/2$ with positive prior probability.

To show that the second term in (\ref{eq:KLbase}) is negligible, we will demonstrate that the conditions in Lemma 3 of \cite{wu2008kullback} is satisfied. Using similar arguments as in \cite{wu2008kullback} as well as the proof of Lemma 1 in \cite{canale2017posterior}, it is possible to show that the weak support of $\Pi^*$  contains any compactly supported $\mathcal{P}$. Since  $P_{\epsilon}$ is compactly supported by definition, it belongs to the weak support of $\Pi^*$. Next, condition (A7) of Lemma 3 in \cite{wu2008kullback} requires $\log(f_{P_\epsilon})$ and $\log\inf\limits_{A_1,\ldots,A_K,\Sigma} \phi_{\Sigma}\big({\bf x}_t - \sum_{k=1}^K A_k{\bf x}_{t-k}\big)$ to be $f_0-$ integrable. Note that for $||{\bf x}_t ||<m$, $\log\inf\limits_{A_1,\ldots,A_K,\Sigma} \phi_{\Sigma}\big({\bf x}_t - \sum_{k=1}^K A_k{\bf x}_{t-k}\big)$ is bounded, and for $||{\bf x}_t ||>m$,
$\inf\limits_{A_1,\ldots,A_K,\Sigma}\phi_{\Sigma}({\bf x}_t - \sum_{k=1}^K A_k{\bf x}_{t-k})\le \bar{M}^{-\nOutcome}\phi(\exp\{-\frac{4||{\bf x}_t ||^2}{2m^{-\eta}} \})$, which is $f_0$ integrable. A similar upper bound can be applied for bounding $|\log(f_{P_\epsilon})|$ for $||{\bf x}_t ||>m$ that implies that $|\log(f_{P_\epsilon})|$ is $f_0$ integrable, which satisfies condition (A7) in Lemma 3 of \cite{wu2008kullback}. Note that the above statements hold point-wise for all $X_{1:(t-1)}$.

Condition (A8) of Lemma 3 in \cite{wu2008kullback} corresponding to the second term in (\ref{eq:KLbase}) is clearly satisfied since the multivariate normal kernel $\phi_{\Sigma}\big({\bf x}_t - \sum_{k=1}^K A_k{\bf x}_{t-k}\big)$ is bounded away from zero for ${\bf x}_t$ in a compact set of $\Re^d$ and $(A_1,\ldots,A_K,\Sigma)\in \mathcal{D}^*_1\times \mathcal{D}^*_2$. To show condition (A9) in Lemma 3 in \cite{wu2008kullback}, we will need to show that the kernel $\phi_{\Sigma}({\bf x}_t - \sum_{k=1}^K A_k {\bf x}_{t-k})$ is equicontinuous as a family of functions of $\{A_1,\ldots,A_K,\Sigma \}$ for ${\bf x}_t$ lying on a compact subset of $\Re^\nOutcome$ and conditional on given values of $X_{1:(t-1)}$. Note that for two distinct sets of parameters $(\bfT,\Sigma)$ and $(\bfT',\Sigma')$,
\begin{align}
&\small \bigg|\phi_{\Sigma}\big({\bf x}_t - \sum_{k=1}^K A_k {\bf x}_{t-k}\big) - \phi_{\Sigma'_t}\big({\bf x}_t - \sum_{k=1}^K A'_k {\bf x}_{t-k}\big)\bigg|\le 
\bigg|\phi_{\Sigma}\big({\bf x}_t - \sum_{k=1}^K A_k {\bf x}_{t-k}\big) - \nonumber \\
&\phi_{\Sigma'}\big({\bf x}_t - \sum_{k=1}^{K-1} A_k {\bf x}_{t-k} - A'_K{\bf x}_{t-K} \big)\bigg| + \bigg|\phi_{\Sigma'}\big({\bf x}_t - \sum_{k=1}^{K-1} A_k {\bf x}_{t-k} - A'_K{\bf x}_{t-K} \big) - \phi_{\Sigma'}\big({\bf x}_t - \sum_{k=1}^K A'_k {\bf x}_{t-k}\big)\bigg|. \label{eq:equicont}
\end{align}
The first term in (\ref{eq:equicont}) can be shown to be arbitrarily small when $(A'_K,\Sigma')$ lies within a small neighborhood of $(A_K,\Sigma)$ (or equivalently $ A_K {\bf x}_{t-K}$ is in a neighborhood of $A'_K {\bf x}_{t-K}$ pointwise for ${\bf x}_{t-K}$, and $\Sigma$ lies in a neighborhood of $\Sigma'$) for ${\bf x}_t\in C\subset \Re^\nOutcome$, where $C$ is a compact subset of $\Re^\nOutcome$, using arguments similar to the last part of the proof of Theorem 2 in \cite{wu2008kullback}. The second term in (\ref{eq:equicont}) can be decomposed using similar and repeated iterative steps, and hence shown to be arbitrarily small. Hence the equicontinuity condition holds, and all conditions of Lemma 3 in \cite{wu2008kullback} are satisfied, point-wise for all $X_{1:(t-1)}$. This implies that the second term in (\ref{eq:KLbase}) is less than or equal to $\epsilon/2$. Combining this with the upper bound on the first term in (\ref{eq:KLbase}), it is possible to show that $KL(f_0,f_{P_{\epsilon}})({\bf x}_t \mid X_{1:(t-1)})<\epsilon$, point-wise almost everywhere for $X_{1:(t-1)}$. Combining the above results and using the expression in (\ref{eq:KLsum}), we have $KL(f_0(X),f(X))\le \epsilon$. {\color{black}Finally, we note that this bound holds with positive prior probability under the PDPM-VAR, lgPDPM-VAR and rgPDPM-VAR models, thus yielding the desired result in Theorem 3.}

%%%%%%%%%%%%%%%%%%%%%%%%%%%%%%%%%%%%%%%%
%%%%%%%%%%%%%%%%%%%%%%%%%%%%%%%%%%%%%%%

\vskip 12pt

{\noindent \bf Proof of Theorem 4:} The proof of this result is based on modifications of the arguments in the proof of Proposition 2 in \cite{shen2013adaptive}, and Lemma 2 in \cite{canale2017posterior}. We will show that for every $f_P\in \mathcal{F}_{n,{\bf jl}}$, it is possible to find another density $f_{\hat{P}}$ belonging to $\hat{\mathcal{G}}$ (the $\epsilon$-net over $\mathcal{F}_{n,{\bf jl}}$) such that $||f_P - f_{\hat{P}} ||_1 \le \epsilon$. Since $N(\epsilon,\mathcal{F}_{n,{\bf jl}},|| \cdot||_1)$ is the minimum cardinality of the $\epsilon$-net over $\mathcal{F}_{n,{\bf jl}}$, we will be able to obtain a desired upper bound on the entropy if the number of balls required to cover the $\epsilon$-net $\hat{\mathcal{G}}$ is bounded. Let us consider $P_1=\sum\limits_{h_1\ge 1}\sum\limits_{\hsig\ge 1}\pi^{(1)}_{h_1}\pi^{(1)}_{\sigma,\hsig}\delta_{\big(\Theta^{(1)}_{h_1},\Sigma^{(1)}_{\hsig}\big)}$ and $P_2=\sum\limits_{h_1\ge 1}\sum\limits_{\hsig\ge 1}\pi^{(2)}_{h_1}\pi^{(2)}_{\sigma,\hsig}\delta_{\big(\Theta^{(2)}_{h_1},\Sigma^{(2)}_{\hsig}\big)}$. Using Lemma 4 (elaborated in the sequel), the distance between the corresponding densities can be expressed as 
\begin{eqnarray}
\small
&& ||f_{P_1} - f_{P_2} ||_1 \le 2\epsilon^2 + \sum_{h_1,\hsig<H_n}|\pi^{(1)}_{h_1}\pi^{(1)}_{\sigma,\hsig}-\pi^{(2)}_{h_1}\pi^{(2)}_{\sigma,\hsig}| + 4\epsilon \nonumber \\
&&+ \sum_{h_1,\hsig\le H_n}\pi^{(1)}_{h_1}\pi^{(1)}_{\sigma,\hsig}\bigg\|\prod_{t=1}^T\phi_{\Sigma^{(1)}_{\hsig}}\big({\bf x}_t - \sum_{k=1}^K A^{(1)}_{k,h_1}{\bf x}_{t-k}\big) - \prod_{t=1}^T\phi_{\Sigma^{(2)}_{\hsig}}\big({\bf x}_t - \sum_{k=1}^K A^{(2)}_{k,h_1}{\bf x}_{t-k}\big) \bigg\|_1. \label{eq:enet}
\end{eqnarray}

Let us first investigate the upper bound for the last term on the right hand side of (\ref{eq:enet}).

 Note that 
\begin{eqnarray}
&&\bigg\|\prod_{t=1}^T\phi_{\Sigma^{(1)}_{\hsig}}\big({\bf x}_t - \sum_{k=1}^K A^{(1)}_{k,h_1}{\bf x}_{t-k}\big) - \prod_{t=1}^T\phi_{\Sigma^{(2)}_{\hsig}}\big({\bf x}_t - \sum_{k=1}^K A^{(2)}_{k,h_1}{\bf x}_{t-k}\big) \bigg\|_1 \nonumber \\
&\le& \bigg\|\prod_{t=1}^T\phi_{\Sigma^{(2)}_{\hsig}}\big({\bf x}_t - \sum_{k=1}^K A^{(1)}_{k,h_1}{\bf x}_{t-k}\big) - \prod_{t=1}^T\phi_{\Sigma^{(2)}_{\hsig}}\big({\bf x}_t - \sum_{k=1}^K A^{(2)}_{k,h_1}{\bf x}_{t-k}\big) \bigg\|_1 \nonumber \\
&+& \bigg\|\prod_{t=1}^T\phi_{\Sigma^{(1)}_{\hsig}}\big({\bf x}_t - \sum_{k=1}^K A^{(1)}_{k,h_1}{\bf x}_{t-k}\big) - \prod_{t=1}^T\phi_{\Sigma^{(2)}_{\hsig}}\big({\bf x}_t - \sum_{k=1}^K A^{(1)}_{k,h_1}{\bf x}_{t-k}\big) \bigg\|_1, \label{eq:term1bound}
\end{eqnarray}
using the triangle inequality. The first term on the right hand side of (\ref{eq:term1bound}) can be bounded above using the following steps. In particular,  the first term is $\le $
\begin{align}
&\bigg\| \bigg\{\phi_{\Sigma^{(2)}_{\hsig}}\big({\bf x}_T - \sum_{k=1}^K A^{(1)}_{k,h_1}{\bf x}_{T-k}\big) -\phi_{\Sigma^{(2)}_{\hsig}}\big({\bf x}_T - \sum_{k=1}^K A^{(2)}_{k,h_1}{\bf x}_{T-k}\big) \bigg\}\prod_{t=1}^{T-1}\phi_{\Sigma^{(2)}_{\hsig}}\big({\bf x}_t - \sum_{k=1}^K A^{(1)}_{k,h_1}{\bf x}_{t-k}\big)
\bigg\|_1 \nonumber \\
&+ \bigg\|\phi_{\Sigma^{(2)}_{\hsig}}\big({\bf x}_T - \sum_{k=1}^K A^{(2)}_{k,h_1}{\bf x}_{T-k}\big)\bigg\{ \prod_{t=1}^{T-1}\phi_{\Sigma^{(2)}_{\hsig}}\big({\bf x}_t - \sum_{k=1}^K A^{(1)}_{k,h_1}{\bf x}_{t-k}\big) - \prod_{t=1}^{T-1}\phi_{\Sigma^{(2)}_{\hsig}}\big({\bf x}_t - \sum_{k=1}^K A^{(2)}_{k,h_1}{\bf x}_{t-k}\big) \bigg\}\bigg\|_1. \label{eq:term2bound}
\end{align}
The first term in (\ref{eq:term2bound}) may be written as 
$\small \int\bigg\{\int\bigg| \bigg\{\phi_{\Sigma^{(2)}_{\hsig}}\big({\bf x}_T - \sum_{k=1}^K A^{(1)}_{k,h_1}{\bf x}_{T-k}\big) - \phi_{\Sigma^{(2)}_{\hsig}}\big({\bf x}_T - \sum_{k=1}^K A^{(2)}_{k,h_1}{\bf x}_{T-k}\big)\bigg\} \bigg|d{\bf x}_T \bigg\}$ $\prod_{t=1}^{T-1}\phi_{\Sigma^{(2)}_{\hsig}}\bigg({\bf x}_t - \sum_{k=1}^K A^{(1)}_{k,h_1}{\bf x}_{t-k}\bigg)
d{\bf x}_{T-1}\ldots d{\bf x}_1 $, which is 
$\le \frac{2}{\pi\sqrt{\lambda_\nOutcome(\Sigma^{(2)}_{\hsig})}} 
\times \int\ldots\int\bigg\{\bigg\|\sum_{k=1}^K\bigg( A^{(1)}_{k,h_1} - A^{(2)}_{k,h_1}\bigg){\bf x}_{T-k} \bigg\| \times \prod_{t=1}^{T-1}\phi_{\Sigma^{(2)}_{\hsig}}\big({\bf x}_t - \sum_{k=1}^K A^{(1)}_{k,h_1}{\bf x}_{t-k}\big)\bigg\}d{\bf x}_{T-1}\ldots d{\bf x}_1,$
where we have used the well-known result $\small || \phi_{\Sigma }\big({\bf x}_T -  \bfmu_1\big) - \phi_{\Sigma}\big({\bf x}_T - \bfmu_2\big)||_1\le \frac{2}{\pi\sqrt{\lambda_\nOutcome(\Sigma)}}||\bfmu_1 - \bfmu_2 ||$.

One can write
$\bigg\|\sum_{k=1}^K\bigg( A^{(1)}_{k,h_1} - A^{(2)}_{k,h_1}\bigg){\bf x}_{T-k} \bigg\|$
=$\sqrt{\sum_{k=1}^K\sum_{l=1}^\nOutcome\bigg\{\sum_{l'=1}^\nOutcome\big( A^{(1)}_{k,h_1}(l,l') - A^{(2)}_{k,h_1}(l,l')\big)x_{T-k,l'}\bigg\}^2  } $
$\le \sqrt{\sum_{k=1}^K\sum_{l=1}^\nOutcome\bigg(\sum_{l'=1}^\nOutcome\big( A^{(1)}_{k,h_1}(l,l') - A^{(2)}_{k,h_1}(l,l')\big)^2\bigg) ||{\bf x}_{T-k}||^2  } = \sqrt{\sum_{k=1}^K ||vec\big( A^{(1)}_{k,h_1} - A^{(2)}_{k,h_1}\big)||^2 \times ||{\bf x}_{T-k}||^2} $
$\le \sum_{k=1}^K ||vec(A^{(1)}_{k,h_1}-A^{(2)}_{k,h_1})||\times ||{\bf x}_{T-k}||$,
where the second to last inequality is obtained using  Cauchy-Schwartz inequality, and the last inequality uses the fact that $\sum_{k=1}^K (a^*)^2_k\le (\sum_{k=1}^K |a^*|_k)^2$. Hence, the first term in (\ref{eq:term2bound}) has an upper bound $\frac{2}{\pi\sqrt{\lambda_d(\Sigma^{(2)}_{\hsig})}}\times \mathcal{K}^*,$ where 
\begin{align}
\small
\mathcal{K}^* &= \sum_{k=1}^K ||vec(A^{(1)}_{k,h_1}-A^{(2)}_{k,h_1})||
\int\bigg\{ ||{\bf x}_{T-k}||  \prod_{t=1}^{T-1}\phi_{\Sigma^{(2)}_{\hsig}}\big({\bf x}_t - \sum_{k'=1}^K A^{(1)}_{k',h_1}{\bf x}_{t-k'}\big)\bigg\}d{\bf x}_{T-1}\ldots d{\bf x}_1 \nonumber \\
&= \sum_{k=1}^K ||vec(A^{(1)}_{k,h_1}-A^{(2)}_{k,h_1})||\int\bigg\{ ||{\bf x}_{T-k}||\prod_{t=1}^{T-K} \phi_{\Sigma^{(2)}_{\hsig}}\big({\bf x}_{t} - \sum_{k'=1}^K A^{(1)}_{k',h_1}{\bf x}_{t-k'}\big)\bigg\}d{\bf x}_{T-k}\ldots d{\bf x}_1 \nonumber \\
&\le \sum_{k=1}^K ||vec(A^{(1)}_{k,h_1}-A^{(2)}_{k,h_1})||\int\bigg\{ ||{\bf x}_{T-k}||\prod_{t=1}^{T-k}(u_{\hsig,l})^{(d-1)/2} \phi_{\lambda_d(\Sigma^{(2)}_{\hsig})I_\nOutcome}\big({\bf x}_{t} - \sum_{k'=1}^K A^{(1)}_{k',h_1}{\bf x}_{t-k'}\big)\bigg\}d{\bf x}_{T-k}\ldots d{\bf x}_1, \label{eq:uppchi}
\end{align}
where the last inequality uses the relationship $ \phi_\Sigma\big({\bf x}\big)\le\big(\frac{\lambda_1(\Sigma)}{\lambda_\nOutcome(\Sigma)}\big)^{(\nOutcome-1)/2} \phi_{\lambda_1(\Sigma)I_\nOutcome}\big({\bf x}\big)$ and the fact that $\bigg(\frac{\lambda_1(\Sigma^{(2)}_{\hsig})}{\lambda_\nOutcome(\Sigma^{(2)}_{\hsig})} \bigg) \le u_{\hsig,l}$ for all densities belonging to $\mathcal{F}_{n,{\bf jl}}$. Note that $||{\bf x}_{T-k}||^2$ is distributed as a $\chi^2(\nOutcome)$ variable, and hence $\small \bigint\bigg\{ ||{\bf x}_{T-k}||\prod_{t=1}^{T-k} \phi_{\lambda_\nOutcome(\Sigma^{(2)}_{\hsig})I_\nOutcome}\big({\bf x}_{t} - \sum_{k'=1}^K A^{(1)}_{k',h_1}{\bf x}_{t-k'}\big)\bigg\}d{\bf x}_{T-k}\ldots d{\bf x}_1$ is bounded. To see this, note that when $k=1$, the integral may be written as 
\begin{eqnarray*}
  &&\int\bigg\{ ||{\bf x}_{T-1}||\prod_{t=1}^{T-1} \phi_{\lambda_\nOutcome(\Sigma^{(2)}_{\hsig})I_\nOutcome}\big({\bf x}_{t} - \sum_{k'=1}^K A^{(1)}_{k',h_1}{\bf x}_{t-k'}\big)d{\bf x}_{T-1}\bigg\}d{\bf x}_{T-2}\ldots d{\bf x}_1\\
  &&=
  \int_{||{\bf x}_{T-1}||\le 1}\bigg\{ ||{\bf x}_{T-1}||\prod_{t=1}^{T-1} \phi_{\lambda_\nOutcome(\Sigma^{(2)}_{\hsig})I_\nOutcome}\big({\bf x}_{t} - \sum_{k'=1}^K A^{(1)}_{k',h_1}{\bf x}_{t-k'}\big)d{\bf x}_{T-1}\bigg\}d{\bf x}_{T-2}\ldots d{\bf x}_1 \\
  &&+ 
  \int_{||{\bf x}_{T-1}||> 1}\bigg\{ ||{\bf x}_{T-1}||\prod_{t=1}^{T-1} \phi_{\lambda_\nOutcome(\Sigma^{(2)}_{\hsig})I_\nOutcome}\big({\bf x}_{t} - \sum_{k'=1}^K A^{(1)}_{k',h_1}{\bf x}_{t-k'}\big)d{\bf x}_{T-1}\bigg\}d{\bf x}_{T-2}\ldots d{\bf x}_1\\
  &&\le \int_{||{\bf x}_{T-1}||\le 1}\bigg\{ \prod_{t=1}^{T-1} \phi_{\lambda_\nOutcome(\Sigma^{(2)}_{\hsig})I_\nOutcome}\big({\bf x}_{t} - \sum_{k'=1}^K A^{(1)}_{k',h_1}{\bf x}_{t-k'}\big)d{\bf x}_{T-1}\bigg\}d{\bf x}_{T-2}\ldots d{\bf x}_1\\
  &&+ \int ||{\bf x}_{T-1}||^2\bigg\{ \prod_{t=1}^{T-1} \phi_{\lambda_\nOutcome(\Sigma^{(2)}_{\hsig})I_\nOutcome}\big({\bf x}_{t} - \sum_{k'=1}^K A^{(1)}_{k',h_1}{\bf x}_{t-k'}\big)d{\bf x}_{T-1}\bigg\}d{\bf x}_{T-2}\ldots d{\bf x}_1.
\end{eqnarray*}
The first term is bounded while the second term can be simplified using the expectation of a Chi-squared variable:
\begin{eqnarray*}
&&\int \bigg\{\nOutcome + \frac{1}{\lambda_\nOutcome\big(\Sigma^{(2)}_{\hsig})}\sum_{l'=1}^\nOutcome\bigg(\sum_{k'=1}^K A^{(1)}_{k',h_1}(l',\cdot){\bf x}_{T-1-k'}\bigg)^2 \bigg\}\prod_{t=1}^{T-2} \phi_{\lambda_\nOutcome(\Sigma^{(2)}_{\hsig})I_\nOutcome}\big({\bf x}_{t} - \sum_{k'=1}^K A^{(1)}_{k',h_1}{\bf x}_{t-k'}\big)d{\bf x}_{T-2}\ldots d{\bf x}_1 \\
%&&\le
%\int \bigg\{\nOutcome+\frac{1}{\lambda_\nOutcome(\Sigma^{(2)}_{\hsig})}\sum_{l'=1}^\nOutcome\sum_{k'=1}^K ||A^{(1)}_{kh_1}(l',\cdot)||^2||{\bf x}_{T-1-k'}||^2 \bigg\}\prod_{t=1}^{T-2} \phi_{\lambda_\nOutcome(\Sigma^{(2)}_{\hsig})I_\nOutcome}\big({\bf x}_{t} - \sum_{k'=1}^K A^{(1)}_{kh_1}{\bf x}_{t-k'}\big)d{\bf x}_{T-2}\ldots d{\bf x}_1\\
&&\le \nOutcome +\frac{1}{\lambda_\nOutcome(\Sigma^{(2)}_{\hsig})}\bar{a}^2_{h_1,j} \int \big(\sum_{k'=1}^K||{\bf x}_{T-1-k'}||^2\big)\prod_{t=1}^{T-2} \phi_{\lambda_\nOutcome(\Sigma^{(2)}_{\hsig})I_\nOutcome}\big({\bf x}_{t} - \sum_{k'=1}^K A^{(1)}_{k',h_1}{\bf x}_{t-k'}\big)d{\bf x}_{T-2}\ldots d{\bf x}_1.
\end{eqnarray*}
Noting that the above integrand is a function of the first moment of a Chi-squared random variable, one can use the same set of steps as above to derive an upper bound for the first term in (\ref{eq:term2bound}). Given that the number of terms in the above expression can be derived via a decision tree, the upper bound is given by 
\begin{eqnarray*}
&&c(\nOutcome,T,K) + \frac{\bar{a}^2_{h_1,j}}{\lambda_\nOutcome(\Sigma^{(2)}_{\hsig})} [K(\frac{\bar{a}^2_{h_1,j}}{\lambda_\nOutcome(\Sigma^{(2)}_{\hsig})})^{T-2} + K^2(\frac{\bar{a}^2_{h_1,j}}{\lambda_\nOutcome(\Sigma^{(2)}_{\hsig})})^{T-3} + \ldots K^{T-3}\frac{\bar{a}^2_{h_1,j}}{\lambda_\nOutcome(\Sigma^{(2)}_{\hsig})} + K^{T-2}]\\
&&\le c(\nOutcome,T,K) +\big(\frac{\bar{a}^2_{h_1,j}}{\underline{\sigma}^2_n} \big)K^{T-2}\times [1+\big(\frac{\bar{a}^2_{h_1,j}}{\underline{\sigma}^2_n} \big)+\big(\frac{\bar{a}^2_{h_1,j}}{\underline{\sigma}^2_n} \big)^2+\ldots + \big(\frac{\bar{a}^2_{h_1,j}}{\underline{\sigma}^2_n} \big)^{T-2}]\\
&&= c(\nOutcome,T,K) +\big(\frac{\bar{a}^2_{h_1,j}}{\underline{\sigma}^2_n} \big)K^{T-2}\times \frac{\big(\frac{\bar{a}^2_{h_1,j}}{\underline{\sigma}^2_n} \big)^{T-2}-1}{\big(\frac{\bar{a}^2_{h_1,j}}{\underline{\sigma}^2_n} \big)-1} \approx c(\nOutcome,T,K) + K^{T-2}\big(\frac{\bar{a}^2_{h_1,j}}{\underline{\sigma}^2_n} \big)^{T-2}  ,
\end{eqnarray*}
where $c(\nOutcome,T,K)$ is a bounded function when $\nOutcome,T,K,$ is fixed. Hence the upper bound in (\ref{eq:uppchi}) is given by $\small \frac{2}{\pi\underline{\sigma}_n}\times \bigg\{\sum_{k=1}^K ||vec(A^{(1)}_{k,h_1}-A^{(2)}_{k,h_1})||(u_{\hsig,l})^{(T-k)(\nOutcome-1)/2}
 \times \bigg(1+ c(\nOutcome,T,K) + K^{T-2}\big(\frac{\bar{a}^2_{h_1,j}}{\underline{\sigma}^2_n} \big)^{T-2} \bigg) \bigg\}$. The second term in (\ref{eq:term2bound}) is equal to 
$\bigg\|\phi_{\Sigma^{(2)}_{\hsig}}\big({\bf x}_T - \sum_{k=1}^K A^{(2)}_{k,h_1}{\bf x}_{T-k}\big)\bigg\{ \prod_{t=1}^{T-1}\phi_{\Sigma^{(2)}_{\hsig}}\big({\bf x}_t - \sum_{k=1}^K A^{(1)}_{k,h_1}{\bf x}_{t-k}\big) - \prod_{t=1}^{T-1}\phi_{\Sigma^{(2)}_{\hsig}}\big({\bf x}_t - \sum_{k=1}^K A^{(2)}_{k,h_1}{\bf x}_{t-k}\big) \bigg\}\bigg\|_1$ 
$ =\bigg\| \prod_{t=1}^{T-1}\phi_{\Sigma^{(2)}_{\hsig}}\big({\bf x}_t - \sum_{k=1}^K A^{(1)}_{k,h_1}{\bf x}_{t-k}\big) - \prod_{t=1}^{T-1}\phi_{\Sigma^{(2)}_{\hsig}}\big({\bf x}_t - \sum_{k=1}^K A^{(2)}_{k,h_1}{\bf x}_{t-k}\big) \bigg\|_1
 $, since $\bigg\|\phi_{\Sigma^{(2)}_{\hsig}}\big({\bf x}_T - \sum_{k=1}^K A^{(2)}_{k,h_1}{\bf x}_{T-k}\big)\bigg\|_1=1$.
Hence the first term in the upper bound in (\ref{eq:term1bound}) can be written as 
\begin{eqnarray}
&&\mathcal{K}^* \lesssim\frac{2\sum_{k=1}^K ||vec(A^{(1)}_{k,h_1}-A^{(2)}_{k,h_1})||(u_{\hsig,l})^{(T-k)(\nOutcome-1)/2}}{\pi\underline{\sigma}_n}\times \bigg\{
 c(\nOutcome,T,K) + K^{T-2}\big(\frac{\bar{a}^2_{h_1,j}}{\underline{\sigma}^2_n} \big)^{T-2} \bigg\} \nonumber \\
&& +  \bigg\|\prod_{t=1}^{T-1}\phi_{\Sigma^{(2)}_{\hsig}}\big({\bf x}_t - \sum_{k=1}^K A^{(1)}_{k,h_1}{\bf x}_{t-k}\big) - \prod_{t=1}^{T-1}\phi_{\Sigma^{(2)}_{\hsig}}\big({\bf x}_t - \sum_{k=1}^K A^{(2)}_{k,h_1}{\bf x}_{t-k}\big) \bigg\|_1\nonumber \\
&& \lesssim T\sum_{k=1}^K ||vec(A^{(1)}_{k,h_1}-A^{(2)}_{k,h_1})||(u_{\hsig,l})^{(T-k)(\nOutcome-1)/2}\times \Bigg(\frac{\pi\underline{\sigma}_n/2 }{ 
 c(\nOutcome,T,K) + K^{T-2}\big(\frac{\bar{a}^2_{h_1,j}}{\underline{\sigma}^2_n} \big)^{T-2} } \Bigg)^{-1}\label{eq:kappa}
%&+& \bigg\|\prod_{t=1}^T\phi_{\Sigma^{(1)}_{h_2}}\big({\bf x}_t - \sum_{k=1}^K A^{(1)}_{kh_1}{\bf x}_{t-k}\big) - \prod_{t=1}^T\phi_{\Sigma^{(2)}_{h_2}}\big({\bf x}_t - \sum_{k=1}^K A^{(1)}_{kh_1}{\bf x}_{t-k}\big) \bigg\|_1.
\end{eqnarray}
using similar steps to obtain an upper bound for  $\bigg\|\prod_{t=1}^{T-1}\phi_{\Sigma^{(2)}_{\hsig}}\big({\bf x}_t - \sum_{k=1}^K A^{(1)}_{k,h_1}{\bf x}_{t-k}\big) - \prod_{t=1}^{T-1}\phi_{\Sigma^{(2)}_{\hsig}}\big({\bf x}_t - \sum_{k=1}^K A^{(2)}_{k,h_1}{\bf x}_{t-k}\big) \bigg\|_1$. %The upper bound $\mathcal{K}^*$ for the first term in (\ref{eq:term1bound}) will satisfy  
%\begin{eqnarray}
%\mathcal{K}^*\lesssim \frac{2}{\pi\sqrt{\underline{\sigma}^2}}\times \sum_{t=1}^T\sum\limits_{k=1}\limits^{\min\{t-1,K\}} ||vec(A^{(1)}_{kh_1}-A^{(2)}_{kh_1})|| (u_{h_2,l})^{0.5(d-1)(t-k)}. \label{eq:kappa}
%\end{eqnarray}
For the second term in (\ref{eq:term1bound}), note that 
\begin{eqnarray}
&&\bigg\|\prod_{t=1}^T\phi_{\Sigma^{(1)}_{\hsig}}\big({\bf x}_t - \sum_{k=1}^K A^{(1)}_{k,h_1}{\bf x}_{t-k}\big) - \prod_{t=1}^T\phi_{\Sigma^{(2)}_{\hsig}}\big({\bf x}_t - \sum_{k=1}^K A^{(1)}_{k,h_1}{\bf x}_{t-k}\big) \bigg\|_1 \le \nonumber \\
&&\bigg\|\prod_{t=1}^T\phi_{\tilde{\Sigma}_{\hsig}}\big({\bf x}_t - \sum_{k=1}^K A^{(1)}_{k,h_1}{\bf x}_{t-k}\big) - \prod_{t=1}^T\phi_{\Sigma^{(2)}_{\hsig}}\big({\bf x}_t - \sum_{k=1}^K A^{(1)}_{k,h_1}{\bf x}_{t-k}\big) \bigg\|_1 + \nonumber \\
&&\bigg\|\prod_{t=1}^T\phi_{\tilde{\Sigma}_{\hsig}}\big({\bf x}_t - \sum_{k=1}^K A^{(1)}_{k,h_1}{\bf x}_{t-k}\big) - \prod_{t=1}^T\phi_{\Sigma^{(1)}_{\hsig}}\big({\bf x}_t - \sum_{k=1}^K A^{(1)}_{k,h_1}{\bf x}_{t-k}\big) \bigg\|_1, \label{eq:term3bound}
\end{eqnarray}
% JL - I ADDED SOME COMMAS HERE BETWEEN t and what was h_2 in the O term
where $\small \tilde{\Sigma}_{\hsig}= \bigg(O^{(2)}_{t, \hsig}\Lambda^{(1)}_{t, \hsig}(O^{(2)}_{t\hsig})'\bigg)^{-1}$, and $\Sigma^{(j)}_{\hsig} = \bigg(O^{(j)}_{t, \hsig}\Lambda^{(j)}_{t, \hsig}(O^{(j)}_{t, \hsig})' \bigg)^{-1}, \mbox{ } j=1,2$. Using Csiszar's inequality the first term in (\ref{eq:term3bound}) $\le$
\begin{eqnarray}
&& \small \sqrt{2\int\ldots\int\log\bigg(\prod_{t=1}^T\frac{\phi_{\tilde{\Sigma}_{\hsig}}\big({\bf x}_t - \sum_{k=1}^K A^{(1)}_{k,h_1}{\bf x}_{t-k}\big)}{ \phi_{\Sigma^{(2)}_{\hsig}}\big({\bf x}_t - \sum_{k=1}^K A^{(1)}_{k,h_1}{\bf x}_{t-k}\big)}\bigg)\bigg\{\prod_{t=1}^T\phi_{\tilde{\Sigma}_{\hsig}}\big({\bf x}_t - \sum_{k=1}^K A^{(1)}_{k,h_1}{\bf x}_{t-k}\big)\bigg\} d{\bf x}_T\ldots d{\bf x}_1} \nonumber \\
&&= \sqrt{2\times \frac{1}{2}\sum_{t=1}^T\bigg\{  \log\det\big(\tilde{\Sigma}^{-1}_{\hsig}\Sigma^{(2)}_{\hsig}\big) + tr\big((\Sigma^{(2)}_{\hsig})^{-1}\tilde{\Sigma}_{\hsig}\big) - \nOutcome \bigg\} }\nonumber \\
&&=
\sqrt{\sum_{t=1}^T\sum_{d'=1}^\nOutcome \bigg\{\frac{\lambda_{d'}(\Sigma^{(1)}_{\hsig})}{\lambda_{d'}(\Sigma^{(2)}_{\hsig})} - \log\big(\frac{\lambda_{d'}(\Sigma^{(1)}_{\hsig})}{\lambda_{d'}(\Sigma^{(2)}_{\hsig})}\big) -1 \bigg\} }. \label{eq:csis1}
\end{eqnarray}
Similarly, the second term in R.H.S. of (\ref{eq:term3bound}) $\small\le \sqrt{\sum_{t=1}^T\bigg\{  \log\det\big(\tilde{\Sigma}^{-1}_{\hsig}\Sigma^{(1)}_{\hsig}\big) + tr\big((\Sigma^{(1)}_{\hsig})^{-1}\tilde{\Sigma}_{\hsig}\big) - \nOutcome \bigg\} }$. Using similar steps as in the proof of Lemma 2 in \cite{canale2017posterior}, it is possible to show that the second term in the R.H.S. of  (\ref{eq:term3bound}) (and hence the last term in the R.H.S. of  (\ref{eq:term1bound})), has an upper bound given by $\sqrt{2T || O^{(1)}_{\hsig} - O^{(2)}_{\hsig} ||_2\frac{\lambda_1(\Sigma^{(1)}_{\hsig})}{\lambda_\nOutcome(\Sigma^{(1)}_{\hsig})} }$. Finally, we need to establish an upper bound for the term $\sum_{h_1,\hsig<H_n}|\pi^{(1)}_{h_1}\pi^{(1)}_{\sigma,\hsig}-\pi^{(2)}_{h_1}\pi^{(2)}_{\sigma,\hsig}|$ in (\ref{eq:enet}), which is given by {\color{black} Lemma 5} as 
$ \sum_{h_1,\hsig<H_n}\bigg|\tilde{\pi}^{(1)}_{h_1}\tilde{\pi}^{(1)}_{\sigma,\hsig}- \pi^{(2)}_{h_1}\pi^{(2)}_{\sigma,\hsig}\bigg| + \bigg|1-(1-\epsilon)^2 \bigg|, \mbox{ where } \tilde{\pi}=\frac{\pi_{h}}{(1-\sum_{h>H}\pi_{h})}.$

For a given $f_P\in \mathcal{F}_{n,{\bf jl}}$ with $P=\sum\limits_{h_1\ge 1}\sum\limits_{\hsig\ge 1}\pi^{(1)}_{h_1}\pi^{(1)}_{\sigma,\hsig}\delta_{\big(\Theta^{(1)}_{h_1},\Sigma^{(1)}_{\hsig}\big)}$ and $\Sigma_{h}=(O_{h}\Lambda_{h}O^T_{h})^{-1}$ where $\Lambda_{h}=diag(\lambda_{h,1},\ldots,\lambda_{h,\nOutcome})$, we will construct another density $f_{\hat{P}}$ with $\hat{P}= \sum\limits_{h_1\ge 1}\sum\limits_{\hsig\ge 1}\pi^{(1)}_{h_1}\pi^{(1)}_{\sigma,\hsig}\delta_{\big(\hat{\Theta}^{(1)}_{h_1},\hat{\Sigma}^{(1)}_{\hsig}\big)}$ within the $\epsilon$-net and then compute the cardinality of the $\epsilon$-net set to derive an upper bound for the entropy of sieve $\mathcal{F}_{n,{\bf jl}}$. 
To construct such a density, we will choose:\\

 {\noindent \bf 1. } $\hat{A}_{k,h_1}\in \mathcal{\hat{R}}_{h_1},h_1=1,\ldots,H,$ where $\mathcal{\hat{R}}_{h_1}$ is a $\epsilon^*$-net of $\mathcal{R}_{h_1}:=\{A\in \Re^{\nOutcome\times \nOutcome}:\underline{a}_{h_1,j}\le ||vec(A) || \le \bar{a}_{h_1,j} \}$, such that $||vec(A)_{k,h_1} - vec(\hat{A})_{k,h_1} ||\le \epsilon^*$, $k=1,\ldots,K,$ where\\ $\epsilon^*=\pi\underline{\sigma}_n\epsilon\times \frac{1}{2\big\{T (u_{\hsig,l})^{(T-k)(\nOutcome-1)/2} \big\} \times   \big\{ c(\nOutcome,T,K) + K^{T-2}\big(\frac{\bar{a}^2_{h_1,j}}{\underline{\sigma}^2_n} \big)^{T-2} \big\} }$, using (\ref{eq:kappa}).\\
 
 % which will reduce the upper bound for $\mathcal{K}^*$ in (\ref{eq:kappa}) to $\lesssim \epsilon$.
 {\noindent \bf 2. }      $\{\hat{\pi}_{h_1}\hat{\pi}_{\hsig}, h_1,\hsig\le H_n \}\in \hat{\Delta}$, where $\hat{\Delta}$ is a $\epsilon$-net of a $H^2_n$ dimensional probability simplex such that $\sum_{h_1,\hsig\le H_n} |\tilde{\pi}_{h_1}\tilde{\pi}_{\hsig} - \hat{\pi}_{h_1}\hat{\pi}_{\hsig} | \le \epsilon$, and $\tilde{\pi}_h=\frac{\pi_h}{\sum_{h\le H_n}\pi_h}, h\le H_n$.\\
   
 {\noindent \bf 3. }     $\hat{O}_{h}\in \mathcal{\hat{O}}_{h}$, where $\mathcal{\hat{O}}_{h}$ is a $\delta_h$-net of the set $\mathcal{O}_{h}$ defined as the set of $\nOutcome\times \nOutcome$ orthogonal matrices with respect to the spectral norm $||\cdot ||_2$ with $\delta_h=\epsilon^2/(2\nOutcome u_{h,l})$ such that $|| O_{h} - \hat{O}_{h}||_2\le T\delta_h$.\\
    
 {\noindent \bf 4. }   $(m_{h,1},...,m_{h_\nOutcome})\in \{1,...,M\}^\nOutcome ,h = 1,...,H,$ such that $\hat{\lambda}_{h,l} = \{\underline{\sigma}^2(1 + \epsilon\sqrt{\nOutcome})^{m_{h,l}-1} \}^{-1}$ will satisfy $1\le \hat{\lambda}_{h,l}/\lambda_{h,l} < (1 + \epsilon/\sqrt{\nOutcome})$.

Using this construction, the term in (\ref{eq:csis1}) is shown to be bounded by $\sqrt{T\sum_{d'=1}^\nOutcome \bigg\{\big(\frac{\hat{\lambda}_{h,\nOutcome-d'+1}}{\lambda_{h,\nOutcome-d'+1}} -1 \big)^2 \bigg\} }$. Moreover under this construction, it can be shown that $||f_P - f_{\hat{P}} ||_1 < C^*\epsilon$ for some constant $C^*$, by employing some additional algebra and the above arguments. Further, the cardinality of the $\epsilon$-net can be computed by noting that 
$\#(\hat{\Delta})\lesssim \epsilon^{-H^2_n} \mbox{ for } j=1,2,\#(\mathcal{\hat{O}}_{h})\lesssim\delta_h^{-\nOutcome(\nOutcome-1)/2},\#(\mathcal{\hat{R}}_{k,h})\lesssim [(\frac{\bar{a}_{h}}{\epsilon^*}+1)^{\nOutcome^2} - (\frac{\underline{a}_{h}}{\epsilon^*}-1)^{\nOutcome^2} ]$. Using these quantities, one can write the upper bound for the exponential of the entropy bound as $\lesssim (M)^{\nOutcome H_n}\epsilon^{-H^2_n}\times \mathcal{K}^*$, where 
\begin{eqnarray*}
&&\mathcal{K}^* = \prod_{h_1\le H_n} \big\{(\frac{\bar{a}_{h_1,j}}{\epsilon^*}+1)^{\nOutcome^2} - (\frac{\underline{a}_{h_1,j}}{\epsilon^*}-1)^{\nOutcome^2}  \big\}^K \prod_{\hsig\le H_n} \big\{\frac{2\nOutcome u_{\hsig,l}}{\epsilon^2} \big\}^{\nOutcome(\nOutcome-1)/2} \\
&&\lesssim \prod_{\hsig\le H_n} \big\{\frac{2 \nOutcome u_{\hsig,l}}{\epsilon^2} \big\}^{\nOutcome(\nOutcome-1)/2}  \prod_{h_1\le H_n}\Bigg\{\bigg(\frac{C^*_{h_{1,j},h_{\sigma,l}}\bar{a}_{h_1,j}}{\underline{\sigma}_n\epsilon} +1\bigg)^{\nOutcome^2} - \bigg(\frac{C^*_{h_{1,j},h_{\sigma,l}}\underline{a}_{h_1,j}}{\underline{\sigma}_n\epsilon} -1 \bigg)^{\nOutcome^2} \Bigg\}^K 
% &\approx& (M)^{dH_n}\epsilon^{-C_1H_n} \prod_{h_2\le H_n} \big\{\frac{2d u_{h_2,l}}{\epsilon^2} \big\}^{d(d-1)/2}  \times \prod_{h_1\le H_n}\bigg(\frac{\bar{a}_{h_1}^{d^2} - \underline{a}_{h_1}^{d^2}}{\epsilon} \bigg)^{K} \times \prod_{h_2\le H_n}\prod_{h_1\le H_n} \sqrt{\lambda_d(\Sigma^{(2)}_{h_2})}
%\lesssim (M)^{dH_n}\epsilon^{-C_1H_n}\prod_{h_1\le H_n} \big\{(\frac{\bar{a}_{h_1}}{\epsilon^*}+1)^{d^2} - (\frac{\bar{a}_{h_1}}{\epsilon^*}-1)^{d^2}  \big\}^{K} \prod_{h_2\le H_n} \big(2d u_{h_2,l} \big)^{d(d-1)/2},
\end{eqnarray*}
and $C^*_{h_{1,j},h_{\sigma,l}}= \frac{2}{\pi}\big\{T (u_{\hsig,l})^{(T-1)(\nOutcome-1)/2} \big\} \times  \big\{ c(\nOutcome,T,K) + K^{T-2}\big(\frac{\bar{a}^2_{h_1,j}}{\underline{\sigma}^2_n} \big)^{T-2} \big\}$.

\vskip 12pt

{\noindent \bf Proof of Corollary 1:}
For computing the entropy bound corresponding to sieves in the {\color{black}lgPDPM-VAR} model, note that using similar calculations as in (\ref{eq:enet}), one can show that the densities satisfy $|f_{P_1} - f_{P_2} ||_1=$
\begin{eqnarray}
%=  ||\sum\limits_{h_{11},\ldots,h_{1K}\ge 1}\sum_{\hsig\ge 1}\pi^{(1)}_{h_1}\pi^{(1)}_{\hsig}\prod_{t=1}^T \phi_{\Sigma^{(1)}_{\hsig}}\big({\bf x}_t - \sum_{k=1}^K A^{(1)}_{kh_1}{\bf x}_{t-k}\big) - \sum\limits_{h_1,\hsig\ge 1}\pi^{(2)}_{h_1}\pi^{(2)}_{\sigma,\hsig}\prod_{t=1}^T \phi_{\Sigma^{(2)}_{\hsig}}\big({\bf x}_t - \sum_{k=1}^K A^{(2)}_{kh_1}{\bf x}_{t-k}\big)||_1\\
&& \sum_{h_{11}}\ldots \sum_{h_{1K}}\sum_{\hsig<H_n}\big(\prod_{k=1}^K\pi^{(1)}_{k,h_{1k}}\big)\pi^{(1)}_{\sigma,\hsig}\bigg\|\prod_{t=1}^T\phi_{\Sigma^{(1)}_{\hsig}}\big({\bf x}_t - \sum_{k=1}^K A^{(1)}_{k,h_{1k}}{\bf x}_{t-k}\big) -  \prod_{t=1}^T\phi_{\Sigma^{(2)}_{\hsig}}\big({\bf x}_t - 
\sum_{k=1}^K A^{(2)}_{k,h_{1k}}{\bf x}_{t-k}\big) \bigg\|_1 \nonumber \\
&& + \sum_{h_{11}}\ldots \sum_{h_{1K}}\sum_{\hsig<H_n}\big|\big(\prod_{k=1}^K\pi^{(1)}_{k,h_{1k}}\big)\pi^{(1)}_{\sigma,\hsig}- \big(\prod_{k=1}^K\pi^{(2)}_{k,h_{1k}}\big)\pi^{(2)}_{\sigma,\hsig}\big| + K\epsilon_1^K+ L\epsilon_1,
\label{eq:enet2}
\end{eqnarray}
for some constant $L$.

Using similar calculations as under the PDPM-VAR case, %upper bound for the exponential of the entropy
$N(\epsilon_1,\mathcal{F}_{n,{\bf jl}},|| \cdot||_1)$ 
$\lesssim \bigg(\frac{M^\nOutcome}{\epsilon_1^{K}}\bigg)^{H_n}\prod_{k=1}^K\prod_{h_{1k}\le H_n}$ $\big\{(\frac{\bar{a}_{h_{1k},j}}{\epsilon_1^*}+1)^{\nOutcome^2} - (\frac{\underline{a}_{h_{1k},j}}{\epsilon_1^*}-1)^{\nOutcome^2}  \big\} \prod_{\hsig\le H_n} \big\{\frac{2\nOutcome u_{\hsig,l}}{\epsilon_1^2} \big\}^{\nOutcome(\nOutcome-1)/2}$ $\lesssim \bigg(\frac{M^{\nOutcome}}{\epsilon_1^{C^*_1}}\bigg)^{H_n}$  
 $\times$ $\prod_{\hsig\le H_n} 
\big\{\frac{2 \nOutcome u_{\hsig,l}}{\epsilon_1^2} \big\}^{\nOutcome(\nOutcome-1)/2} $ $\prod_{k=1}^K\prod_{h_{1k}\le H_n}\Big\{\big(\frac{C^{**}_{h_{1,j},h_{\sigma,l}}\bar{a}_{h_{1k},j}}{\underline{\sigma}_n\epsilon_1} +1\big)^{\nOutcome^2} - \big(\frac{C^{**}_{h_{1,j},h_{\sigma,l}}\underline{a}_{h_{1k},j}}{\underline{\sigma}_n\epsilon_1} -1 \big)^{\nOutcome^2} \Big\}$,
where $C^{**}_{h_{1,j},h_{\sigma,l}}= \frac{2}{\pi}\bigg\{T (u_{\hsig,l})^{(T-1)(\nOutcome-1)/2} \bigg\}$ $\times   \bigg\{
 c^{**}(\nOutcome,T,K) +  K^{T-2}\bigg(\frac{\max\big\{\bar{a}^2_{h_{11},j},\ldots,\bar{a}^2_{h_{1K},j}\big\}}{\underline{\sigma}_n} \bigg)^{T-2} \bigg\}$.

%except the part related to stick-breaking weights. In particular, the bounds derived in (\ref{eq:stickprob}) under the PDPM approach will change when considering the sieves in (\ref{eq:sieves-gPDPM}).

Similarly, for computing the entropy bound corresponding to sieves in the {\color{black}rgPDPM-VAR} model, note that using similar calculations as before, one can show that the densities satisfy $||f_{P_1} - f_{P_2} ||_1=$
\begin{eqnarray}
&& \sum\limits_{h_{11}\le H_n}\cdots \sum\limits_{h_{1D}\le H_n}\sum_{\hsig<H_n}\bigg|\big(\prod_{d'=1}^\nOutcome \pi^{*(1)}_{d',h_{1d'}}\big)\pi^{(1)}_{\sigma,\hsig}- \big(\prod_{d'=1}^\nOutcome \pi^{*(2)}_{d',h_{1d'}}\big)\pi^{(2)}_{\sigma,\hsig}\bigg| + \nonumber \\
&&
\sum\limits_{h_{11}\le H_n}\cdots \sum\limits_{h_{1D}\le H_n}\sum_{\hsig\le H_n}\big(\prod_{d'=1}^\nOutcome \pi^{*(1)}_{d',h_{1d'}}\big)\pi^{(1)}_{\sigma,\hsig} \times 
\bigg\|\prod_{t=1}^T\phi_{\Sigma^{(1)}_{\hsig}}\big({\bf x}_t - \sum_{k=1}^K A^{(1)}_{k,h_{11},\ldots,h_{1d}}{\bf x}_{t-k}\big) - \nonumber \\
&& \prod_{t=1}^T\phi_{\Sigma^{(2)}_{\hsig}}\big({\bf x}_t - 
\sum_{k=1}^K A^{(2)}_{k,h_{11},\ldots,h_{1d}}{\bf x}_{t-k}\big) \bigg\|_1 + K\epsilon_2^K+ L^*_2\epsilon_2, \quad \quad \label{eq:enet3}
\end{eqnarray}
 for constant $L^*_2$. Similar calculations as before yield the bound for exponential of the entropy $N(\epsilon_2,\mathcal{F}_{n,{\bf jl}},|| \cdot||_1))$ as:
 \begin{eqnarray}
&&\lesssim \bigg(\frac{M^\nOutcome}{\epsilon_2^{\nOutcome}}\bigg)^{H_n}\prod_{\hsig\le H_n} \big\{\frac{2\nOutcome u_{\hsig,l}}{\epsilon_2^2} \big\}^{\nOutcome(\nOutcome-1)/2}\prod_{d'=1}^\nOutcome\prod_{h_{1d'}\le H_n} \big\{(\frac{\bar{a}_{h_{1d'},j}}{\epsilon_2^*}+1)^{\nOutcome} - (\frac{\underline{a}_{h_{1d'},j}}{\epsilon_2^*}-1)^{\nOutcome}  \big\}^K \lesssim \bigg(\frac{M^{\nOutcome}}{\epsilon_2^{C^*_1}}\bigg)^{H_n}  \nonumber   \\
&& \times \prod_{\hsig\le H_n} \big\{\frac{2\nOutcome u_{\hsig,l}}{\epsilon_2^2} \big\}^{\nOutcome(\nOutcome-1)/2}  \prod_{d'=1}^\nOutcome\prod_{h_{1d'}\le H_n}  \Bigg\{\bigg(\frac{\tilde{C}^*_{h_{1,j},h_{\sigma,l}}\bar{a}_{h_{1d'},j}}{\underline{\sigma}_n\epsilon_2} +1\bigg)^{\nOutcome} - \bigg(\frac{\tilde{C}^*_{h_{1,j},h_{\sigma,l}}\underline{a}_{h_{1d'},j}}{\underline{\sigma}_n\epsilon_2} -1 \bigg)^{\nOutcome} \Bigg\}^K, \label{eq:ent-rgPDPM}
\end{eqnarray}
 where $\tilde{C}^*_{h_{1,j},h_{\sigma,l}}= \frac{2}{\pi}\bigg\{T (u_{\hsig,l})^{(T-1)(\nOutcome-1)/2} \bigg\} \times   \bigg\{
 \tilde{c}^*(\nOutcome,T,K) + K^{T-2}\bigg(\frac{\max\{\bar{a}^2_{h_{11},j},\ldots,\bar{a}^2_{h_{1d},j}\}}{\underline{\sigma}_n} \bigg)^{T-2} \bigg\}$.
 
\vskip 12pt

{\noindent \bf Proof of Theorem 5:} The proof follows using Theorem 2 and the entropy bounds established in Theorem 3. For the case of PDPM-VAR, consider the sieves
\begin{eqnarray}
\small
&&\mathcal{F}_{n,{\bf j,l}} = \bigg\{f_p\in \mathcal{F}_n: \mbox{for } h_1,\hsig\le H_n, 
n^{H^2_n}(j_{h_1} -1)\le ||vec(A_{k,h_1}) ||\le n^{H^2_n}j_{h_1},\forall k, \mbox{ } \nonumber\\
&& n^{l_{\hsig}-1} <\frac{\lambda_1(\Sigma_{\hsig})}{\lambda_\nOutcome(\Sigma_{\hsig})}\le n^{l_{\hsig}}  \bigg\}, \mbox{ }
\mathcal{F}_n = \bigg\{f_p: P = \sum\limits_{h_1\ge 1}\sum\limits_{\hsig\ge 1}\pi_{h_1}\pi_{\sigma,\hsig}\delta_{\Theta_{h_1},\Sigma_{\hsig}}: \sum_{h_1>H_n}\pi_{h_1}<\epsilon, \nonumber \\ &&\sum_{\hsig>H_n}\pi_{\sigma,\hsig}<\epsilon, \mbox{for }
 \hsig\le H_n, \mbox{ } \underline{\sigma}_n^2\le \lambda_\nOutcome, \lambda_1 \le \underline{\sigma}_n^2(1 + \epsilon/\sqrt{\nOutcome})^{M_n}, \mbox{ } 1<\frac{\lambda_1}{\lambda_\nOutcome}\le  n^{H_n} \bigg\},\mbox{ } \label{eq:sieves-PDPM1.1}
\end{eqnarray}
where $M_n=\underline{\sigma}^{-2c_2}=n$ and $H_n=\lfloor Cn\epsilon^2/\log(n) \rfloor$ for some positive constant $C$, 
and clearly $\mathcal{F}_n\subset \cup_{{\bf j,l}} \mathcal{F}_{n,{\bf j,l}}$. Comparing to the notations used in the manuscript, 
we note that $\underline{a}_{h_1,j}= n^{H_n}(j_{h_1}-1), \bar{a}_{h_1,j}= n^{H_n}j_{h_1}$, and $u_{\hsig,l}=n^{l_{\hsig}}$, for integers $(j_1,\ldots,j_{H_n})\in \mathbb{N}$ and $(l_1,\ldots,l_{H_n})\in \{1,\ldots, H_n \}$. 

Using Lemma 6 in the Appendix, it is clear that condition (2A) holds for PDPM-VAR. Next, we will derive the entropy bounds and the complement of the prior probability for the sieves, and illustrate that the summability condition {\it(2B)} in Theorem 2 holds.

Now, using Theorem 4, the upper bound on the entropy term is  $\lesssim (M^\nOutcome\epsilon^{-C_1})^{H_n} \times \\\prod_{\hsig=1} ^{H_n} \big\{\frac{2\nOutcome u_{\hsig,l}}{\epsilon^2} \big\}^{\nOutcome(\nOutcome-1)/2} \times \mathcal{K}^*$, $\small \mathcal{K}^*= \prod_{h_1=1}^ {H_n}\Big\{\big(\frac{C^*_{h_{1,j},h_{\sigma,l}}\bar{a}_{h_1,j}}{\underline{\sigma}_n\epsilon} +1\big)^{\nOutcome^2} - \big(\frac{C^*_{h_{1,j},h_{\sigma,l}}\underline{a}_{h_1,j}}{\underline{\sigma}_n\epsilon} -1 \big)^{\nOutcome^2} \Big\}^K $,  
\begin{eqnarray}
%&&=  (M^d\epsilon^{-C_1})^{H_n}  \prod_{h_2\le H_n} \big\{\frac{2d u_{h_2,l}}{\epsilon^2} \big\}^{d(d-1)/2}   \prod_{h_1\le H_n}\big(C^*_{h_{1,j},h_{\sigma,l}} \big)^{Kd^2} \Bigg\{\bigg(\frac{\bar{a}_{h_1,j}}{\underline{\sigma}_n\epsilon} + \frac{1}{C^*_{h_{1,j},h_{\sigma,l}}}\bigg)^{d^2} - \bigg(\frac{\underline{a}_{h_1,j}}{\underline{\sigma}_n\epsilon} - \frac{1}{C^*_{h_{1,j},h_{\sigma,l}}} \bigg)^{d^2} \Bigg\}^K \nonumber\\
&& \mbox{ i.e. } \mathcal{K}^* \approx   \prod_{h_1\le H_n}\big(C^*_{h_{1,j},h_{\sigma,l}} \big)^{K\nOutcome^2}  \Bigg\{\bigg(\frac{\bar{a}_{h_1,j}}{\underline{\sigma}_n\epsilon} + o_n(1)\bigg)^{\nOutcome^2} - \bigg(\frac{\underline{a}_{h_1,j}}{\underline{\sigma}_n\epsilon} - o_n(1)\bigg)^{\nOutcome^2} \Bigg\}^K \nonumber \\
&&\lesssim    \prod_{h_1\le H_n}\big(C^*_{h_{1,j},h_{\sigma,l}} \big)^{K \nOutcome^2} \Bigg\{\bigg(\frac{\bar{a}_{h_1,j}}{\underline{\sigma}_n\epsilon} + 1\bigg)^{\nOutcome^2} - \bigg(\frac{\underline{a}_{h_1,j}}{\underline{\sigma}_n\epsilon} - 1\bigg)^{\nOutcome^2} \Bigg\}^K \lesssim \big\{ (u_{\hsig,l})^{(T-1)(\nOutcome-1)/2} \big\}^{K\nOutcome^2H_n}  \nonumber \\
&&\times  \prod_{h_1\le H_n} \bigg[\frac{2T}{\pi}\bigg\{
 c(\nOutcome,T,K) + K^{T-2}\big(\frac{\bar{a}^2_{h_1,j}}{\underline{\sigma}_n} \big)^{T-2}  \bigg\}\bigg]^{K\nOutcome^2} \Bigg\{\bigg(\frac{\bar{a}_{h_1,j}}{\underline{\sigma}_n\epsilon} + 1\bigg)^{\nOutcome^2} - \bigg(\frac{\underline{a}_{h_1,j}}{\underline{\sigma}_n\epsilon} - 1\bigg)^{\nOutcome^2} \Bigg\}^K, \quad \quad \label{eq:num_en}
\end{eqnarray}
where  $o_n(1)$ is a vanishing term with increasing sample size. Using similar steps as in the proof of Theorem 2 in Canale and De Blasi (2017), it is possible to show that 
  $\bigg[\big(\frac{\bar{a}_{h_1,j}}{\underline{\sigma}_n\epsilon/2}+1 \big)^{\nOutcome^2} - \big(\frac{\underline{a}_{h_1,j}}{\underline{\sigma}_n\epsilon/2}-1 \big)^{\nOutcome^2}  \bigg]^K\lesssim \bigg[\frac{n^{(H_n+\frac{1}{2c_2})\nOutcome^2}j_{h1}^{\nOutcome^2-1}}{(\epsilon)^{\nOutcome^2}} \bigg]^K $,  when $n$ and $j_{h_1}$ is large. Hence, when $n$ is large enough, %where $c_4=\frac{1}{2}+ \frac{1}{2c_2}$,
  \begin{eqnarray*}
&&  \bigg[\frac{2T}{\pi}\bigg\{
 c(\nOutcome,T,K) + \big(\frac{\bar{a}^2_{h_1,j}}{\underline{\sigma}_n)} \big)^{T-2} \big(\frac{K^{T-1}-1}{K-1}\big) \bigg\}\bigg]^{K\nOutcome^2} \Bigg\{\bigg(\frac{\bar{a}_{h_1,j}}{\underline{\sigma}_n\epsilon} + 1\bigg)^{\nOutcome^2} - \bigg(\frac{\underline{a}_{h_1,j}}{\underline{\sigma}_n\epsilon} - 1\bigg)^{\nOutcome^2} \Bigg\}^K \\
 &&\le \bigg\{\frac{n^{(H^2_n+\frac{1}{2c_2})\nOutcome^2} j_{h1}^{\nOutcome^2-1}}{(\epsilon)^{\nOutcome^2}} \bigg\}^K  \times \bigg[\frac{2T}{\pi}\bigg\{
 c(\nOutcome,T,K) + \big( n^{2H^2_n + 1/(2c_2)}j^2_{h_1}\big)^{T-2} \bigg\}\bigg]^{K\nOutcome^2} \\
 &&\approx \mathcal{C}\bigg(1 + o_n(1)\bigg)^{K\nOutcome^2}\big( n^{2H^2_n + 1/(2c_2)}j^2_{h_1}\big)^{K\nOutcome^2(T-2)}\bigg\{\frac{n^{(H^2_n+\frac{1}{2c_2})\nOutcome^2}j_{h1}^{\nOutcome^2-1}}{(\epsilon)^{\nOutcome^2}} \bigg\}^K\le  \mathcal{C}\bigg(1 + o_n(1)\bigg)^{K\nOutcome^2}  \\
 &&\times \exp\bigg\{{H^2_nK\nOutcome^2(2T-3)\log(n) + \frac{1}{c_2}K\nOutcome^2(T-1)}\log(n) \bigg\}\times \big(j_{h1} \big)^{2K\nOutcome^2(T-2) + K(\nOutcome^2-1)} \big(\frac{1}{\epsilon}\big)^{K\nOutcome^2},
  \end{eqnarray*}
 where $\mathcal{C}$ is a constant independent of $n$. Further, 
  \begin{eqnarray*}
 &&  \big\{\frac{2\nOutcome u_{\hsig,l}}{\epsilon^2} \big\}^{\nOutcome(\nOutcome-1)/2} \big\{ (u_{\hsig,l})^{(T-1)(\nOutcome-1)/2} \big\}^{K\nOutcome^2H_n} = \mathcal{C}_1 (u_{\hsig,l})^{\nOutcome(\nOutcome-1)/2 + K\nOutcome^2H_n(T-1)(\nOutcome-1)/2} \\
 &&\approx \mathcal{C}_1 \exp\bigg\{ \bigg(\frac{\nOutcome(\nOutcome-1)}{2} + K\nOutcome^2H_n(T-1)\frac{\nOutcome-1}{2}\bigg) \log(n^{l_{\hsig}})\bigg\}\times \big(\frac{1}{\epsilon}\big)^{\nOutcome(\nOutcome-1)},
 \end{eqnarray*}
 for large $n$, where the constant $\mathcal{C}_1$ that does not depend on $n$.

  %Moreover $C^*_{h_{1,j},h_{2,l}}$ can be simplified as $C^*_{h_{1,j},h_{2,l}}\approx \mathcal{C}\bigg(\big(n^{2^{l_{h_2}}} \big)^{\frac{d-1}{2}} n^{1 + 1/(2c_2)}j^2_{h_1} \bigg)^{T-2}$ when $n$ is large enough, for some constant $\mathcal{C}>0$.
  
 Hence we have 
\begin{eqnarray}
&&N(\epsilon,\mathcal{F}_{n,{\bf jl}},|| \cdot||_1)\lesssim  \mathcal{C}^{H_n}_1\big(\frac{1}{\epsilon}\big)^{C_1H_n} \bigg\{\mathcal{C}\bigg(1 + o_n(1)\bigg)^{K\nOutcome^2}\bigg\}^{H_n} \exp\bigg\{\nOutcome H_n\log(M) + H_n\log \frac{C_1}{\epsilon} \bigg\}  \nonumber \\
%\exp\bigg\{dH_n\log(M) + H_n\log \frac{C_1}{\epsilon} + K\sum\limits_{h_1\le H_n}\log\bigg[\frac{n^{c_4d^2}j_{h1}^{d^2-1}}{(\epsilon)^{d^2}} \bigg]  + \frac{d(d-1)}{2}\sum_{h_2\le H_n}\log\big( \frac{2d n^{2^{l_{h_2}}}}{\epsilon^2}\big) \\
%&&+ Kd^2(T-2) \bigg(\sum\limits_{h_1\le H_n} \log\big( n^{1 + 1/(2c_2)}j^2_{h_1}\big) + \frac{(d-1)}{2}\sum\limits_{h_2\le H_n} \log(n^{2^{l_{h_2}}})  \bigg)\bigg\}.
%&\times& \exp\bigg\{dH_n\log(M) + H_n\log \frac{C_1}{\epsilon} \bigg\}\times \prod_{h_2\le H_n} (n^{l_{h_2}})^{\frac{d(d-1)}{2} + Kd^2H_n(T-1)\frac{d-1}{2}}\\
%+\frac{H_n}{2}\bigg(\frac{d(d-1)}{2} + Kd^2H_n(T-1)\frac{d-1}{2}\bigg) \log(n^{l_{\hsig}})\bigg\}\\
&\times&   \exp\bigg\{H^3_n{(2K\nOutcome^2(T-2) + K\nOutcome^2)\log(n) + \frac{H_n}{c_2}K\nOutcome^2(T-1)}\log(n) \bigg\} \nonumber \\
%&\times& \exp\bigg\{ \bigg(\frac{d(d-1)}{2} + Kd^2H_n(T-1)\frac{d-1}{2}\bigg) \log(n^{l_{\hsig}})\bigg\} \\
&\times&
 \bigg\{ \prod_{\hsig\le H_n} (n^{l_{\hsig}})^{\frac{\nOutcome(\nOutcome-1)}{2} + K\nOutcome^2H_n(T-1)\frac{\nOutcome-1}{2}}\bigg\}\bigg\{ \prod_{h_1\le H_n}(j_{h1})^{2K\nOutcome^2(T-2) + K(\nOutcome^2-1)}\bigg\}\times \bigg(\frac{1}{\epsilon}\bigg)^{H_n(K\nOutcome^2 + \nOutcome(\nOutcome-1))}.\label{eq:fullentropy}
\end{eqnarray}

%\exp\bigg\{dH_n\log(M) + H_n\log \frac{C_1}{\epsilon}\bigg\}\times
%\mathcal{C}_1 \exp\bigg\{ \bigg(\frac{d(d-1)}{2} + Kd^2H_n(T-1)\frac{d-1}{2}\bigg) \log(\sqrt{n}l_{h_2})\bigg\}\times \big(\frac{1}{\epsilon}\big)^{d(d-1)}\times \mathcal{C}\bigg(1 + o_n(1)\bigg)^{Kd^2}  \exp\bigg\{{(2Kd^2(T-2) + Kd^2)H_n + \frac{1}{c_2}Kd^2(T-1)}\log(n) \bigg\}\times \frac{1}{\epsilon^{Kd^2}}

Further, under the specification $P^*_1(A_1,\ldots,A_K)=\prod_{k=1}^K P^*_1(A_k)$, we have
\begin{eqnarray}
&&\Pi(\mathcal{F}_{n,{\bf jl}})\le \prod\limits_{h_1\le H_n} P^*_1\big(||vec(A_k) ||>n^{H^2_n}(j_{h_1} -1), \mbox{ } \forall k\big)\prod_{\hsig\le H_n}P^*_2\big( \lambda_1(\Sigma)/\lambda_\nOutcome(\Sigma)>n^{(l_{\hsig}-1)}\big), \nonumber\\
&\lesssim& \prod_{h_1\le H_n}\big\{\big( n^{H^2_n}(j_{h_1}-1)\big)^{-1_{(j_{h_1}\ge 2)}2(r+1)}\big\}^{K}\times\prod_{\hsig\le H_n} (n^{(l_{\hsig}-1)})^{-1_{(l_{\hsig}\ge 1)}\kappa} \nonumber\\
&\approx&  \bigg\{n^{-2H^3_n(r+1)K} \prod_{h_1\le H_n} \big(j_{h_1}-1\big)^{-1_{(j_{h_1}\ge 2)}2K(r+1)} \bigg\}\times \bigg\{\prod_{\hsig\le H_n} (n^{\kappa(l_{\hsig}-1)})^{-1_{(l_{\hsig}\ge 1)}} \bigg\}, \mbox{ for large } n. \quad \quad \label{eq:sieve-prob}
\end{eqnarray}

Under Lemma 7 in the Appendix, we can show that for PDPM-VAR,\\ $\sum_{j_{h_1}\in N}\sum_{1\le l_{\hsig\le H_n}}\sqrt{N(\epsilon,\mathcal{F}_{n,j_{h_1}l_{\hsig}})\Pi(\mathcal{F}_{n,j_{h_1}l_{\hsig}})} e^{-(4-c)n\epsilon^2}\to 0$ as $n\to\infty$ for a suitable choice of constants. Hence the condition $(2B)$ in Theorem 2 is satisfied and the strong posterior consistency result is proved corresponding to the PDPM-VAR model.

To prove the summability result for the {\color{black}lgPDPM-VAR} approach, consider the sieves defined in the manuscript as 
\begin{eqnarray}
\small
&&\mathcal{F}_n = \bigg\{f_p: P = \sum_{h_{11}=1}^\infty\ldots\sum_{h_{1K}=1}^\infty \sum_{\hsig=1}^\infty \pi_{\sigma,\hsig} \big(\prod_{k=1}^K\pi_{k,h_{1k}}\big)\delta_{\Theta_{h_{1k}},\Sigma_{\hsig}}: \sum_{h_{1,1k}>H_n}\pi_{h_{1,1k}}<\epsilon_1,1\le k\le K,\nonumber \\
&&\sum_{h_{\sigma}>H_n}\pi_{\sigma,h_{\sigma}}<\epsilon_2, 
 \underline{\sigma}_n^2\le \lambda_\nOutcome(\Sigma_{\hsig}), \lambda_1(\Sigma_{\hsig}) \le \underline{\sigma}_n^2(1 + \epsilon/\sqrt{\nOutcome})^{M_n}, \mbox{ } 1<\frac{\lambda_1(\Sigma_{\hsig})}{\lambda_\nOutcome(\Sigma_{\hsig})}\le  n^{H_n}, \mbox{ } \hsig\le H_n  \bigg\}, \mbox{ }  \nonumber  \\
&&\mathcal{F}_{n,{\bf j}, {\bf l}}= \bigg\{f_p\in \mathcal{F}_n:  \mbox{ for } h_{11},\ldots,h_{1K}\le H_n, n^{H^2_n}(j_{h_{1k}}-1)\le ||vec(A_{kh_{1k}}) ||\le n^{H^2_n}j_{h_{1k}},  \\ \nonumber 
&&\mbox{ and for } \hsig \le H_n, n^{l_{\hsig}-1}\le \frac{\lambda_1(\Sigma_{\hsig})}{\lambda_\nOutcome(\Sigma_{\hsig})}\le n^{l_{\hsig}}  \bigg\} \quad \label{eq:sieves-gPDPM1.2}
\end{eqnarray}

Now note that the prior probability on the sieve $\mathcal{F}_n$  satisfies
\begin{eqnarray*}
&&\Pi(\mathcal{F}_n^c)\le Pr\bigg( \sum_{h_{\sigma}>H_n}\pi_{\sigma,\hsig}>\epsilon_1\bigg) + \sum_{k=1}^K Pr\bigg( \sum_{h_{1k}>H_n}\pi_{k,h_{1k}}>\epsilon_1\bigg) +  H_nP^*_2\bigg( \lambda_\nOutcome(\Sigma) \le \underline{\sigma}^2_n\bigg) + \\
&& H_n P^*_2\bigg(\lambda_1(\Sigma)>\underline{\sigma}_n^2(1+ \epsilon_1/\sqrt{\nOutcome})^{M_n} \bigg) + H_n P_2^*\bigg( \frac{\lambda_1(\Sigma_{\hsig})}{\lambda_\nOutcome(\Sigma_{\hsig})}> n^{H_n}\bigg) \lesssim e^{-b^*n},
\end{eqnarray*}
using similar techniques used to derive (\ref{eq:prior-prob1}) in Lemma 6 of the Appendix. Hence, condition $(2A)$ in Theorem 2 holds. Further, one can rewrite (\ref{eq:sieve-prob}) as
\begin{eqnarray*}
&&\Pi(\mathcal{F}_{n,{\bf jl}})\lesssim
 \bigg\{n^{-2H^3_n(r+1)K} \prod_{k=1}^K \prod_{h_{1k}\le H_n} \big(j_{h_{1k}}-1\big)^{-1_{(j_{h_{1k}}\ge 2)}2(r+1)} \bigg\}\times \bigg\{\prod_{\hsig\le H_n} (n^{\kappa(l_{\hsig}-1)})^{-1_{(l_{\hsig}\ge 1)}} \bigg\}, \label{eq:sieve-prob2}
\end{eqnarray*}
for large $n$. Moreover, using similar steps as in the proof for Lemma 7 corresponding to the PDPM-VAR model,  it is possible to show that the entropy bound in Corollary 1 satisfies $\sqrt{N(\epsilon_1,\mathcal{F}_{n,{\bf j}_{h_1}{\bf l}_{\hsig}},|| \cdot||_1)\Pi(\mathcal{F}_{n,{\bf j}_{h_1}{\bf l}_{\hsig}}})$
\begin{eqnarray*}
 &&\lesssim \sqrt{\mathcal{C}^{H_n}_1\times \bigg\{\mathcal{C}\bigg(1 + o_n(1)\bigg)^{K\nOutcome^2}\bigg\}^{H_n}}\exp\bigg\{H_n\log( \frac{C^*_1}{\epsilon_1})+\frac{H_n}{2}(K\nOutcome^2+\nOutcome(\nOutcome-1)\log(\frac{1}{\epsilon_1}\bigg\} \nonumber \\
 &\times& \exp\bigg\{\frac{1}{2}\big(\nOutcome H_n\log(M)  + \frac{H_n}{2c_2}K\nOutcome^2(T-1)\log(n) ) \bigg\}\times n^{KH^3_n\big(\nOutcome^2(T-2) + \frac{1}{2}\nOutcome^2- r\big)-KH^3_n}\nonumber \\
 &\times& \bigg\{\prod_{k=1}^K \prod_{h_{1k}\le H_n} (j_{h_{1k}})^{\nOutcome^2(T-2) + \frac{1}{2}(\nOutcome^2-1)}\bigg\} \bigg\{ \prod_{k=1}^K\prod_{h_{1k}\le H_n}\big(j_{h_{1k}}-1\big)^{-1_{(j_{h_{1k}}\ge 2)}(r+1)} \bigg\} \nonumber \\
 &\times& \prod_{\hsig\le H_n}
\bigg\{  n^{l_{\hsig}}\bigg\}^{\frac{\nOutcome(\nOutcome-1)}{4} + K\nOutcome^2H_n(T-1)\frac{\nOutcome-1}{4}} \times 
\bigg\{\prod_{\hsig\le H_n} (n^{\kappa(l_{\hsig}-1)/2})^{-1_{(l_{\hsig}\ge 1)}} \bigg\}.\label{eq:sumbound2}
\end{eqnarray*}
Using the above expressions and similar arguments as in (\ref{eq:final-bound}), it is straightforward to show that condition $(2B)$ in Theorem 2 holds. Hence the result is proved.

The proof for {\color{black} rgPDPM-VAR} proceeds in a similar fashion by noting that the prior probability for the complement of sieves defined in the manuscript can be written as
\begin{eqnarray*}
&&\Pi(\mathcal{F}_n^c)\le   H_n P^*_2\bigg(\lambda_1(\Sigma)>\underline{\sigma}_n^2(1+ \epsilon/\sqrt{\nOutcome})^{M_n} \bigg) 
+ H_n P_2^*\bigg( \frac{\lambda_1(\Sigma_{\hsig})}{\lambda_\nOutcome(\Sigma_{\hsig})}> n^{H_n}\bigg) + H_n P^*_2\bigg( \lambda_\nOutcome(\Sigma) \le \underline{\sigma}^2_n\bigg)\\
&&+Pr\bigg(\sum_{\hsig>H_n} \pi_{\sigma,\hsig}>\epsilon_2 \bigg) + Pr\bigg( \sum_{h_{11}>H_n}\pi^*_{1,h_{11}}>\epsilon_2\bigg) + \cdots + Pr\bigg( \sum_{h_{1D}>H_n}\pi^*_{D,h_{1D}}>\epsilon_2\bigg) \lesssim e^{-b^*n},
\end{eqnarray*}
using similar techniques used to derive (\ref{eq:prior-prob1}) in Lemma 6 in the Appendix. Hence, condition $(2A)$ in Theorem 2 holds. 
Also define $\mathcal{F}_{n,{\bf j}, {\bf l}}$ such that $\mathcal{F}_n\subset \cup_{{\bf j,l}}\mathcal{F}_{n,{\bf j}, {\bf l}}$ and 
\begin{eqnarray}
\small
&&\mathcal{F}_{n,{\bf j,l}}= \bigg\{f_p\in \mathcal{F}_n: \mbox{for } h_{11},\ldots,h_{1\nOutcome}, n^{H^2_n}(j_{h_{1d'}}-1)\le ||vec(A_{kh_{1k}}) ||\le n^{H^2_n}j_{h_{1d'}},  \nonumber \\
&& d'=1,\ldots,\nOutcome, \mbox{ and for } \hsig\le H_n, \quad  n^{l_{\hsig}-1}\le \frac{\lambda_1(\Sigma_{\hsig})}{\lambda_\nOutcome(\Sigma_{\hsig})}\le n^{l_{\hsig}}  \bigg\} \quad \label{eq:sieves-rgPDPM1.2}.
\end{eqnarray}
Given the entropy bound in (\ref{eq:ent-rgPDPM}) and using similar calculations as in Lemma 7 for the PDPM-VAR case, it is possible to show that condition $(2B)$ in Theorem 2 holds. The strong consistency result follows under rgPDPM-VAR once conditions $(2A)$-$(2B)$ in Theorem 2 are satisfied.

%{\noindent \bf Proof of Lemma 1:} The proof of Lemma 1 involving multivariate density estimation under a product of DP priors, follow the principles of the Kullback-Leibler result in \cite{wu2008kullback}. In particular, the results Theorems 1-2 and Lemmas 1-3 in that paper hold for certain kernel mixtures of the form $\int K(x; \Theta)dP(\Theta)$ satisfying certain conditions, which includes the kernel mixtures under product of DP priors considered in this article. Once the compact support for $P_\epsilon$ is taken as $ [-\mu^*,\mu^*]^D \times \{\Sigma\in \mathcal{S}: \underline{\sigma}^2 < \lambda_l(\Sigma) < \bar{\sigma}^2, 1\le l\le p \},$ for some constants $\mu^*>0$ and $0<\underline{\sigma}<\bar{\sigma}$, and eigen values denoted as $\lambda_l,l=1,\ldots,p$, the rest of the proof can proceed along similar lines as Lemma 1 in \cite{canale2017posterior}.

\vskip 12pt

{\noindent \bf Proof of Lemma 2:} For the case with independent double exponential priors involving shrinkage parameter $\lambda$, note that $P\bigg(a^2(k,l)\le \frac{(x^*)^2}{\nOutcome^2}, \mbox{for all } 1\le k,l\le \nOutcome\bigg)\le P(||vec(A_k)||\le x^*)$. Further for large positive $x^*>\nOutcome$, and denoting $\pi(s\mid \lambda)=\frac{1}{2}\sqrt{\lambda}\exp(-\sqrt{\lambda}\frac{s}{2})$
\begin{eqnarray*}
&&P^*_1\bigg(a^2(k,l)> \frac{(x^*)^2}{\nOutcome^2}\bigg) = P\bigg(|a(k,l)| > \frac{(x^*)}{\nOutcome}\bigg) =2\int_{x^*/\nOutcome}^\infty \int \frac{1}{\sqrt{2\pi s}}\exp(-\frac{1}{2s} a^2) \pi(s\mid \lambda) \mbox{ ds da}\\
&& \le 2\int \frac{1}{\sqrt{2\pi}s(x^*/\nOutcome)}\exp(-\frac{1}{2s}(x^*/\nOutcome)^2)\pi(s\mid \lambda) ds \le 2\int \frac{1}{\sqrt{2\pi}s}\exp(-\frac{1}{2s}(x^*/d)^2)\pi(s\mid \lambda) ds\\
&& = 2\exp(-\lambda|x^*/\nOutcome|) = 2\exp(-\lambda x^*/\nOutcome) \le (x^*/\nOutcome)^{-\lambda}.
\end{eqnarray*}
The above implies that $1-(x^*/\nOutcome)^{-\lambda} \le P^*_1\bigg(a^2(k,l)\le  \frac{(x^*)^2}{\nOutcome^2}\bigg) $ and further $(1-(x^*/\nOutcome)^{-\lambda})^{\nOutcome^2}\le P^*_1\bigg(a^2(k,l)\le \frac{(x^*)^2}{\nOutcome^2}, \mbox{for all } 1\le k,l\le \nOutcome\bigg) \le P^*_1(||vec(A_k)||\le x^*)$. This implies that $P^*_1(||vec(A_k)||>x^*)> 1 - (1-(x^*/\nOutcome)^{-\lambda})^{\nOutcome^2}$. For large $x^*$ and choosing $\lambda$ large enough, one can use the binomial expansion to write (when $\nOutcome$ is even)
{\small
\begin{eqnarray*}
&& 1 - (1-(x^*/\nOutcome)^{-\lambda})^{\nOutcome^2} = 1 - \bigg\{(1 - \nOutcome^2 (x^*/\nOutcome)^{-\lambda}) + \frac{\nOutcome^2(\nOutcome^2-1)}{2}(x^*/\nOutcome)^{-2\lambda}\bigg(1-\frac{\nOutcome^2-2}{3}(x^*/\nOutcome)^{-\lambda}\bigg) +\\  
&&\frac{\nOutcome^2(\nOutcome^2-1)(\nOutcome^2-2)(\nOutcome^2-3)}{4!}(x^*/\nOutcome)^{-4\lambda}\bigg(1-\frac{\nOutcome^2-4}{5}(x^*/\nOutcome)^{-\lambda}\bigg) + \ldots +\\
&& \frac{\nOutcome^2(\nOutcome^2-1)\ldots\{\nOutcome^2-(\nOutcome^2-1)\}}{(\nOutcome^2)!}(x^*/\nOutcome)^{-(\nOutcome^2)\lambda}\bigg\} = \nOutcome^2 (x^*/\nOutcome)^{-\lambda} - \kappa^* \lesssim (x^*)^{-\lambda},
\end{eqnarray*} }
for large  $x^*$ and $\lambda$ such that $1-\nOutcome^2(x^*/\nOutcome)^{-\lambda}>0$, and for some positive constant $\kappa^*$. Similar calculations hold for odd $\nOutcome$. 

Further, when  $\small P^*_1(vec(A_k))= N_{\nOutcome^2}(vec(A_k);\bfmu,\Lambda)$, $\small \Lambda\sim IW(\Lambda_0,\nu_\lambda)$, the resulting distribution follows a multivariate t-distribution. Hence one can write $P^*_1(||vec(A_k)||>x^*) \le (x^*)^{(\nu_\lambda - \nOutcome^2 + 1)/2}$, using arguments similar to those in \cite{canale2017posterior}.

\section{Additional Lemmas}

{\noindent \bf Lemma 4:} {\it 
The distance between densities $f_{P_1}$ and $f_{P_2}$ under PDPM-VAR can be expressed as }
\begin{eqnarray*}
\small
&& ||f_{P_1} - f_{P_2} ||_1 \le 2\epsilon^2 + \sum_{h_1,\hsig<H_n}|\pi^{(1)}_{h_1}\pi^{(1)}_{\sigma,\hsig}-\pi^{(2)}_{h_1}\pi^{(2)}_{\sigma,\hsig}| + 4\epsilon \nonumber \\
&& + \sum_{h_1,\hsig\le H_n}\pi^{(1)}_{h_1}\pi^{(1)}_{\sigma,\hsig}\bigg\|\prod_{t=1}^T\phi_{\Sigma^{(1)}_{\hsig}}\big({\bf x}_t - \sum_{k=1}^K A^{(1)}_{k,h_1}{\bf x}_{t-k}\big) - \prod_{t=1}^T\phi_{\Sigma^{(2)}_{\hsig}}\big({\bf x}_t - \sum_{k=1}^K A^{(2)}_{k,h_1}{\bf x}_{t-k}\big) \bigg\|_1.
\end{eqnarray*}

{\noindent \bf Proof:}  $||f_{P_1} - f_{P_2} ||_1=$
\begin{eqnarray*}
\small
&&  ||\sum\limits_{h_1,\hsig\ge 1}\pi^{(1)}_{h_1}\pi^{(1)}_{\sigma,\hsig}\prod_{t=1}^T \phi_{\Sigma^{(1)}_{\hsig}}\big({\bf x}_t - \sum_{k=1}^K A^{(1)}_{k,h_1}{\bf x}_{t-k}\big) 
- 
\sum\limits_{h_1,\hsig\ge 1}\pi^{(2)}_{h_1}\pi^{(2)}_{\sigma,\hsig}\prod_{t=1}^T \phi_{\Sigma^{(2)}_{\hsig}}\big({\bf x}_t - \sum_{k=1}^K A^{(2)}_{k,h_1}{\bf x}_{t-k}\big)||_1 \nonumber \\
&=& \bigg\|\sum_{h_1,\hsig>H_n}\pi^{(1)}_{h_1}\pi^{(1)}_{\sigma,\hsig}\prod_{t=1}^T\phi_{\Sigma^{(1)}_{\hsig}}\big({\bf x}_t - \sum_{k=1}^K A^{(1)}_{k,h_1}{\bf x}_{t-k}\big) - \sum_{h_1,\hsig>H_n}\pi^{(2)}_{h_1}\pi^{(2)}_{\sigma,\hsig}\prod_{t=1}^T\phi_{\Sigma^{(2)}_{\hsig}}\big({\bf x}_t - \sum_{k=1}^K A^{(2)}_{k,h_1}{\bf x}_{t-k}\big)  \nonumber \\
&+& \sum_{h_1,\hsig\le H_n}\pi^{(1)}_{h_1}\pi^{(1)}_{\sigma,\hsig}\big\{\prod_{t=1}^T\phi_{\Sigma^{(1)}_{\hsig}}\big({\bf x}_t - \sum_{k=1}^K A^{(1)}_{k,h_1}{\bf x}_{t-k}\big) - \prod_{t=1}^T\phi_{\Sigma^{(2)}_{\hsig}}\big({\bf x}_t - \sum_{k=1}^K A^{(2)}_{k,h_1}{\bf x}_{t-k}\big) \big\}  \nonumber \\
&+&\sum_{h_1,\hsig<H_n}\big(\pi^{(1)}_{h_1}\pi^{(1)}_{\sigma,\hsig}-\pi^{(2)}_{h_1}\pi^{(2)}_{\sigma,\hsig}\big)\prod_{t=1}^T\phi_{\Sigma^{(2)}_{\hsig}}\big({\bf x}_t - \sum_{k=1}^K A^{(2)}_{k,h_1}{\bf x}_{t-k}\big) - \sum_{j'=1,2}\bigg\{\sum_{h_1\le H_n}\sum_{\hsig> H_n}\pi^{(j')}_{h_1}\pi^{(j')}_{\sigma,\hsig}\nonumber \\
&&\prod_{t=1}^T\phi_{\Sigma^{(j')}_{\hsig}}\big({\bf x}_t - \sum_{k=1}^K A^{(j')}_{k,h_1}{\bf x}_{t-k}\big)
+
\sum_{h_1>H_n}\sum_{\hsig\le H_n}\pi^{(j')}_{h_1}\pi^{(j')}_{\sigma,\hsig}\prod_{t=1}^T\phi_{\Sigma^{(j')}_{\hsig}}\big({\bf x}_t - \sum_{k=1}^K A^{(j')}_{k,h_1}{\bf x}_{t-k}\big)\bigg\}\bigg\|_1 . \nonumber 
\end{eqnarray*}

The upper bound for the right hand side of the above equation may be further written as
\begin{eqnarray*}
\small
&&\sum_{j'=1,2}\bigg\{\sum_{h_1,\hsig>H_n}\pi^{(j')}_{h_1}\pi^{(j')}_{\sigma,\hsig}\bigg\|\prod_{t=1}^T\phi_{\Sigma^{(1)}_{\hsig}}\big({\bf x}_t - \sum_{k=1}^K A^{(j')}_{k,h_1}{\bf x}_{t-k}\big)\bigg\|_1  \bigg\} \nonumber \\
&+& \sum_{h_1,\hsig\le H_n}\pi^{(1)}_{h_1}\pi^{(1)}_{\sigma,\hsig}\bigg\|\prod_{t=1}^T\phi_{\Sigma^{(1)}_{\hsig}}\big({\bf x}_t - \sum_{k=1}^K A^{(1)}_{k,h_1}{\bf x}_{t-k}\big) - \prod_{t=1}^T\phi_{\Sigma^{(2)}_{\hsig}}\big({\bf x}_t - \sum_{k=1}^K A^{(2)}_{k,h_1}{\bf x}_{t-k}\big) \bigg\|_1  \nonumber \\
&+&\sum_{h_1,\hsig<H_n}\big(\pi^{(1)}_{h_1}\pi^{(1)}_{\sigma,\hsig}-\pi^{(2)}_{h_1}\pi^{(2)}_{\sigma,\hsig}\big)\bigg\|\prod_{t=1}^T\phi_{\Sigma^{(2)}_{\hsig}}\big({\bf x}_t - \sum_{k=1}^K A^{(2)}_{k,h_1}{\bf x}_{t-k}\big)\bigg\|_1 + \sum_{j'=1,2}\bigg\{\sum_{h_1\le H_n}\sum_{\hsig> H_n}\pi^{(j')}_{h_1}\pi^{(j')}_{\sigma,\hsig}\nonumber \\
&&\bigg\|\prod_{t=1}^T\phi_{\Sigma^{(j')}_{\hsig}}\big({\bf x}_t - \sum_{k=1}^K A^{(j')}_{k,h_1}{\bf x}_{t-k}\big)\bigg\|_1
+
\sum_{h_1>H_n}\sum_{\hsig\le H_n}\pi^{(j')}_{h_1}\pi^{(j')}_{\sigma,\hsig}\bigg\|\prod_{t=1}^T\phi_{\Sigma^{(j')}_{\hsig}}\big({\bf x}_t - \sum_{k=1}^K A^{(j')}_{k,h_1}{\bf x}_{t-k}\big)\bigg\|_1 \bigg\} \nonumber 
\end{eqnarray*}

The right hand side of the above can be further bounded as 
\begin{eqnarray*}
 &&\sum_{h_1,\hsig\le H_n}\pi^{(1)}_{h_1}\pi^{(1)}_{\sigma,\hsig}\bigg\|\prod_{t=1}^T\phi_{\Sigma^{(1)}_{\hsig}}\big({\bf x}_t - \sum_{k=1}^K A^{(1)}_{k,h_1}{\bf x}_{t-k}\big) - \prod_{t=1}^T\phi_{\Sigma^{(2)}_{\hsig}}\big({\bf x}_t - \sum_{k=1}^K A^{(2)}_{k,h_1}{\bf x}_{t-k}\big) \bigg\|_1  \nonumber \\
 &+&\bigg|\sum_{h_1,\hsig>H_n} \pi^{(1)}_{h_1}\pi^{(1)}_{\sigma,\hsig} + \sum_{h_1,\hsig>H_n} \pi^{(2)}_{h_1}\pi^{(2)}_{\sigma,\hsig} \bigg| +
 \sum_{h_1,\hsig<H_n}\bigg|\pi^{(1)}_{h_1}\pi^{(1)}_{\sigma,\hsig}-\pi^{(2)}_{h_1}\pi^{(2)}_{\sigma,\hsig}\bigg| \nonumber \\
&+& \sum_{j'=1,2}\bigg\{\sum_{h_1\le H_n}\sum_{\hsig> H_n}\pi^{(j')}_{h_1}\pi^{(j')}_{\sigma,\hsig} +
\sum_{h_1>H_n}\sum_{\hsig\le H_n}\pi^{(j')}_{h_1}\pi^{(j')}_{\sigma,\hsig} \bigg\}\nonumber \\
&\le& \sum_{h_1,\hsig\le H_n}\pi^{(1)}_{h_1}\pi^{(1)}_{\sigma,\hsig}\bigg\|\prod_{t=1}^T\phi_{\Sigma^{(1)}_{\hsig}}\big({\bf x}_t - \sum_{k=1}^K A^{(1)}_{k,h_1}{\bf x}_{t-k}\big) - \prod_{t=1}^T\phi_{\Sigma^{(2)}_{\hsig}}\big({\bf x}_t - \sum_{k=1}^K A^{(2)}_{k,h_1}{\bf x}_{t-k}\big) \bigg\|_1 \nonumber \\
&+& 2\epsilon^2 + \sum_{h_1,\hsig<H_n}|\pi^{(1)}_{h_1}\pi^{(1)}_{\sigma,\hsig}-\pi^{(2)}_{h_1}\pi^{(2)}_{\sigma,\hsig}| + 4\epsilon,
\end{eqnarray*}
using the fact that $\sum_{h_1\le H_n}\sum_{\hsig> H_n}\pi^{(j')}_{h_1}\pi^{(j')}_{\sigma,\hsig} = (\sum_{h_1\le H_n} \pi^{(j')}_{h_1} )(\sum_{\hsig> H_n} \pi^{(j')}_{\sigma,\hsig})\le \epsilon$, since $\sum_{h_1>H_n}\pi_{h_1}<\epsilon,\sum_{\hsig>H_n}\pi_{\sigma,\hsig}<\epsilon$. 

\vskip 12pt

{\noindent \bf Lemma 5:} {\it The upper bound for $\sum_{h_1,\hsig<H_n}|\pi^{(1)}_{h_1}\pi^{(1)}_{\sigma,\hsig}-\pi^{(2)}_{h_1}\pi^{(2)}_{\sigma,\hsig}|$ is given by $ \sum_{h_1,\hsig<H_n}\bigg|\tilde{\pi}^{(1)}_{h_1}\tilde{\pi}^{(1)}_{\sigma,\hsig}- \pi^{(2)}_{h_1}\pi^{(2)}_{\sigma,\hsig}\bigg| + \bigg|1-(1-\epsilon)^2 \bigg|, \mbox{ where } \tilde{\pi}=\frac{\pi_{h}}{(1-\sum_{h>H}\pi_{h})}.$}

{\noindent \bf Proof:} Note that $\sum_{h_1,\hsig<H_n}|\pi^{(1)}_{h_1}\pi^{(1)}_{\sigma,\hsig}-\pi^{(2)}_{h_1}\pi^{(2)}_{\sigma,\hsig}|$ 
\begin{eqnarray}
&&\sum_{h_1,\hsig<H_n}|\pi^{(1)}_{h_1}\pi^{(1)}_{\sigma,\hsig}-\pi^{(2)}_{h_1}\pi^{(2)}_{\sigma,\hsig}| \le  \sum_{h_1,\hsig<H_n}\bigg|\pi^{(1)}_{h_1}\pi^{(1)}_{\sigma,\hsig}-\bigg(1-\sum_{h_1>H}\pi^{(1)}_{h_1}\bigg)\bigg(1-\sum_{\hsig>H}\pi^{(2)}_{\sigma,\hsig}\bigg)\pi^{(2)}_{h_1}\pi^{(2)}_{\sigma,\hsig}\bigg| \nonumber \\
&+& \sum_{h_1,\hsig<H_n}\bigg| \bigg(1-\sum_{h_1>H}\pi^{(1)}_{h_1}\bigg)\bigg(1-\sum_{\hsig>H}\pi^{(2)}_{\sigma,\hsig}\bigg)\pi^{(2)}_{h_1}\pi^{(2)}_{\sigma,\hsig} - \pi^{(2)}_{h_1}\pi^{(2)}_{\sigma,\hsig} \bigg| \nonumber\\
&=& \bigg(1-\sum_{h_1>H}\pi^{(1)}_{h_1}\bigg)\bigg(1-\sum_{\hsig>H}\pi^{(2)}_{\sigma,\hsig}\bigg) \sum_{h_1,\hsig<H_n}\bigg|\tilde{\pi}^{(1)}_{h_1}\tilde{\pi}^{(1)}_{\hsig}- \pi^{(2)}_{h_1}\pi^{(2)}_{\sigma,\hsig}\bigg| \nonumber\\
&+& \bigg| \bigg(1-\sum_{h_1>H}\pi^{(1)}_{h_1}\bigg)\bigg(1-\sum_{\hsig>H}\pi^{(2)}_{\sigma,\hsig}\bigg) - 1 \bigg|\bigg(\sum_{h_1,\hsig<H_n}\pi^{(2)}_{h_1}\pi^{(2)}_{\sigma,\hsig}\bigg)\nonumber\\
&\le& \sum_{h_1,\hsig<H_n}\bigg|\tilde{\pi}^{(1)}_{h_1}\tilde{\pi}^{(1)}_{\sigma,\hsig}- \pi^{(2)}_{h_1}\pi^{(2)}_{\sigma,\hsig}\bigg| + \bigg|1-(1-\epsilon)^2 \bigg|, \mbox{ where } \tilde{\pi}=\frac{\pi_{h}}{(1-\sum_{h>H}\pi_{h})}. \label{eq:stickprob}
\end{eqnarray}

{\noindent \bf Lemma 6:} {\it For PDPM-VAR the prior tail condition (2A) in Theorem 2 is satisfied.}

{\noindent \bf Proof:}
For PDPM-VAR, note that the prior probability on the sieve $\mathcal{F}_n$ defined in (\ref{eq:sieves-PDPM1.1}) satisfies
\begin{eqnarray*}
\Pi(\mathcal{F}_n^c)&\le& Pr\bigg( \sum_{\hsig>H_n}\pi_{\sigma,\hsig}>\epsilon\bigg) + Pr\bigg( \sum_{h_1>H_n}\pi_{h_1}>\epsilon\bigg) + H_nP^*_2\bigg( \lambda_\nOutcome(\Sigma) \le \underline{\sigma}^2_n\bigg) + \\
&& H_n P^*_2\bigg(\lambda_1(\Sigma)>\underline{\sigma}_n^2(1+ \epsilon/\sqrt{\nOutcome})^{M_n} \bigg) + H_n P_2^*\bigg( \frac{\lambda_1(\Sigma_{\hsig})}{\lambda_\nOutcome(\Sigma_{\hsig})}> n^{H_n}\bigg),
\end{eqnarray*}
using the fact that $\small P[(A\cap B \cap C)^c]=P[A^c\cup B^c \cup C^c]\le P(A^c) + P(B^c) + P(C^c)$. Using the stick-breaking representation for DP for the first term, and the tail conditions for the priors, and following similar steps as in the proof of Proposition 2 in \cite{shen2013adaptive}, one has $\Pi(\mathcal{F}_n^c)$
\begin{eqnarray}
 &\lesssim& \sum_{m=1,2}\big\{\frac{e\alpha_m}{H_n}\log(1/\epsilon) \big\}^{H_n} + H_n  \big\{ e^{-c_1\underline{\sigma}_n^{-2c_2}} + \underline{\sigma}_n^{-2c_3}\big( 1+\epsilon/\sqrt{\nOutcome}\big)^{-c_3 M_n} + (n^{\frac{1}{\log n}})^{-\kappa Cn\epsilon^2}\big\} \nonumber \\
&\lesssim& 2\big( Cn\epsilon^2/\log(n)\big)^{-Cn\epsilon^2/\log(n)} + \big( Cn\epsilon^2/\log(n)\big)\bigg( e^{-c_1n} + n^{c_3/c_2}(1+\epsilon/\sqrt{\nOutcome})^{-c_3n} +  e^{-\kappa Cn\epsilon^2}\bigg) \nonumber \\
&\lesssim& e^{-bn}, \mbox{ } \quad \label{eq:prior-prob1} 
\end{eqnarray}
 since $n^{\frac{1}{\log n}}=e$ and due to the fact that $\big( Cn\epsilon^2/\log(n)\big)\log\{-Cn\epsilon^2/\log(n)\}>Cn\epsilon^2 $ for large $n$, where $0<b<\min\{C\epsilon^2/2,c_1,c_3\log(1+\epsilon/\sqrt{\nOutcome}),\kappa C\epsilon^2 \}$. Hence the first condition $(2A)$ in Theorem 2 is satisfied.

\vskip 12pt

{\noindent \bf Lemma 7:} For PDPM-VAR, we have $\sum_{j_{h_1}\in N}\sum_{1\le l_{\hsig\le H_n}}\sqrt{N(\epsilon,\mathcal{F}_{n,j_{h_1}l_{\hsig}})\Pi(\mathcal{F}_{n,j_{h_1}l_{\hsig}})} e^{-(4-c)n\epsilon^2}\to 0$ as $n\to\infty$ for a suitable choice of constants.

{\noindent \bf Proof: }
Hence we can write $\sqrt{N(\epsilon,\mathcal{F}_{n,{\bf jl}},|| \cdot||_1)\Pi(\mathcal{F}_{n,{\bf jl}})}$
\begin{eqnarray*}
&\lesssim&  \sqrt{\mathcal{C}^{H_n}_1\big(\frac{1}{\epsilon}\big)^{C_1H_n}\times \bigg\{\mathcal{C}\bigg(1 + o_n(1)\bigg)^{K\nOutcome^2}\bigg\}^{H_n}\times \big(\frac{1}{\epsilon}\big)^{H_n K\nOutcome^2+\nOutcome(\nOutcome-1)}}\nonumber \\
&\times& \exp\bigg\{\frac{1}{2}\big(\nOutcome H_n\log(M) + H_n\log \frac{C_1}{\epsilon}) \bigg\}\times n^{H^3_n{(K\nOutcome^2(T-2) + \frac{1}{2}K\nOutcome^2) + \frac{H_n}{2c_2}K\nOutcome^2(T-1)}}  \nonumber \\
&\times&  \bigg\{n^{-H^3_n(r+1)K} \prod_{h_1\le H_n} (j_{h_1}-1)\big)^{-1_{(j_{h_1}\ge 2)}K(r+1)} \bigg\} \times \bigg\{ \prod_{h_1\le H_n}(j_{h1})^{2K\nOutcome^2(T-2) + K(\nOutcome^2-1)}\bigg\} \nonumber \\
&\times& \bigg\{ \prod_{\hsig\le H_n} (n^{l_{\hsig}})^{\frac{\nOutcome(\nOutcome-1)}{2} + K\nOutcome^2H_n(T-1)\frac{\nOutcome-1}{2}}\bigg\}
\times \bigg\{\prod_{\hsig\le H_n} (n^{\frac{\kappa}{2}(l_{\hsig}-1)})^{-1_{(l_{\hsig}\ge 1)}} \bigg\}.
\end{eqnarray*}

The above can be simplified further as
\begin{eqnarray}
 &&\sqrt{\mathcal{C}^{H_n}_1\times \bigg\{\mathcal{C}\bigg(1 + o_n(1)\bigg)^{K\nOutcome^2}\bigg\}^{H_n}}\exp\bigg\{H_n\log( \frac{C_1}{\epsilon})+\frac{H_n}{2}(K\nOutcome^2+\nOutcome(\nOutcome-1)\log(\frac{1}{\epsilon}\bigg\} \nonumber \\
 &\times& \exp\bigg\{\frac{1}{2}\big(\nOutcome H_n\log(M)  + \frac{H_n}{2c_2}K\nOutcome^2(T-1)\log(n) ) \bigg\}\times n^{KH^3_n\big(\nOutcome^2(T-2) + \frac{1}{2}\nOutcome^2- r\big)-KH^3_n}\nonumber \\
 &\times& \bigg\{ \prod_{h_1\le H_n}(j_{h1})^{K\nOutcome^2(T-2) + \frac{K}{2}(\nOutcome^2-1)}\bigg\} \bigg\{\prod_{h_1\le H_n} (j_{h_1}-1)\big)^{-1_{(j_{h_1}\ge 2)}K(r+1)} \bigg\} \nonumber \\
 &\times& \prod_{\hsig\le H_n}
\bigg\{  n^{l_{\hsig}}\bigg\}^{\frac{\nOutcome(\nOutcome-1)}{4} + K\nOutcome^2H_n(T-1)\frac{\nOutcome-1}{4}} \times 
\bigg\{\prod_{\hsig\le H_n} (n^{\kappa(l_{\hsig}-1)/2})^{-1_{(l_{\hsig}\ge 1)}} \bigg\},\label{eq:sumbound}
\end{eqnarray}
Note that $n^{KH^3_n\big(\nOutcome^2(T-2) + \frac{1}{2}\nOutcome^2- r\big)}$ is bounded when $r>\nOutcome^2(T-2) + \frac{1}{2}\nOutcome^2$. Further, looking at the terms involving $j$ in the last line of (\ref{eq:sumbound}), one can sum over $j$ (for a fixed $h_1$) to have
\begin{eqnarray*}
\sum_{j_{h_1\ge 2}} \big\{(j_{h1})^{K\nOutcome^2(T-2) + \frac{K}{2}(\nOutcome^2-1)}  (j_{h_1}-1)\big)^{-1_{(j_{h_1}\ge 2)}K(r+1)} \big\}
\approx \sum_{j_{h_1\ge 2}}(j_{h1})^{K\nOutcome^2(T-2) + \frac{K}{2}(\nOutcome^2-1)-K(r+1)} = \bigg( 1+\mathcal{B}\bigg),
\end{eqnarray*}
where $\mathcal{B}$ is a suitable finite constant that does not depend on $n$ when $r$ is large enough such that $r>\nOutcome^2(T-2) + \frac{1}{2}\nOutcome^2$, and where the approximation holds for $n$ large enough.

Similarly, looking at the terms involving $l$, one can sum over $l$ (for a fixed $\hsig$) to have
$\small \sum_{l_{\hsig\ge 1}}
\bigg\{(n^{l_{\hsig}})^{\frac{\nOutcome(\nOutcome-1)}{4} -(\kappa/2)} \bigg\}$ $\times  \bigg\{(n^{l_{\hsig}})^{K\nOutcome^2H_n(T-1)\frac{\nOutcome-1}{4}}\bigg\} \\
\le \sqrt{\sum_{l_{\hsig\ge 1}}
\bigg\{(n^{l_{\hsig}})^{\frac{\nOutcome(\nOutcome-1)}{2} -(\kappa/2)} \bigg\}}$ $\times \sqrt{\sum_{l_{\hsig\ge 1}}\bigg\{(n^{l_{\hsig}})^{K\nOutcome^2H_n(T-1)\frac{\nOutcome-1}{2}}\bigg\} }$
%&= \big( 1+\mathcal{B}_1\big)^{1/2}\times\sqrt{\sum_{l_{\hsig\ge 2}}\bigg\{(l_{\hsig})^{Kd^2H_n(T-1)\frac{d-1}{2}}\bigg\} },$
where the inequality is using Cauchy-Schwartz, and the first term in the upper bound (denoted by $\mathcal{B}_1$ ) is finite when $\kappa>\nOutcome(\nOutcome-1)/2$.

 Hence the upper bound on the combined terms in the last two lines in (\ref{eq:sumbound}) is given by 
\begin{eqnarray}
&&\big( 1+\mathcal{B}\big)^{H_n}\big( 1+\mathcal{B}_1\big)^{H_n/2}n^{KH^3_n\big(\nOutcome^2(T-2) + \frac{1}{2}\nOutcome^2- r\big)-KH^3_n}\times\prod_{\hsig\le H_n} \sqrt{\sum_{l_{2\le \hsig\le H_n}}\bigg\{(n^{l_{\hsig}})^{K\nOutcome^2H_n(T-1)\frac{\nOutcome-1}{2}}\bigg\} }\nonumber \\
&=& \big( 1+\mathcal{B}\big)^{H_n}\big( 1+\mathcal{B}_1\big)^{H_n/2}n^{KH^3_n\big(\nOutcome^2(T-2) + \frac{1}{2}\nOutcome^2- r\big)}\times\prod_{\hsig\le H_n}\sqrt{\sum_{l_{2\le \hsig \le H_n }}\bigg\{\frac{(n^{l_{\hsig}})^{K\nOutcome^2H_n(T-1)\frac{\nOutcome-1}{2}}}{n^{2KH^2_n}}\bigg\}}. \label{eq:sumbound1}
\end{eqnarray}

The sum within the square root is given as 
\begin{eqnarray*}
\sum_{l_{2\le \hsig \le H_n }}\bigg\{\frac{(n^{l_{\hsig}})^{K\nOutcome^2H_n(T-1)\frac{\nOutcome-1}{2}}}{n^{2KH^2_n}}\bigg\}= \sum_{\hsig\le H_n} \big(n ^{K\nOutcome^2H_n(T-1)\frac{\nOutcome-1}{2} -2KH^2_n} \big)^{l_{\hsig}} = \bigg(\frac{1-(r^*)^{H_n}}{1-r^*} \bigg)<1,
\end{eqnarray*}
where the equality is obtained by summing $H_n$ terms in geometric progression, and $r^*=n ^{K\nOutcome^2H_n(T-1)\frac{\nOutcome-1}{2} -2KH^2_n}<1$ since $K\nOutcome^2H_n(T-1)\frac{\nOutcome-1}{2} -2KH^2_n<0$ when $n$ is large enough, and recalling that $1\le l_{\hsig}=\hsig\le H_n$. Hence using (\ref{eq:sumbound1}), $\sum_{j_{h_1}\in N}\sum_{1\le l_{\hsig\le H_n}}\sqrt{N(\epsilon,\mathcal{F}_{n,j_{h_1}l_{\hsig}})\Pi(\mathcal{F}_{n,j_{h_1}l_{\hsig}})}$
\begin{eqnarray}
&&\lesssim  \bigg\{\mathcal{C}\bigg(1 + o_n(1)\bigg)^{K\nOutcome^2}\bigg\}^{H_n/2}\bigg( 1+\max\{\mathcal{B}, \mathcal{B}_1\}\bigg)^{H_n}n^{KH^3_n\big(\nOutcome^2(T-2) + \frac{1}{2}\nOutcome^2- r\big)} \bigg(\frac{1-(r^*)^{H_n}}{1-r^*}\bigg)^{H_n} \nonumber\\
&&\times \exp\bigg\{H_n\log( \frac{C_1}{\epsilon})+\frac{H_n}{2}(K\nOutcome^2+\nOutcome(\nOutcome-1)\log(\frac{1}{\epsilon})\bigg\} \times \exp\bigg\{\frac{1}{2}\big(\nOutcome  + \frac{1}{2c_2}K\nOutcome^2(T-1)\big)Cn\epsilon^2 ) \bigg\} \nonumber\\
&&\lesssim \bigg(\frac{\mathcal{K}^*}{n^{KH^2_n\big(r-\nOutcome^2(T-2) + \frac{1}{2}\nOutcome^2\big)}}\bigg)^{H_n} \times \exp\bigg\{\frac{1}{2}\big(\nOutcome  + \frac{1}{2c_2}K\nOutcome^2(T-1)\big)Cn\epsilon^2 ) \bigg\}, \quad \label{eq:final-bound}
% &&\times  \exp\bigg\{\frac{1}{2}\big(dH_n\log(M) + H_n\log \frac{C_1}{\epsilon}) + \frac{H_n}{2c_2}Kd^2(T-1)\log(n) +\frac{H_n}{2}(Kd^2+d(d-1)\log(\frac{1}{\epsilon}) \bigg\}\nonumber
\end{eqnarray}
where $r^*<1$, $r-\nOutcome^2(T-2) + \frac{1}{2}\nOutcome^2>0$ $H_n\log(M_n)=Cn\epsilon^2$, and $\mathcal{K}^*>0$ is some finite constant that is a function of $K,\nOutcome,\epsilon$, and other constants but does not depend on $n$.

Therefore, $\sum_{j_{h_1}\in N}\sum_{1\le l_{\hsig\le H_n}}\sqrt{N(\epsilon,\mathcal{F}_{n,j_{h_1}l_{\hsig}})\Pi(\mathcal{F}_{n,j_{h_1}l_{\hsig}})} e^{-(4-c)n\epsilon^2}\to 0$ as $n\to\infty$ for a suitable choice of $C$ such that $\frac{1}{2}\big(\nOutcome  + \frac{1}{2c_2}K\nOutcome^2(T-1)\big)C<1$. Hence the condition $(2B)$ in Theorem 2 is satisfied and the strong posterior consistency result is proved corresponding to the PDPM-VAR model.

\section{Posterior Computation Steps}

\subsection{Residual Covariance Updates:} %Following \cite{ghosh2009default}, we use a parameter expanded approach to facilitate sampling parameters $(\FactorLoadings_i,\LowRankDiagCov_i)$ corresponding to $\Sigma_i $. Denoting $\RedundantTerm_i = \text{diag} \{   \redundantTerm_{i,1}, \ldots,  \redundantTerm_{i,\nLatentFactor} \} $, we use the following working model:
%\begin{align*}
%	\mathbf{x}_{i,t} &= \sum_{k=1}^{\min\{t-1, K\}} A_{ik} \mathbf{x}_{i,t-k} + \FactorLoadings_i^* \SubjectLatentFactor_{i,t}^* + \boldsymbol{\epsilon}^*_{i,t}, \;
%	\SubjectLatentFactor_{i,t}^* \sim N(0, \RedundantTerm_i), \;
%	\boldsymbol{\epsilon}^*_{i,t} \sim N(0, \LowRankDiagCov_i),
%\end{align*}
%where $\FactorLoadings_i^*$ is a lower triangular matrix without constraints. The residual covariance $\Sigma_i = \FactorLoadings_i  \FactorLoadings_i'  + \LowRankDiagCov_i$ can be directly recovered from this working model as described in \cite{ghosh2009default} by transforming: $\FactorLoadings_{i, d',b} = \text{sign}(\FactorLoadings_{i,b,b}^*) \FactorLoadings_{i,d',b}^* \redundantTerm_{i,b}^{1/2}$ and $\subjectLatentFactor_{i, t, b} = \text{sign}(\FactorLoadings_{i,b,b}^*)  \redundantTerm_{i,b}^{-1/2} \subjectLatentFactor_{i, t, b}$, where $\FactorLoadings_{i, d',b}$ is the $d'$th row, $b$th element of $\FactorLoadings_{i}$ and $\subjectLatentFactor_{i, t, b}$ is the $b$th element of $\SubjectLatentFactor_{i, t}$.

Under the low rank representation, we impose DP mixture priors on $(\FactorLoadings_i^*$, $\RedundantTerm_i$, $\LowRankDiagCov_i$) leading to a mixture prior on $\Sigma_i$. This corresponds to the prior $\Sigma_i \sim \sum_{h_\sigma=1}^{\infty} \pi_{\sigma,h_\sigma}\delta_{(\FactorLoadings^*_{h_\sigma},\RedundantTerm_{h_\sigma},\LowRankDiagCov_{h_\sigma})}$, where $(\FactorLoadings^*_{h_\sigma},\RedundantTerm_{h_\sigma},\LowRankDiagCov_{h_\sigma})\sim P^*_2\equiv P_{\FactorLoadings^*} \times P_{\RedundantTerm} \times P_{\LowRankDiagCov}$. Here $P_{\FactorLoadings^*}$ is a product of independent standard normal distributions, $P_{\RedundantTerm}$ is a product of independent $Gamma(1/2, 1/2)$ distributions yielding a half-Cauchy prior on the diagonal elements of $\FactorLoadings$ and a Cauchy prior on the lower-off-diagonal elements of  $\FactorLoadings$ as in \cite{ghosh2009default}, and the inverse of the diagonal elements of $\LowRankDiagCov$ have independent $Gamma(\alpha_\sigma, \beta_\sigma)$ priors. Note that here $\FactorLoadings_i^*$ is not a square matrix, and by diagonal elements we refer to elements $\Gamma_{i, 1, 1}, \ldots, \Gamma_{i, B, B}$. %This combination of priors induces a half Cauchy prior for the diagonal elements of $\FactorLoadings$ and Cauchy priors for its lower-triangular elements - {\it \color{red} requires more explanation}.

Under the stick-breaking representation of  \cite{sethuraman1994constructive}, we can write $\pi_{\sigma,h_\sigma} =\nu_{\sigma,h_\sigma} \prod_{l_\sigma = 1}^{h_\sigma-1} (1 - \nu_{\sigma,l_\sigma})$, $\nu_{l_\sigma} \sim Beta(1, \alpha_2)$.  We use the slice sampling approach of \cite{walker2007sampling} to facilitate sampling. This approach introduces a cluster membership indicator, $V$, with $V_i = h_\sigma$ when subject $i$ belongs to cluster $h_\sigma$, and let $\mathcal{V}_{h_\sigma} = \{i: V_i = h_\sigma \}$ be the indices of all subjects belonging to covariance cluster $h_\sigma$ and let $n_{\sigma,h_\sigma}$ be the cardinality of this set. Let $g_i$ be a  uniformly distributed latent variable used to reduce the stick-breaking representation of the DPM to a finite sum. Our sampler updates $\nu_{\sigma,h_\sigma}$, and $g_i$ as:
$\small 
\nu_{\sigma,h_\sigma} | \{V_1,\ldots,V_N\}  \sim Beta\left(1 + n_{\sigma,h_\sigma}, \alpha_2 + \sum_{i=1}^N I_{(V_i > \nu_{\sigma,h_\sigma})} \right), \mbox{ }
g_i | V_i \sim U(0,  \pi_{\sigma, V_i }).$
%p(V_i = h_\sigma | -) &\sim\frac{ I_{g_i < \pi_{\sigma,h_\sigma} } \prod_{t=1}^{T_i} \phi\left(x_{i,t}; \boldsymbol{\mu}_{i,t}, \Sigma_{h_\sigma} \right) }{ \sum_{h_\sigma'=1}^{h_\sigma^*} \left\{ I_{g_i < \pi_{\sigma,h_\sigma'}} \prod_{t=1}^{T_i} \phi\left(x_{i,t};  \boldsymbol{\mu}_{i,t}, \Sigma_{h_\sigma'} \right)  \right\} },  \nonumber

The cluster membership indicators $V_i$ are then sampled from a multinomial distribution with posterior probabilities $\big(p(V_i = 1 | -),\ldots,p(V_i = h^*_\sigma | -)\big)$ expressed as, 
    %p(V_i = h_\sigma | -) \sim\frac{ I_{g_i < \pi_{\sigma,h_\sigma} } \prod_{t=1}^{T_i} \phi\left(x_{i,t}; A_{i1}, \ldots, A_{iK}, V_i = h_\sigma \right) }{ \sum_{h_\sigma'=1}^{h_\sigma^*} \left\{ I_{g_i < \pi_{\sigma,h_\sigma'}} \prod_{t=1}^{T_i} \phi\left(x_{i,t}; A_{i1}, \ldots, A_{iK}, V_i = h_\sigma' \right)  \right\} },  
 $ \small p(V_i = h_\sigma | -) \sim\frac{ I_{(g_i < \pi_{\sigma,h_\sigma} )} \prod_{t=1}^{T_i} \phi_{\Sigma_{h_\sigma}}\left(x_{i,t}; A_{i1}, \ldots, A_{iK} \right) }{ \sum_{h_\sigma'=1}^{h_\sigma^*} \left\{ I_{(g_i < \pi_{\sigma,h_\sigma'})} \prod_{t=1}^{T_i} \phi_{\Sigma_{h_\sigma'}}\left(x_{i,t}; A_{i1}, \ldots, A_{iK} \right)  \right\} }, $
where $h_\sigma^* = \min \{ h_\sigma: g_i > 1 - \sum_{h_\sigma'=1}^{h_\sigma} \pi_{\sigma,h_\sigma}, \text{for all } i \}$. %and $\phi\left(\cdot; A_{i1}, \ldots, A_{iK}, V_i = h_\sigma \right)$ denotes the Gaussian density function with mean and variance determined by $A_{i1}, \ldots, A_{iK}$ and membership in covariance cluster $h_\sigma$. 
Conditioned on the cluster memberships, it is straightforward to update the variables in the low rank representation of $\Sigma$ using similar steps as in \cite{ghosh2009default}.  We start by sampling the elements of $\FactorLoadings_{h_{\sigma}^*}$ one row at a time from their full conditionals: 
$\small 
\FactorLoadings_{h_\sigma, d'}^* | - \sim N\left( \mu_{\FactorLoadings_{h_\sigma, d'}^*}, \Sigma_{\FactorLoadings_{h_\sigma, d'}^*} \right), \mbox{ where }
\Sigma_{\FactorLoadings_{h_\sigma, d'}^*} = \left(  \sigma_{h_\sigma, d'}^{-2} \sum_{i \in \mathcal{V}_{h_\sigma}} \sum_{t = 1}^{T_i} (\mathcal{E}^*_{\eta,itd'})' (\mathcal{E}^*_{\eta,itd'}) + I_{\min\{d', B\}}    \right)^{-1},  \nonumber \\
  \mu_{\FactorLoadings_{h_\sigma, d'}^*} = \Sigma_{\FactorLoadings_{h_\sigma, d'}^*} \left(  \sigma_{h_\sigma, d'}^{-2} \sum_{i \in \mathcal{V}_{h_\sigma}} \sum_{t = 1}^{T_i}  \mathcal{E}^*_{\eta,itd'} \left[ x_{i,t}^{(d')} - A_{i, d', \bullet} \mathbf{z}_{i,t} \right] \right), \mbox{ } \mathcal{E}^*_{\eta,itd'}=\left(\subjectLatentFactor_{i, t, 1}^*, \ldots, \subjectLatentFactor_{i, t,  \min\{d', B\} }^*\right)',
$
 $x_{i,t}^{(d)}$ is the response for the $i$th subject at the $d'$th node and $t$th time point, and $A_{ik,d' \bullet}'$ is the transpose of the $d'$th row of $A_{ik}$, $A_{i,  d' \bullet}$ is the $DK \times 1$ vector formed by stacking the $A_{ik,d' \bullet}'$ across all lags, and $\mathbf{z}_{i,t} = [\mathbf{x}_{i,t-1}', \ldots, \mathbf{x}_{i,t-K}']'$ is the $DK \times 1$ vector of previous outcomes used to predict the outcome at time $t$, padded with zeros for the case that $t-k < 1$.

{\noindent The} conditionals for the remaining terms in the lower rank representation for $\Sigma_i$ are:
$\small 
	\subjectLatentFactor_{i,t}^* |-  \sim N\bigg{(}  \mu_{\subjectLatentFactor_{i,t}^*} $,  $\Sigma_{\subjectLatentFactor_{i,t}^*}  \bigg{)} , \mbox{ }
	 \redundantTerm_{h_\sigma,b} | - \sim Gamma\left(   \frac{1 + N_{h_\sigma} }{2} , \frac{1}{2} \left[1 + \sum_{i \in \mathcal{V}_{h_\sigma}} \sum_{t = 1}^{T_i} \subjectLatentFactor_{i,t,b}^{*^2}  \right] \right), \nonumber \\
	 \sigma_{h_\sigma, d'}^{-2} | - \sim Gamma\left(  a_\sigma + \frac{N_{h_\sigma} }{2} , b_\sigma + \frac{1}{2} \sum_{i \in \mathcal{V}_{h_\sigma}} \sum_{t = 1}^{T_i}  \left[ x_{i,t}^{(d')} -     A_{i, d', \bullet} \mathbf{z}_{i,t}  - \FactorLoadings_{i,d'}^* \SubjectLatentFactor_{i,t}^* \right]^2    \right)
\small
 $ where $N_{h_\sigma}$ is equal to the total number of time points across all subjects in covariance cluster $h_\sigma$, and $ \Sigma_{\subjectLatentFactor_{i,t}^*} = \left( \RedundantTerm_{V_i}^{-1} + \FactorLoadings_{V_i}' \LowRankDiagCov_{V_i} \FactorLoadings_{V_i}  \right)^{-1}$ and $
	 \mu_{\subjectLatentFactor_{i,t}^*} = \Sigma_{\subjectLatentFactor_{i,t}^*} \FactorLoadings_{V_i}^{*'} \LowRankDiagCov_{V_i} \left[ x_{i,t}^{(d')} - A_{i, d', \bullet} \mathbf{z}_{i,t} \right]$.

\subsection{Autocovariance Parameter Updates}

%We outline the steps for the PDPM-VAR model in detail below, and then discuss the modifications required for the lgPDPM-VAR and rgPDPM-VAR.

 %We outline the posterior computation steps for the autocovariance parameters using the base measure that corresponds to independent double exponential prior with shrinkage parameter $\lambda$ on the elements of the autocovariance matrices. 

\subsubsection{Computation Steps for PDPM-VAR}

As with the covariance terms, we use the stick-breaking representation \citep{sethuraman1994constructive} of the Dirichlet process to enable posterior computation under the DPM priors. For the autocovariance terms, we can express the prior as, $
A_{i} | P_{\Theta} \sim P_{\Theta} , \;\; P_{\Theta}  = \sum_{h_1=1}^{\infty} \piLagPDPM \delta_{ A_{h} } $,
where $\piLagPDPM = \nuLagPDPM \prod_{l_1 < \nuLagPDPM} (1 - \nuAltLagPDPM )$, $\nuLagPDPM \sim Beta(1, \concParLagPDPM )$, and $A_h \sim P_{1}^*$, where $P_{1}^*$ is a multivariate normal distribution with mean $\mathbf{0}$ and variance $\text{diag}\{ \boldsymbol{\tau}^2 \}$.  The prior for the individual $\tau^2$ terms is given by $p(\tau_{k,d'}^2) = \frac{\lambda^2}{2} \exp \{ -\frac{1}{2}\lambda^2 \tau_{k,d'}^2 \}$, which implies a double exponential base measure for modeling the autocovariance terms \citep{park2008bayesian}. Furthermore, we place a conjugate $Gamma(r, \delta)$ hyperprior on $\lambda^2$ to facilitate Gibbs sampling. Throughout our applications we fix $r = 1.0$ and $\delta = 2.0$, which yield good performance in a wide range of settings.  As with the residual covariance, we use the slice sampling approach of  \cite{walker2007sampling} to facilitate sampling from this infinite mixture. Let $\clustMembLagPDPM$ be a cluster membership indicator, where $\clustMembSubjLagPDPM = \clustIndLagPDPM$ if subject $i$ belongs to the $\clustIndLagPDPM$th autocovariance matrix cluster. Let $\SetclustMembLagPDPM = \{i : \clustMembSubjLagPDPM = \clustIndLagPDPM \}$ be the indices of all subjects belonging to autocovariance cluster $\clustIndLagPDPM$, with $n_{\clustIndLagPDPM}$ being the cardinality of this set. The sampler proceeds by introducing a latent uniform variable $u_i$, relating the cluster memberships to the stick breaking representation of the DPM. The sampler proceeds by iteratively sampling $\nuLagPDPM$ and $\uiLagPDPM$ from their full conditionals,
$\small 
	\nuLagPDPM | \{\clustMembSubjLagPDPM\} \sim Beta\left(1 + n_{\clustIndLagPDPM},  \concParLagPDPM + \sum_{i=1}^N I_{(\clustMembSubjLagPDPM > \clustIndLagPDPM)} \right), \; \uiLagPDPM | \nuLagPDPM, \clustMembSubjLagPDPM \sim U(0, \piSubjClustLagPDPM). $
The cluster memberships, $\clustMembSubjLagPDPM$, are then sampled from a multinomial distribution with posterior probabilities $	\left( P(\clustMembSubjLagPDPM = 1 | -), \ldots, 	P(\clustMembSubjLagPDPM = \clustIndLagPDPM^* | -) \right)$ given by:\\
$\small 
	P(\clustMembSubjLagPDPM = h_1 | -) = \frac{ I_{(\uiLagPDPM < \piLagPDPM)} \prod_{t=1}^{T_i} \phi_{\Sigma_{V_i}}\left(\mathbf{x}_{i, t}-  \sum_{k=1}^K A_{k, \clustIndLagPDPM} \mathbf{x}_{i, t-k} \right) }{ \sum_{\clustIndLagPDPM'=1}^{\clustIndLagPDPM^*} \left\{ I_{(\uiLagPDPM < \pi_{\clustIndLagPDPM'})} \prod_{t=1}^{T_i} \phi_{\Sigma_{V_i}}\left(\mathbf{x}_{i, t} - \sum_{k=1}^K A_{k, \clustIndLagPDPM'} \mathbf{x}_{i, t-k} \right)  \right\} }
	%P( \clustMembSubjLagPDPM = h_1 \mid -) &\sim \frac{ I_{\uiLagPDPM < \piLagPDPM,} \prod_{t=1}^{T_i} \phi\left(\mathbf{x}_{i, t};  \clustMembSubjLagPDPM \right) = \clustIndLagPDPM, V_i \right) }{ \sum_{\clustIndLagPDPM'=1}^{\clustIndLagPDPM^*} \left\{ I_{\uiLagPDPM < \pi_{\clustIndLagPDPM'},} \prod_{t=1}^{T_i} \phi\left(\mathbf{x}_{i, t};  \clustMembSubjLagPDPM = \clustIndLagPDPM', V_i \right)  \right\} },
$
where $\clustIndLagPDPM^* = \min \{ \clustIndLagPDPM: \uiLagPDPM > 1 - \sum_{\clustIndLagPDPM'=1}^\clustIndLagPDPM \pi_{A, \clustIndLagPDPM'}, \text{for all } i \}$.%, and $\phi\left(\cdot; \clustMembSubjLagPDPM = \clustIndLagPDPM, V_i \right)$ denotes the Gaussian density function with mean and variance determined by memberships in autocovariance cluster $\clustIndLagPDPM$ and residual covariance cluster $V_i$.

Conditioned on the cluster memberships, we sample the autocovariance matrices across all lags one outcome at a time. The full conditional for $A_{\clustIndLagPDPM,  d' \bullet}$ is given by $A_{\clustIndLagPDPM,  d' \bullet} | - \sim N(\boldsymbol{\mu}_{A_\clustIndLagPDPM, d'}^*,  \boldsymbol{\Sigma}_{A_\clustIndLagPDPM,d'}^*)$ with variance and mean:
$	\boldsymbol{\Sigma}_{A_\clustIndLagPDPM,d'}^* = \left(  \sum_{i \in \mathcal{H}_\clustIndLagPDPM} \sum_{t = 1}^{T_i}   \sigma_{i,d'}^{-2} \mathbf{z}_{i,t} \mathbf{z}_{i,t}'    + \boldsymbol{\Lambda}_D^{-1} \right)^{-1}$, $
	\boldsymbol{\mu}_{A_\clustIndLagPDPM, d'}^* = \boldsymbol{\Sigma}_{A_\clustIndLagPDPM,d'}^* \left\{  \sum_{i \in \mathcal{H}_\clustIndLagPDPM} \sum_{t = 1}^{T_i} \left[  \sigma_{i,d'}^{-2} \mathbf{z}_{i,t}\left( x_{i, t}^{(d')} - \FactorLoadings_{i,d'}^* \SubjectLatentFactor_{i,t}^* \right)     \right]   \right\}$, respectively. 
%where $\mathcal{H}_\clustIndLagPDPM}$ is the set of indices s.t. $i \in \mathcal{H}_\clustIndLagPDPM}$ if subject $i$'s autocovariance matrix belongs to cluster $\clustIndLagPDPM$.

Finally, the parameters of the double exponential base measure can be updated using the approach outlined in \cite{park2008bayesian}. The variance term in the base measure can be sampled using  $\tau_{k,d'}^{-2} \sim InverseGaussian\left(  \sqrt{\frac{\lambda^2}{C \bar{A}_{k, d'}^2 }}, \lambda^2 \right),$ for $k=1, \ldots, K$ and $d' = 1, \ldots, D^2$, where $C$ is the number of clusters. The posterior distribution for the lasso parameter is a gamma distribution, $\lambda^2 | \boldsymbol{\tau}^2 \sim Gamma(KD^2 + r, \delta + \sum_{k=1}^{K} \sum_{d'=1}^{D^2} \frac{\tau_{k,d'}^2} {2})$.

\subsubsection{Computation Steps for rgPDPM-VAR}

The rgDPM-VAR requires some modification to the slice sampling approach. In particular, the sampler for the rgDPM-VAR extends the latent terms in the slice sampler along the outcome dimension. Let $\clustMembLagrgPDPM$ be the vector of autocovariance cluster indices for outcome $d'$, with $\clustMembSubjrgLagPDPM = \clustIndLagrgPDPM$ when subject $i$ belongs to outcome $d'$ cluster $\clustIndLagrgPDPM$, and let $\SetclustMembLagrgPDPM = \{i : \clustMembSubjrgLagPDPM = \clustIndLagrgPDPM \}$ be the indices of all subjects belonging to outcome $d'$ cluster $\clustIndLagrgPDPM$, with $n_{\clustIndLagrgPDPM}$ being the cardinality of this set. Then we have the following full conditionals:
$\small
	\nuLagrgPDPM | \{ \clustMembSubjrgLagPDPM \} \sim Beta\left(1 + n_{d',h},  \concParLagrgPDPM + \sum_{i=1}^N I_{(\clustMembSubjrgLagPDPM > \clustIndLagrgPDPM)} \right), \mbox{ } u_{1d',i} | \nuLagrgPDPM, \clustMembSubjrgLagPDPM \sim U(0, \piSubjClustLagrgPDPM ) $.
The cluster memberships, $\clustMembSubjrgLagPDPM$, are then sampled from a multinomial distribution with posterior probabilities $	\left( P(\clustMembSubjrgLagPDPM = 1 | -), \ldots, 	P(\clustMembSubjrgLagPDPM = \clustIndLagrgPDPM^* | -) \right)$ given by:\\
$\small 
P\left( \clustMembSubjrgLagPDPM = \clustIndLagrgPDPM | - \right) = \frac{ I_{(u_{1d',i} < \piLagrgPDPM)} \prod_{t=1}^{T_i} \phi_{\Sigma_{V_i}} \left(x_{i,t}^{(d')} - \sum_{k=1}^K A_{k, h_{1d'}} \mathbf{x}_{t-k} \right)  }{ \sum_{\clustIndLagrgPDPM'=1}^{\clustIndLagrgPDPM^*} \left\{ I_{(u_{1d',i} < \altpiLagrgPDPM)} \prod_{t=1}^{T_i} \phi_{\Sigma_{V_i}} \left(x_{i,t}^{(d')} -  \sum_{k=1}^K A_{k, \clustIndLagrgPDPM'} \mathbf{x}_{t-k} \right)  \right\} },
%\clustMembSubjrgLagPDPM = \clustIndLagrgPDPM | - &\sim \frac{ I_{u_{1d',i} < \piLagrgPDPM} \prod_{t=1}^{T_i} \phi\left(x_{i,t}^{(d')}; \clustMembSubjrgLagPDPM = \clustIndLagrgPDPM, V_i \right) }{ \sum_{\clustIndLagrgPDPM'=1}^{\clustIndLagrgPDPM^*} \left\{ I_{u_{1d',i} < \altpiLagrgPDPM} \prod_{t=1}^{T_i} \phi\left(x_{i,t}^{(d')}; \clustMembSubjrgLagPDPM = \clustIndLagrgPDPM', V_i \right)  \right\} }
$
where $\clustIndLagrgPDPM^* = \min \{ h: u_{1d',i} > 1 - \sum_{\clustIndLagrgPDPM'=1}^h \altpiLagrgPDPM, \text{for all } i \}$. Conditioned on the cluster memberships, the autocovariance terms can be updated in an identical manner to the PDPM-VAR.  %, after making appropriate modifications to the prior variance. %, and $\phi\left(\cdot; \clustMembSubjrgLagPDPM = \clustIndLagrgPDPM, V_i \right)$ denotes the Gaussian density function with mean and variance determined by memberships in autocovariance cluster $\clustIndLagrgPDPM$ and residual covariance cluster $V_i$. 
When updating the parameters of the double exponential base measure the variance terms can be sampled from inverse Gaussian distributions: $\tau_{k,d', d^*}^{-2} \sim InverseGaussian\left(  \sqrt{\frac{\lambda_{d'}^2}{C_{d'} \bar{A}_{k, d', d^*}^2 }},  \lambda_{d'}^2 \right)$  for $k=1, \ldots, K$, $d^* = 1, \ldots, D$ and $d' = 1, \ldots, D$, where $C_{d'}$ is the number of autocovariance clusters for outcome $d'$ and $\tau_{k,d', d^*}^{-2}$ is the variance term corresponding to the $d^*$th element of the $d'$th row of $A_k$, and $\bar{A}_{k, d', d^*}$ is the average of element $d^*$ of the $d'$th row of $A_k$ across the $C_{d'}$ clusters. The outcome-specific lasso parameters  have gamma posteriors: $\lambda_{d'}^2 | \boldsymbol{\tau_{k, d', d^*}}^2 \sim Ga(DK + r, \delta + \sum_{k=1}^{K} \sum_{d^*=1}^{D} \frac{\tau_{k,d',d^*}^2} {2})$ for $d'=1, \ldots, D$.

\subsubsection{Computation Steps for lgPDPM-VAR}

The sampling steps under the lgPDPM-VAR model proceeds in a similar manner as the other variants outlined in the manuscript, and are omitted here for space constraints.

\section{Additional Simulation Results}

\begin{figure}
\centering
\includegraphics[width=1\linewidth]{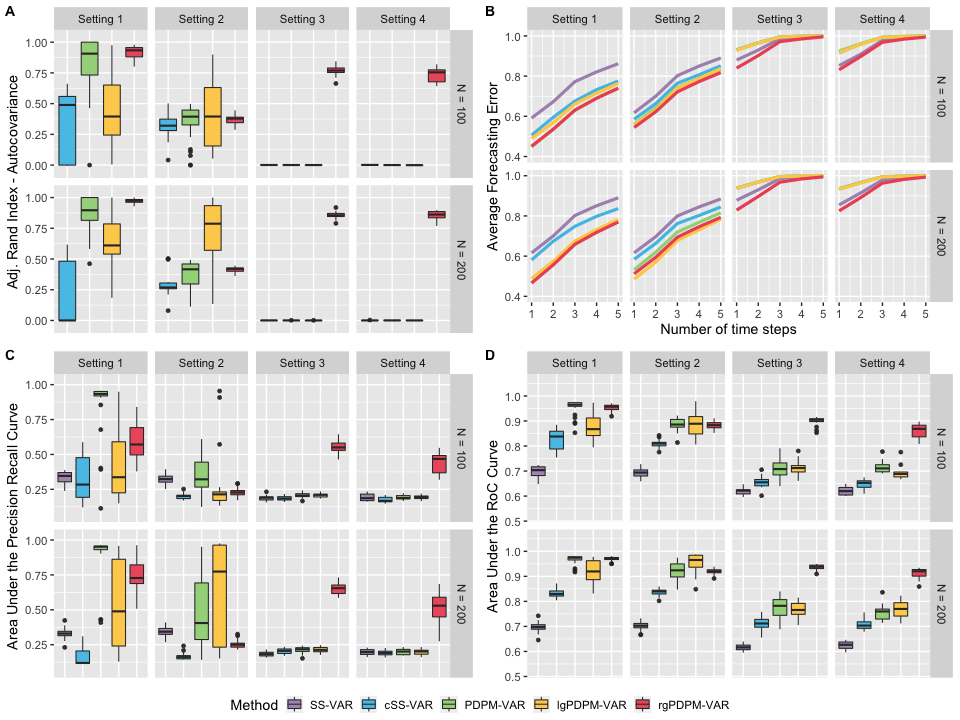}
\caption{Simulation results for the $D=100$, $T=250$ case with sparsity level $0.9$. Panel A displays the adjusted Rand index for clustering the autocovariance matrices. Panel B displays the forecasting error at 1--5 time steps. Panels C and D display the area under the PR and RoC curves for identifying non-zero elements of the autocovariance matrices.}
\label{sim:ARIPRCROC_D100_T250_Sp9}
\end{figure}

\begin{figure}
\centering
\includegraphics[width=1\linewidth]{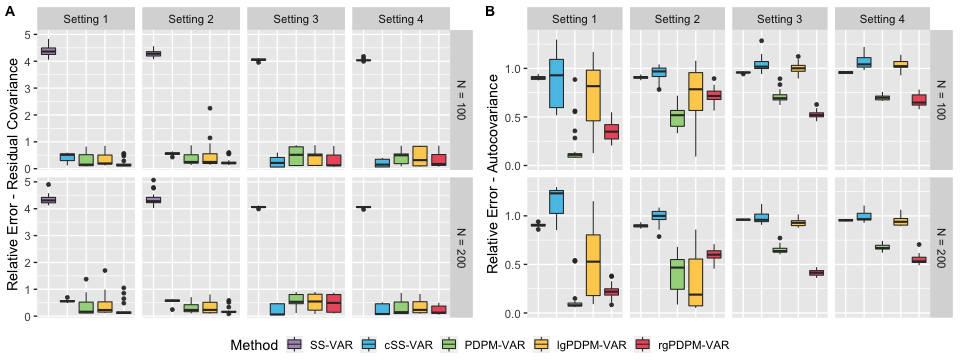}
\caption{Simulation results for the $D=100$, $T=250$ case with sparsity level $0.9$. Panel A and B display the relative L1 error for estimating the residual covariance and the subject-specific autocovariance matrices, respectively.}
\label{sim:L1_D100_T250_Sp9}
\end{figure}

%Simulation results for the $D=40$, $T=250$ case with 75\% sparsity are presented in Figures \ref{sim:ARIPRCROC_D40_T250_Sp75}--\ref{sim:L1_D40_T250_Sp75}. Results for the $D=40$, $T=250$ case with 90\% sparsity are presented in Figures \ref{sim:ARIPRCROC_D40_T250_Sp9}--\ref{sim:L1_D40_T250_Sp9}. %Results for the $D=40$, $T=250, 350$ case with 75\% sparsity are presented in Figures \ref{sim:ARIPRCROC_D40_T250350_Sp75}--\ref{sim:L1_D40_T250350_Sp75}.
%Results for the $D=40$, $T=250, 350$ case with 90\% sparsity are presented in Figures \ref{sim:ARIPRCROC_D40_T250350_Sp9}--\ref{sim:L1_D40_T250350_Sp9}.
Simulation results for the $D=100$, $T=250$ case with 75\% sparsity are presented in the main manuscript. Results for the $D=100$, $T=250$ case with 90\% sparsity are presented in Figures \ref{sim:ARIPRCROC_D100_T250_Sp9}--\ref{sim:L1_D100_T250_Sp9}. %Results for the $D=100$, $T=250, 350$ case with 75\% sparsity are presented in Figures \ref{sim:ARIPRCROC_D100_T250350_Sp75}--\ref{sim:L1_D100_T250350_Sp75}. Results for the $D=100$, $T=250, 350$ case with 90\% sparsity are presented in Figures \ref{sim:ARIPRCROC_D100_T250350_Sp9}--\ref{sim:L1_D100_T250350_Sp9}.

\end{document}